\documentclass{article} 
\usepackage{amsmath, amsthm, amssymb}
\usepackage{mathrsfs}
\usepackage{graphicx}
\usepackage{verbatim}
\usepackage{natbib}
\usepackage{booktabs}
\usepackage{threeparttable}
\usepackage{caption}
\usepackage{subcaption}
\usepackage{fancyvrb}
\usepackage{enumerate}
\usepackage{relsize}
\usepackage{multirow}
\usepackage{makecell}
\usepackage{rotating}
\usepackage{longtable}
\usepackage{lscape}
\usepackage{setspace}
\usepackage{ltablex}
\usepackage{ragged2e}
\usepackage[dvipsnames]{xcolor}
\usepackage[linesnumbered,boxed,commentsnumbered]{algorithm2e}
\usepackage[section]{placeins}
\usepackage{hyperref}
\usepackage[margin=1.4in]{geometry}
\hypersetup{colorlinks,citecolor=blue,urlcolor=blue,linkcolor=blue}

\newcommand\blfootnote[1]{%
  \begingroup
  \renewcommand\thefootnote{}\footnote{#1}%
  \addtocounter{footnote}{-1}%
  \endgroup
}

\usepackage{stefan_tex}

\newcommand{\sampleNumCountries}{34}
\newcommand{\sampleCountries}{Burundi, Benin, Burkina Faso, Bangladesh, Central African Republic, Côte d’Ivoire, Congo - Kinshasa, Colombia, Ethiopia, Ghana, Guinea-Bissau, Indonesia, India, Kenya, Liberia, Madagascar, Mexico, Mali, Malawi, Namibia, Niger, Nigeria, Pakistan, Rwanda, Sudan, Senegal, Sierra Leone, Togo, Timor-Leste, Tanzania, Uganda, Yemen, South Africa, and Zimbabwe}
\newcommand{\sampleShareWorldsPoor}{76}

\newcommand{\sampleRate}{13}
\newcommand{\sampleMinRate}{1}
\newcommand{\sampleMaxRate}{84}
\newcommand{\sampleMinRateCountry}{Mexico}
\newcommand{\sampleMaxRateCountry}{Congo - Kinshasa}
\newcommand{\sampleCostPercentGDP}{16}

\newcommand{\sampleOdaPercentGDP}{7}
\newcommand{\nationalTarget}{1}
\newcommand{\globalTarget}{1}
\newcommand{\nationalPovertyRateForGlobalTarget}{1.6}
\newcommand{\globalPovertyRateForNationalTarget}{0.7}

\newcommand{\headlineGapNationalTarget}{211}
\newcommand{\headlineGapNationalTargetCILower}{207}
\newcommand{\headlineGapNationalTargetCIUpper}{215}
\newcommand{\headlineOracleNationalTarget}{52}

\newcommand{\headlineGapUBIPercentNationalTarget}{19}
\newcommand{\headlineGapUBIVariablePercentNationalTarget}{57}
\newcommand{\headlineGapOracleRatioNationalTarget}{4.0}
\newcommand{\headlineBinaryContPercentIncreaseNationalTarget}{39}

\newcommand{\headlineGapNewPovertyLineNationalTarget}{347}
\newcommand{\percentIncreaseContGapNewOldPovertyLineNationalTarget}{64}
\newcommand{\extrapolationWBCost}{304}

\newcommand{\extrapolationFutureCostPercentDecrease}{17}
\newcommand{\extrapolationWPCFutureCostPercentGDP}{0.23}
\newcommand{\extrapolationWPCFutureCostPercentOECDGDP}{0.37}
\newcommand{\extrapolationWBRSquared}{0.66}

\newcommand{\extrapolationWBOECDGDPPercent}{0.47}
\newcommand{\extrapolationWPCOECDGDPPercent}{0.45}

\newcommand{\extrapolationWBOECDPlusChinaGDPPercent}{0.37}

\newcommand{\extrapolationWBDroppedCountriesGap}{Channel Islands, Cuba, Isle Of Man, Liechtenstein, Monaco, Korea, Dem. People's Rep., and Taiwan}

\newcommand{\extrapolationWBGlobalGDP}{0.28}
\newcommand{\extrapolationWPCGlobalGDP}{0.27}
\newcommand{\extrapolationWBQuadraticGlobalGDP}{0.29}
\newcommand{\oracleWBGlobalGDP}{0.08}

\newcommand{\togoBinaryGapTransferAmount}{1.59}
\newcommand{\oracleFeasibleRatioMin}{1.3}
\newcommand{\oracleFeasibleRatioMax}{14.7}
\newcommand{\togoCovariateDimension}{160}
\newcommand{\togoSampleSize}{6171}
\newcommand{\minDimension}{40}
\newcommand{\maxDimension}{420}
\newcommand{\refugeeGlobalGDP}{0.04}
\newcommand{\refugeePlusExtrapolationGlobalGDP}{0.32}
\newcommand{\refugeeDroppedCountries}{Cuba, Hong Kong SAR China, Liechtenstein, Sint Maarten (Dutch part), Cayman Islands, and Monaco}
\newcommand{\percentDecreaseSampleSize}{1}
\newcommand{\percentDecreaseGaptoWelfare}{1}
\newcommand{\percentIncreaseGeotoSurvey}{45}
\newcommand{\percentIncreaseSatelliteSurvey}{18}
\newcommand{\percentDecreaseSatelliteUBI}{24}
\newcommand{\percentDecreaseSatelliteCombined}{6}

\newcommand{\CODAbsDiscrepancy}{5.2}
\newcommand{\CODDiscrepancyDirection}{higher}
\newcommand{\BDIAbsDiscrepancy}{4.7}
\newcommand{\BDIDiscrepancyDirection}{higher}
\newcommand{\ZAFAbsDiscrepancy}{2.9}
\newcommand{\ZAFDiscrepancyDirection}{lower}
\newcommand{\CAFAbsDiscrepancy}{1.6}
\newcommand{\CAFDiscrepancyDirection}{lower}
\newcommand{\differentSurveysMeanAbsDiscrepancy}{2.4}
\newcommand{\differentSurveysMeanDiscrepancy}{0.1}

\newcommand{\lsmsPerHhCost}{416}
\newcommand{\screeningPerHhCost}{7.93}
\newcommand{\alphaEarthAdminSavingsMillions}{16}
\newcommand{\surveyTransferSavings}{272}

\theoremstyle{plain}
\newtheorem{prop}{Proposition}

\newtheorem{coro}[prop]{Corollary}
\newtheorem{lemm}[prop]{Lemma}
\newtheorem{theo}[prop]{Theorem}

\theoremstyle{definition}
\newtheorem{exam}{Example}

\newtheorem{defi}{Definition}
\newtheorem{assumption}{Assumption}

\theoremstyle{remark}

\counterwithin{subexam}{exam}

\author{
Roshni Sahoo$^{*}$ \and Joshua Blumenstock$^{\dagger}$ \and Paul Niehaus$^{\ddagger}$ \and Leo Selker$^{\dagger}$ \and Stefan Wager$^{\mathsection}$
}

\title{What Would it Cost to End Extreme Poverty? \blfootnote{Corresponding author (email \texttt{r.sahoo@columbia.edu}), Columbia Business School, $^{\dagger}$School of Information, UC Berkeley, $^{\ddagger}$Department of Economics, UC San Diego, $^{\mathsection}$Graduate School of Business, Stanford University. We are grateful for helpful comments from seminar participants at the BREAD/NBER Program Meeting, Brown University Bravo/SNSF Workshop, Cornell University Young Researchers Workshop, Duke University, UChicago AI in Social Science Conference, UChicago Machine Learning in Economics Summer Conference, Harvard Data Science Institute, University of Maryland Workshop on AI and Analytics for Social Good, INFORMS, Stanford Econometrics, Stanford Statistics, and Yale University. We thank Apollinaire Abi, Alia Aghajanian, Vincent Armentano, Anik Ashraf, Oscar Barriga Cabanillas, Lina Cardona-Sosa, S. Chandrasekhar, Grown Chirongwe, Oeindrilla Dube, Elizabeth Foster, Tanzed Hossain, Roy Katayama, Patricia Koka, Arsène Koyangbo, Daniel Mahler, Juan Manuel Monroy Barragan, Jose Mpesayabo, Abhiroop Mukhopadhyay, Fabrice Mukendi Mutombo, Nicolas Ndayishimiye, Noé Nduwabike, Iary Rakotondradany, Khaled Saad, Nozipho Shabalala, Dhiraj Sharma, David Stifel, Jonathan Weigel, and Christina Weiser for their generous assistance accessing and interpreting data; Puneet Bhaskar, Qiqi Chen, Nolan Chu, Michael Cleary, Karl Gorski, Neela Kaushik, Henry Lopez Rodriguez, Matthewos Mesfin, Francesca Vescia, and Victor Zhenyi Wang for help with data preparation; and Aditi Acharya, Erick Rosas Lopez, and Shruthi Ramesh for excellent research assistance. This work was partially supported by NSF grant SES-2242876, a Stanford BGS Research Fund grant, and by gifts from Luke Ding and from the Manglaben Chimanbhai Patel Family Foundation. RS was supported by a Stanford Data Science Fellowship and a Stanford DARE (Diversifying Academia, Recruiting Excellence) Fellowship.}}

\begin{document}
\maketitle
\begin{abstract}

We study poverty minimization via direct transfers, framing this as a statistical learning problem while retaining the information constraints faced by real-world programs. Using nationally representative household consumption surveys from \sampleNumCountries \ countries that together account for \sampleShareWorldsPoor\% of the world's poor, we estimate that reducing the poverty rate to \nationalTarget \% (from a baseline of \sampleRate\%) would cost \$\headlineGapNationalTarget B nominal per year. This is \headlineGapOracleRatioNationalTarget \ times the corresponding reduction in the aggregate poverty gap, but only \headlineGapUBIPercentNationalTarget\% of the cost of universal basic income. Extrapolated globally, the results imply a cost of \extrapolationWBGlobalGDP\% of global GDP to (approximately) end extreme poverty.

\end{abstract}

\section{Introduction}

The share of the world's population living in extreme poverty---defined as living on less than \$2.15 in 2017 PPP dollars per day\footnote{Throughout the paper we refer to extreme poverty as having a material standard of living below some given threshold, a definition which may differ from local communities' concepts of poverty \citep{Alatasetal2012targeting}. We focus primarily on the traditional threshold of \$2.15 in 2017 PPP dollars, but also report results for the newer threshold of \$3.00 in 2021 PPP dollars \citep{Fosteretal2025poverty}.}---has declined dramatically, from an estimated 41\% in 1981 to 8\% in 2024.\footnote{World Bank Poverty \& Inequality Platform, \url{https://pip.worldbank.org/poverty-calculator}, accessed 7 November 2025. See also \citet{ArmentanoNiehausVogl2025poverty} for a more general discussion about the global decline in poverty.} This progress has historically taken place in conjunction with economic growth \citep{Pritchett2024growth}. Yet it may not continue: analyzing current growth trends, \citet{Roser2025progress} concludes that ``progress against extreme poverty will come to a halt.'' In either case, the fact that hundreds of millions of people still live below very low consumption thresholds raises an immediate policy question: could extreme poverty be eliminated more rapidly through redistribution? 

Recent proposals explicitly appeal to this possibility \citep[e.g., the wealth transfers envisioned by][]{kharas_end_2026}. And back-of-the-envelope calculations raise hope that the sums involved might be manageable: If policymakers could identify every household below the poverty line and transfer exactly the income needed to reach it, the total annual cost would equal the global poverty gap, estimated as of 2023 to be just \oracleWBGlobalGDP\% of global GDP.\footnote{Authors' calculations using countries'  2023 poverty gap indices, total population in 2023, 2017 market exchange rate, and PPP conversion factor for all countries in the World Bank PIP, and an estimate of 2023 global GDP of \$107T from \url{https://data.worldbank.org/indicator/NY.GDP.MKTP.CD}.} 

However, such transfers are not feasible. Poverty gap estimates rely on household surveys that typically cover only 0.001\% to 0.01\% of a country's population. Thus, for almost all households in low- and middle-income countries, there are no direct estimates of their standard of living. Collecting that information through regular surveys would be prohibitively expensive \citep{kilic2017costing} and, if used to determine transfer eligibility, would create incentives for households to misreport hard-to-verify quantities like consumption \citep{martinelli_deception_2009,Banerjeeetal2020distortionary}. These cost and incentive considerations help explain why many existing programs instead determine eligibility using proxies for living standards that are cheaper to measure and verify \citep{hanna2018universal}. For example, in the common Proxy Means Test (PMT) approach, administrators use a small amount of survey data to identify a set of proxies that are jointly predictive of consumption (such as housing materials) and then collect those proxies for every household to determine eligibility in the full population \citep{grosh1995proxy}. 

\paragraph{Learning to Target Income Transfers} In this paper, we maintain this same information environment and ask how to design feasible policies that would end (up to a given tolerance) extreme poverty at the lowest possible cost. We formulate poverty targeting as a statistical learning problem, building on recent work that proposes to learn decision rules via loss minimization \citep{athey2021policy,bertsimas2020predictive,kitagawa2018should}. The policymaker learns a transfer policy that maps a household's observable characteristics to a nonnegative transfer amount, minimizing expected loss subject to a budget constraint (or, equivalently, minimizing cost subject to a constraint on loss). We consider both unrestricted transfer policies, which allow each household to receive any nonnegative transfer, and binary policies, in which each household receives either zero or some common positive amount. The domain of the loss function is the household's post-transfer standard of living.\footnote{To be precise, and as we discuss below, this is the standard of living the household \emph{could} attain if it spent the entire transfer on consumption. We abstract from dynamics, but consider households that save or invest a portion of their transfer as revealing that they are at least weakly better-off by doing so. One can thus interpret the exercise as learning policies that raise welfare to at least the level that ending today’s consumption poverty would achieve.}

Choosing the right loss function is a central concern. While the headcount poverty rate is the most easily understood and most widely used metric in public discourse about poverty,%
\footnote{For example, Target 1.1 of the Sustainable Development Goals is ``By 2030, eradicate extreme poverty for all people everywhere, currently measured as people living on less than \$1.25 a day" and the corresponding indicator 1.1.1 is the ``Proportion of the population living below the international poverty line by sex, age, employment status and geographic location (urban/rural)" \citep{statistics2019global}.} 
it is well known that, with full information, minimizing the poverty rate directly can generate inequitable allocations  \citep{atkinson1987measurement, bourguignon1990poverty, sen1976poverty}. Intuitively, the most efficient way to reduce the poverty rate would be to allocate transfers to households just below the poverty line, rather than to the poorest households. We show that this undesirable property generalizes to the imperfect-information case. Minimizing the poverty gap, on the other hand, is equitable in the sense that the resulting policies direct (weakly) greater support to poorer households. Gap minimization also delivers formal guarantees on poverty rate reduction: it is equivalent to minimizing the worst-case (across population subgroups) conditional poverty rate. We therefore emphasize the performance of gap-minimizing policies, while also reporting rate-minimizing results for comparison.

\paragraph{The Cost of Targeted Transfers} We first characterize solutions to the gap- and rate-minimization problems analytically in order to (a) restrict the class of transfer policies that must be considered, and (b) identify specific functionals of the population distribution that must be learned. Gap minimization requires learning conditional quantiles of the distribution of living standards, letting us use modern approaches based on neural networks and large predictor sets, while rate minimization requires learning conditional densities, which require a bespoke semi-parametric approach and smaller predictor sets. We use these results to construct computationally feasible algorithms for implementing both approaches.

We then quantify the aggregate cost of reducing poverty in a sample of \sampleNumCountries \, countries, including essentially all of those which (i) have an extreme poverty rate above 10\% or account for more than 1\% of the world's extreme poor, and for which (ii) there exists a recent, publicly-available, high-quality, nationally representative living standards survey that includes a consumption aggregate and is large enough to cover at least 1,000 poor households. The first condition focuses the analysis on the set of countries likely to influence aggregate costs the most; the second increases the reliability of our living standards measures and helps ensure that we observe enough poor households for effective statistical learning.\footnote{The resulting full list of countries covered is \sampleCountries. See Section \ref{sec:data-sources-uses} for further discussion and Appendix \ref{sec:data} for details of the data sources.} Each could be relaxed in the future. Collectively, this sample accounts for \sampleShareWorldsPoor \% of the world's extreme poor.\footnote{The estimate of the share of the world's poor that our sample accounts for is based on national poverty rates and populations for all countries from the World Bank PIP, accessed 9 April 2026. National poverty rates in the surveys are generally consistent with poverty rate estimates from the World Bank in the corresponding survey year.}

We learn policies that target transfers to households based on observable characteristics that have been used by, or are analogous to others that have been used by, existing real-world PMTs. We omit characteristics that we consider too difficult to verify in practice (e.g., income) or likely to be politically controversial (e.g., ethnicity). That said, one should think of the exercise as characterizing what is technically feasible; if policies need to satisfy additional criteria in order to be feasible politically, this might affect their costs.

We estimate that reducing the \$2.15 2017 PPP poverty rate to \nationalTarget\% in all countries in our sample, using a gap-minimizing (and hence weakly equitable) policy, would cost \$\headlineGapNationalTarget B nominal per year. This is substantially lower than the cost of blunter, uniform policies: it is \headlineGapUBIPercentNationalTarget\% of the cost of providing a Universal Basic Income at the level of the poverty line, and \headlineGapUBIVariablePercentNationalTarget\% of the cost of achieving the same poverty rate using supplemental income at amounts that are universal within countries but vary across them (``Universal Supplemental Income''). These comparisons highlight the benefit of targeting with observable proxies of consumption over not targeting at all. On the other hand, the cost of the feasible gap-minimizing policy is \headlineGapOracleRatioNationalTarget \ times the achieved reduction in the poverty gap, or equivalently \headlineGapOracleRatioNationalTarget \ times the cost of achieving the same gap reduction if all households' living standards were perfectly observable (the ``oracle'' scenario). This comparison quantifies the relative cost of targeting with proxies versus perfect information. Both ratios vary substantially across countries: in countries with low poverty rates the benefits of targeting are very high, for instance, while in some of the poorest countries universal transfers cost little more than the optimal targeted ones.

Real-world transfer policies are often simpler than the optimal gap-minimizing one. If we restrict attention to binary policies, the cost of reducing poverty rates to \nationalTarget\% increases by \headlineBinaryContPercentIncreaseNationalTarget \%. If we allocate a binary transfer using a PMT rather than loss minimization, but continue to learn the optimal size of that transfer from the data, we achieve similar performance. If we use a transfer sized like those used in practice, on the other hand, we cannot reduce poverty very far, as current transfer amounts are too small to raise more than a few households out of poverty. Overall, we conclude that reducing poverty using simpler transfer structures is feasible, but only if transfers are large enough, and at meaningful added cost. 

A more fundamental concern is that minimizing poverty need not maximize welfare. Restricting to equitable, gap-minimizing policies may help here to some extent. But any poverty line is still inherently arbitrary, and gap-minimization does not value gains to households even just slightly above the line. In practice, however, we find that gap-minimizing policies are cost-effective both at reducing the poverty rate, and at increasing welfare. Specifically, policies learned to directly minimize the rate are weakly \emph{more} expensive, for a given reduction in the rate, than policies learned to minimize the gap. And gap-minimizing policies increase mean log consumption by only \percentDecreaseGaptoWelfare\% less than policies learned to directly maximize it. Gap-minimizing policies thus allow a planner to approximately maximize welfare, without sacrificing performance on easy-to-communicate goals like headcount poverty.

We examine comparative statics with respect to several additional parameters of the problem. Defining poverty using the new \$3.00 2021 PPP poverty line, rather than the \$2.15 2017 PPP line, raises costs by \percentIncreaseContGapNewOldPovertyLineNationalTarget \%, reflecting the real shift in objectives this definition implied \citep{Fosteretal2025poverty}. Restricting to purely geographic targeting, which simplifies data requirements and reduces scope for gaming, increases costs by \percentIncreaseGeotoSurvey \%. Alternatively, replacing household survey data with remote-sensed satellite imagery in one country (Togo) for which we can geo-match the two increases costs by \percentIncreaseSatelliteSurvey \%. And reducing the amount of training data available to the algorithm by 10\% increases costs by \percentDecreaseSampleSize\%---a modest elasticity, but still large enough to suggest that collecting \emph{more} data might have high returns, since surveys cost far less than the transfer policies themselves.

We then discuss financing and its implications. For our estimates to be valid, financing for the policies we learn would need to be incremental, not diverted from existing programming that separately affects living standards. A meaningful share of that incremental financing could plausibly come, in many countries, from domestic revenue. But most countries would likely also require international transfers. International transfers at such scale could in turn have substantial macroeconomic effects; we briefly review some pertinent recent evidence.\footnote{Recent work on the general equilibrium effects of cash transfers, for example, has documented substantial expansionary effects \citep{Eggeretal2022ge,GerardNaritomiSilva2024cash}, which we necessarily abstract from here.} We discuss what scale of wealth transfers, as proposed by \citet{kharas_end_2026}, might be required to achieve income gains equivalent to the income transfers we study here. We also discuss the potential administrative costs of implementing policies like those we learn; while non-trivial, these would likely be orders of magnitude smaller than the costs of the transfers themselves.

Finally, we ask what the results suggest about the cost of (approximately) ending poverty globally, using the results from the \sampleNumCountries \ countries where we have data to predict the cost of ending poverty in countries where we do not. This extrapolation implies that reducing the global poverty rate to \globalTarget\% would cost \$\extrapolationWBCost B per year: \extrapolationWBGlobalGDP\% of global GDP, or \extrapolationWBOECDGDPPercent\% of OECD GDP.\footnote{Note that achieving a \globalTarget\% global rate requires reducing country-specific poverty rates to \nationalPovertyRateForGlobalTarget \%, since some countries have essentially no extremely poor people.}

\paragraph{Related Work} Our work sits at the intersection of two long-standing traditions of thought about economic development. The first begins with an ambitious development goal and then seeks to calculate the resources required to reach it. From the 1960s onwards, for example, analysts at development banks used a ``two-gap'' framework based on the Harrod-Domar constant-returns growth model to calculate the capital investment required to achieve a target rate of economic growth \citep{Domar1946capital,Easterly1999ghost,Harrod1939dynamic}. In the early 2000s, as attention shifted to poverty reduction and a broader set of social objectives, the UN Millennium Project estimated that its proposed Millennium Development Goals (MDGs) could be achieved if rich countries increased foreign aid to 0.54\% of their GNI in 2015, from 0.25\% in 2003 \citep{Sachs2005investing}.

The analyses most closely related to ours have focused specifically on the cost of achieving the MDG for extreme poverty. These studies calculated the growth in GDP per capita required to achieve a given poverty rate (assuming that the relative distribution of income stayed fixed) and then inferred the aid needed to generate that growth, using either a ``two-gap'' model \citep[e.g.,][]{DevarajanMillerSwanson2002goals} or coefficients from cross-country aid-on-growth regressions \citep[e.g.,][]{AndersonWaddington2007aid}. As the authors took care to point out, the causal assumptions in these analyses were heroic, and indeed in some cases empirically rejected.\footnote{For example, \citet{DevarajanMillerSwanson2002goals} used the ``two-gap’’ approach despite the fact that it was soundly rejected by the data \citep{Easterly1999ghost} and had been widely criticized, including by one of the authors. \citet{AndersonWaddington2007aid} used coefficients from cross-country growth regressions despite the fact that many would have agreed with \citet{Mankiw1995growth} even a decade earlier that ``using these regressions to decide how to foster growth is ... most likely a hopeless task.’’} This was necessarily, in the words of \citet{DevarajanMillerSwanson2002goals}, ``a highly speculative exercise.'' Our approach, in contrast, is built on the more mechanical idea that giving someone \$1 increases their disposable income by \$1. It is in this sense less speculative (though it may for that very reason yield conservative estimates, if other strategies that reduce poverty through less direct causal mechanisms are also cheaper).

The second literature concerns the optimal design of transfer programs. A natural point of departure is the body of older theoretical work on poverty measurement that assumed full knowledge of the income or consumption distribution \citep[e.g.,][]{foster1984class,greer1986methodology,sen1976poverty}. A logical next question would then be how to extend these ideas to environments where policymakers observe only limited information about households. Along these lines, \citet{kanbur1987measurement} and \citet{ravallion1989targeted} discussed theoretical properties of poverty-minimizing transfer policies under specific, stylized information structures,\footnote{Specifically, they assumed that the population consists of pre-defined, mutually-exclusive groups with known income distributions, and that no further predictors are available.} and \citet{glewwe1992targeting}---in perhaps the closest antecedent to our work---proposed and illustrated learning a transfer policy as a function of a (small) set of observable covariates via the plug-in principle, i.e., by minimizing a given poverty metric in-sample. In practice, however, most subsequent work has focused on the more straightforward problem of classifying households as poor or non-poor to determine their eligibility for a fixed transfer. Common methods for doing so include proxy means testing \citep{Alatasetal2012targeting,brown2018poor,grosh1995proxy,noriega2020algorithmic}, including recent approaches that use non-traditional data \citep{aiken2022machine,smythe_geographic_2022}; community-based targeting \citep{Alatasetal2012targeting,coady2004targeting}; and geographic targeting \citep{baker1994poverty,BigmanFofack2000geographical}. Our approach differs in that we seek to learn a policy that directly minimizes a poverty criterion subject to resource constraints, rather than predicting a binary eligibility outcome. With modern computational tools and high-dimensional statistical learning methods, revisiting \citeauthor{glewwe1992targeting}'s (\citeyear{glewwe1992targeting}) formulation is now feasible.

The methods we develop also contribute to a broader literature on data-driven decision making \citep{athey2021policy,bertsimas2020predictive, bhattacharya2012inferring,kallus2021minimax,kitagawa2018should, manski2004statistical}. The problems we study are ``prediction-policy problems'' in the sense of \citet{kleinberg2015prediction}, and do not require estimates of causal effects, as in recent work on learning treatment assignment policies \citep[e.g.,][]{bhattacharya2012inferring,huang2020estimating,luedtke2016optimal, sun2021empirical,sun2021treatment, sverdrup2023qini, wang2018learning}, though one methodological parallel is that it can be useful to reformulate budget-constrained treatment assignment problems as (fractional) knapsack-style problems \citep{sun2021treatment,sverdrup2023qini} as we do here. \citet{bjorkegren_machine_2022} and \citet{haushofer_targeting_2022} are particularly relevant, as they study welfare-maximizing allocations of fixed-size cash transfers in the presence of treatment effect heterogeneity.\footnote{For an overview of empirical evidence on the effects of cash transfers, see \citet{bastagli_impact_2019} and \citet{crosta_unconditional_2024}.} Future work might combine methods for learning heterogeneous effects with approaches like ours to learn dynamic transfer policies that efficiently minimize poverty over time, taking advantage of the fact that some recipients have predictably higher investment returns (i.e., are more likely to ``graduate’’) in response to a large initial transfer.

\section{Targeting Transfers via Statistical Learning}
\label{sec:setup}

Let $\mathcal{X} :=\mathbb{R}^{d}$ denote the space of observable characteristics and let $\mathcal{Y}:= \mathbb{R}_{+}$ denote the space of consumption values. A country has a population distribution $F$ over these variables; while the results in this section will involve this population distribution, they anticipate the idea that we will be able to draw representative samples from it for learning.

Cash transfer policies are common in low- and middle-income countries and, due to the cost and incentive-compatibility considerations noted above, are typically allocated based on imperfect proxies $X_{i} \neq Y_{i}$ of living standards. Given an available (per-unit) budget $B \in \mathbb{R}_{+}$, define a transfer policy as a mapping from the space of observable characteristics $\mathcal{X}$ and possible budgets $\mathbb{R}_{+}$ to a nonnegative cash-transfer amount, i.e. $t: \mathcal{X} \times \mathbb{R}_{+} \rightarrow \mathbb{R}_{+}$. This must satisfy the budget constraint $\EE[F]{t(X; B)} \leq B$.

We define a unit's post-transfer standard of living as $Y_i + t(X_i; B)$. In effect this means analyzing the consumption the unit \emph{could} attain if they consumed their entire transfer and if $Y_i$ were unaffected by transfers. This formulation is standard in the literature on optimal transfer policies \citep{bourguignon1990poverty, glewwe1992targeting, kanbur1987measurement, ravallion1989targeted} and we view it as conservative in the sense that units who save or invest a portion of their transfer are presumed at least weakly better off by doing so. In practice, reviews by \citet{Banerjeeetal2017debunking} and \citet{crosta_unconditional_2024} find that transfers tend to increase labor supply and earnings from other sources.\footnote{There is also a distinct question whether the \emph{rule} that determines eligibility for transfers might reduce $Y_i$ by disincentivizing effort. Evidence on this point from low income countries is sparse, though \citet{Banerjeeetal2020distortionary} find little evidence that eligibility rules distorted consumption choices in Indonesia.} Note also that this formulation precludes general equilibrium effects; if transfers stimulate the economy, as \citet{Eggeretal2022ge} and \citet{GerardNaritomiSilva2024cash} find, then our approach is conservative in this regard.

Given this setting, there are many possible ways to learn transfer policies from data. As a point of reference we first outline a widely-used approach for allocating cash transfers based on a proxy means test (PMT) \citep{Alatasetal2012targeting,brown2018poor,grosh1995proxy,hanna2018universal}. 

\begin{defi}[Proxy Means Test]
\label{defi:pmt}
In a Proxy Means Test (PMT), observable characteristics, or ``proxies,'' $X_{i}$ are used to define the conditional mean consumption $\mu(x):= \EE[F]{Y_i \mid X_i=x}$. The conditional mean consumption $\mu$ is then thresholded to determine the allocation of transfers to units in the population, i.e.
\begin{equation}
\label{eq:pmt}
t_{\mathrm{PMT}}(x; B) := \begin{cases} \bar{t} & \mu(x) \leq \eta(\bar{t}, B),\, \\ 0 & \mu(x) > \eta(\bar{t}, B),\, \end{cases}
\end{equation}
for some transfer size $\bar{t} \in \mathbb{R}_{+}$ selected by the policymaker (but not necessarily learned from data) and eligibility threshold $\eta(\bar{t}, B)$ chosen so that $\PP[F]{\mu(X) \leq \eta(\bar{t}, B)} = B/ \bar{t}$.
\end{defi}
\noindent Note that the PMT policy \eqref{eq:pmt} is a binary policy: each unit can receive zero or some common amount $\bar{t}.$

The PMT has a connection to \emph{statistical} loss minimization in the sense that $\mu(\cdot)$ is the function that minimizes $\EE[F]{(Y_{i} - \hat{\mu}(X_{i}))^{2}}$, the mean-squared error between $Y_{i}$ and a predictor $\hat{\mu}(X_{i}).$ But this does not imply that the transfer policy it yields minimizes any particular post-transfer loss. More generally, there is no obvious formal link between this procedure and minimization of any \emph{welfare} loss. Given this, it may be possible to improve on a PMT's poverty-reducing performance by explicitly learning policies via minimization of a loss function, chosen to reflect the ultimate poverty reduction goal.

\begin{defi}[Unrestricted Policies via Loss Minimization]
\label{defi:loss_min}
For any loss function $L: \mathbb{R} \rightarrow \mathbb{R},$ an optimal unrestricted transfer policy is one which minimizes the expected post-transfer loss $\EE[F]{L(Y_i + t(X_i))}$ subject to a budget constraint $B$, i.e. which solves
\begin{equation} 
\label{eq:prob} 
\min_{t(\cdot; B): \mathcal{X} \rightarrow \mathbb{R}_{+}} \left\{ \EE[F]{L(Y_i + t(X_i; B))}  : \EE[F]{t(X_i; B)} \leq B \right\}. 
\end{equation}
\end{defi}
Given data $(X_{i}, Y_{i}) \sim F$ i.i.d. for units $i=1, 2, \dots n$ from a representative survey, a policymaker can estimate the solution of \eqref{eq:prob} via empirical minimization of the loss.

The transfer policies obtained in Definition \ref{defi:loss_min} are unrestricted in that they can allocate transfers of any nonnegative amount. In practice many current programs allocate transfers in only a few different sizes, and sometimes only two: zero, and some fixed positive amount. To understand the impact on costs of restricting the support of transfers, we also study the problem of learning such ``binary'' policies.

\begin{defi}[Binary Policies via Loss Minimization]
\label{defi:binary}
For any loss function $L: \mathbb{R} \rightarrow \mathbb{R}$, an optimal binary transfer policy is one which minimizes the expected loss $\EE[F]{L(Y_{i} + t(X_{i}))}$ subject to a budget constraint $B$ and takes on only one non-zero value, i.e. which solves
\begin{equation}
 \label{eq:binary_prob} \min_{\substack{t(\cdot; B): \mathcal{X} \rightarrow \mathbb{R}_{+},\\ \bar{t} \in \mathbb{R}_{+}}} \left\{ \EE[F]{L(Y_i + t(X_i; B))}  : \EE[F]{t(X_i; B)} \leq B,\, \quad  t(x) \in \{0, \bar{t}\right\} \quad \forall x \in \mathcal{X}\}.
 \end{equation}
 Note that this approach requires learning both the optimal transfer size $\bar{t}$ and the transfer policy $t$, which determines who is eligible for the transfer.
\end{defi}

Loss minimization as in Definition \ref{defi:binary}, and Proxy Means Testing as in Definition \ref{defi:pmt}, both yield binary policies; the key difference is that the loss-minimization approach directly minimizes a post-transfer poverty measure. The cash-transfer policy that solves the loss minimization problem \eqref{eq:binary_prob} for a particular post-transfer poverty measure yields the greatest possible reduction in that measure, while the PMT policy does not provide formal poverty reduction guarantees. 

The appropriate loss function depends on one's definition of poverty. This is a classic topic in economics, with seminal contributions from \citet{sen1976poverty},  \citet{foster1984class}, and \citet{greer1986methodology}, among others. One important idea in this literature was that a measure of poverty should be sensitive to changes in the living standards of people below a given poverty line. Nevertheless, the poverty measure that is most easily understood and widely used in public discourse is one that lacks this feature: the poverty rate.

\begin{exam}[Poverty Rate]
\label{exam:rate}
Given a poverty line $c \in \mathbb{R}_{+}$, the pre-transfer poverty rate is the fraction of units in the population whose per-capita consumption falls below the threshold, i.e. $\PP[F]{Y_{i} < c}$. The post-transfer poverty rate is $\PP[F]{Y_i + t(X_i) < c}$.
\end{exam}

We can also consider minimizing other measures such as the poverty gap index, which is sensitive to changes below the poverty line. 

\begin{exam}[Poverty Gap Index]
\label{exam:gap}
Given a poverty line $c \in \mathbb{R}_{+}$, the pre-transfer poverty gap index is the average shortfall from the poverty line relative to the poverty line, i.e. $\EE[F]{(c - Y_i)_{+}/c}$. The post-transfer poverty gap index is, analogously, $\EE[F]{(c  - Y_i - t(X_{i}))_{+}/c}$.
\end{exam}

Both the poverty rate and the poverty gap index are special cases of the class of FGT-$\alpha$ indices studied by \citet{foster1984class} and \citet{greer1986methodology}, defined as follows.
\begin{exam}[FGT Index]
\label{exam:FGT}
Given a poverty line $c \in \mathbb{R}_{+}$ and any $\alpha \geq 0$, the pre-transfer FGT-$\alpha$ index is $\EE[F]{\mathbb{I}(Y_i < c) \cdot (c - Y_i)^{\alpha}}$. The corresponding post-transfer index is 
\[\EE[F]{\mathbb{I}(Y_i + t(X_i) < c) \cdot (c - Y_i - t(X_i))^{\alpha}}.\]
\end{exam}
The FGT-0 index corresponds to the poverty rate and the FGT-1 index corresponds to the poverty gap. 


\subsection{Weakly Equitable Policies}
Due to the emphasis on the poverty rate in public discourse, it may be tempting to learn transfer policies via minimization of the empirical poverty rate as in Example \ref{exam:rate}, as this approach directly seeks the greatest reduction in the poverty rate under a limited budget. However, in the setting where the policymaker has perfect information on consumption, i.e., $X_i = Y_i$, it is well known that rate minimization yields a policy family that prioritizes transfers to the ``richest poor'' \citep{bourguignon1990poverty}.\footnote{In particular, and under the additional condition that the budget is not large enough to lift all units above the poverty line, i.e. $0 < B < \EE[F]{(c - Y_i)_{+}}$, and $Y_{i}$ has positive support under $F$, the rate-minimizing policy is
\begin{equation}
\label{eq:oracle_rate_policy_family}
t(y; B) = (c - y)_{+} \cdot \mathbb{I}(\eta(B) < y < c),
\end{equation}
where $\eta(B) \in (0, c)$ and $\eta(B)$ is decreasing in $B$. Clearly, transfers are not allocated to the worst-off units.} This is because the post-transfer poverty rate can be minimized more efficiently by raising those very close to the poverty line over it than by raising those who are further away. To rule this out we introduce the following notion of equity, which prioritizes transfers to the worst-off units.
\begin{defi}[Weakly Equitable Policy]
\label{defi:weakly_equitable}
A family of policies $t(x; B)$ is weakly equitable with respect to a population distribution $F$ if incremental transfers are monotone decreasing in post-transfer consumption, i.e. let $x, x' \in \mathcal{X}$ and budgets $B, B'$ such that $0 \leq B' \leq B$, then
	\begin{equation} 
	\label{eq:inc_transfer}
	F_{Y + t(X; B') \mid X=x} \preceq_{\text{SD}} F_{Y + t(X; B') \mid X=x'} \implies t(x; B) - t(x; B') \geq t(x'; B) - t(x'; B').
	\end{equation}
\end{defi}

By Definition \ref{defi:weakly_equitable}, a weakly equitable policy family allocates weakly larger funds to the poorer units compared to richer units in terms of post-transfer consumption. This property generalizes the requirement that transfers be monotone decreasing in (pre-transfer) consumption.\footnote{To see this, consider \eqref{eq:inc_transfer} with $B  > 0$ and $B' = 0$; then
\begin{equation}
\label{eq:simple_version_consumption}
F_{Y \mid X=x} \preceq_{\text{SD}} F_{Y \mid X=x'} \implies t(x; B) \geq t(x'; B).
\end{equation}
} In the full information setting, this property ensures both that larger transfers are provided to poorer units and that transfers to poorer units are not so large that the post-transfer consumption of poorer units exceeds that of richer units who receive smaller transfers.\footnote{In the full-information setting, we show that if a policy family $t(y; B)$ is continuous in budget and weakly equitable, then post-transfer consumption is monotone increasing in pre-transfer consumption. This result is stated formally in Appendix \ref{app:oracle} and proven in Appendix \ref{sec:proof_rank_preservation}.}

Clearly, rate minimization is not per se weakly equitable in the perfect information regime, i.e. when $X_{i} =Y_{i}$. This raises the question of whether it is weakly equitable in the more realistic regime where policy is based on imperfect proxies, i.e. there is uncertainty in $Y_{i}$ given $X_{i}$. The following assumption makes concrete this notion of imperfect proxies.
\begin{assumption}[Imperfect Proxies]
\label{assumption:cdf}
For every $x \in \mathcal{X}$, the conditional distribution $F_{Y\mid X=x}$ is continuously differentiable and has positive density on $\mathcal{Y} \subseteq \mathbb{R}_{+}$ that contains $(0, c)$.
\end{assumption}
Theorem \ref{theo:inequity} then demonstrates the negative result that for any nonconvex loss there exists an environment that satisfies Assumption \ref{assumption:cdf} where the optimal policy does not satisfy weak equity:
\begin{theo}
\label{theo:inequity}
Consider Definition \ref{defi:loss_min} with some loss function $L$ that is decreasing, integrable on $\mathbb{R}_{+}$, and bounded on $\mathbb{R}_{+}$. If $L$ is nonconvex, then there exists a population distribution $F$, satisfying Assumption \ref{assumption:cdf}, under which the optimal policy family $t_{L}(x; B)$ is not weakly equitable.
\end{theo}
This result applies in particular to the poverty rate loss function (Example \ref{exam:rate}), which is nonconvex. Fortunately, the opposite also holds for a broad class of relevant poverty measures.
\begin{theo}
\label{theo:simple_equity}
Consider loss minimization in Definition \ref{defi:loss_min}. If (a) $L$ is a loss function that either is decreasing, strictly convex, differentiable, and bounded below by $C > - \infty$ and the budget $B > 0$ or (b) $L$ is proportional to the loss function of an FGT index with $\alpha \geq 1$ (Example \ref{exam:FGT}) and the budget $0 < B < c$, then for any population distribution $F$ that satisfies Assumption \ref{assumption:cdf}, there exists a unique optimal policy $t_{L}$ induced by solving \eqref{eq:prob} and $t_{L}$ is weakly equitable with respect to $F$.
\end{theo}
Theorem \ref{theo:simple_equity} provides the positive result that minimization of a convex poverty measure yields weakly equitable policy families. However, one might worry that this approach yields policies that are cost-ineffective at reducing the more familiar poverty rate metric. To help navigate this tradeoff we next show that minimizing one particular convex loss function---the poverty gap index (Example \ref{exam:gap})---yields guarantees on both weak equity and poverty rate reduction. The first point is immediate, since the poverty gap index satisfies the conditions of Theorem \ref{theo:simple_equity}. In terms of formal guarantees on poverty rate reduction, we first show that gap minimization solves a worst-case version of rate minimization: it is equivalent to minimization of the worst-case conditional poverty rate among covariate-defined subgroups of the population.\footnote{Lemma \ref{lemm:gap} is closely related to an observation of \citet{kanbur1987measurement}, who studies gap minimization in a population consisting of two groups and finds that relatively more should be spent on the group with the higher poverty rate.}
\begin{lemm}
\label{lemm:gap}
Suppose that Assumption \ref{assumption:cdf} holds. Consider gap minimization as specified in Example \ref{exam:gap} with budget $0 < B < c$. The optimal policy that solves \eqref{eq:prob} also solves
\begin{equation} 
\label{eq:wc_rate}
\min_{t: \mathcal{X} \rightarrow \mathbb{R}_{+}} \left\{\max_{x \in \mathcal{X}} \PP[F]{Y + t(X) < c \mid X=x}: \EE[F]{t(X)} \leq B \right\}.
\end{equation}
\end{lemm}
Given the desirable equity and poverty rate reduction guarantees of gap minimization, our empirical analysis will focus on gap-minimizing policies, while also presenting rate-minimizing policies to examine how much more the poverty rate could be reduced if this equity condition were ignored.

\section{Data Sources and Uses}
\label{sec:data-sources-uses}

\subsection{Data Sources and Preparation}

Appendix \ref{subsec:country-selection} describes in full detail the criteria and exceptions by which we select countries for inclusion in the analysis. In brief, we begin with those with poverty rates above 10\% or that account for more than 1\% of the world's poor, to focus attention on the countries that matter most for global poverty. Among these we limit attention to countries with data publicly available from a recent, high-quality, nationally representative LSMS-style survey \citep{GroshGlewwe2000}, including a consumption aggregate, and that we expect (given the sample size and published poverty rate estimates) to contain at least 1,000 poor households. These conditions are meant to ensure that we can reliably learn accurate national-level policy costs. We make a few further adjustments to the resulting list: including Bangladesh because of its size, and omitting Afghanistan, Somalia, and South Sudan because we were not able to obtain reliable currency conversion rates for the relevant years. In total this yields a sample of \sampleNumCountries \ countries that vary meaningfully in their headcount poverty rates, from \sampleMinRate\% (in \sampleMinRateCountry) to \sampleMaxRate\% (in \sampleMaxRateCountry), and that collectively account for \sampleShareWorldsPoor\% of the world's extreme poor.\footnote{As of 9 April 2026, as computed from World Bank poverty rate estimates.}

Table \ref{tab:survey_data_sources} lists the specific household surveys, in the indicated countries and years, for which we learn transfer policies. These are in most cases the surveys from which recent estimates of the headcount poverty rate in the World Bank's Poverty and Inequality Platform (PIP, \url{pip.worldbank.org}) were calculated, and are part of the World Bank's flagship Living Standards Measurement Study (LSMS) program to collect high-quality, standardized household surveys on poverty and welfare \citep{world2014living}. In a few cases, the data underlying PIP estimates are not publicly available, and we therefore use an alternative or earlier source. All surveys are designed to be nationally representative; that said, one caveat is that they are thought to under-represent refugees, internally displaced persons, and stateless populations, particularly when these groups are concentrated in institutional settings such as camps \citep{EGRISS2024}.\footnote{As an upper bound on the magnitude of this potential omission, the cost of providing a UBI of \$2.15 per day in 2017 PPP to \emph{all} refugees and IDPs identified by the UNHCR as of 2023 is \refugeeGlobalGDP\% of global GDP, increasing the total cost of the policy we derive below from \extrapolationWBGlobalGDP \% to \refugeePlusExtrapolationGlobalGDP\%. This calculation excludes refugees residing in \refugeeDroppedCountries \ for which the market exchange rate in 2017 and PPP conversion factor are unavailable from the World Bank. Authors' calculations based on \url{https://www.unhcr.org/refugee-statistics/download}, accessed 7 December 2025.}

Several years have passed since most surveys were conducted, and our estimates should thus be interpreted as the cost of reducing poverty at the time of the underlying survey. More generally, because the estimates draw on information about living standards and predictors collected at the same time, one should think of them as representing the costs of ending poverty if poverty measurement surveys were conducted regularly. One could also use data collected in one year to learn policies administered in future years, but because poverty tends to fluctuate substantially over time \citep{MerfeldMorduch2024poverty,ArmentanoNiehausVogl2025poverty} this would reduce their accuracy, and one would want to learn policies that explicitly account for this, as for example \citet{Aikenetal2025moving} do for proxy means tests.\footnote{See also \citet{Beuermannetal2025moving}.}

All surveys collected detailed information on household expenditure and self-production, and we make use of consumption aggregates constructed from this information. We compute household per capita consumption in 2017 PPP USD by converting the household consumption aggregate into 2017 PPP USD per day and dividing by household size. As is standard, this approach treats all individuals in the household as having equal consumption.\footnote{The methods described in Section \ref{subsec:empirical-methods} would apply equally to individual-level measures of living standards, though these are uncommon.} We compute an analysis weight for each household by multiplying its household-level sampling weight by its size.\footnote{The same adjustment is used by the World Bank to construct poverty estimates for the population of individuals in a country rather than the population of households. See \href{https://datanalytics.worldbank.org/PIP-Methodology/welfareaggregate.html}{here} for details.} The estimated headcount poverty rates this procedure yields are generally close to the World Bank estimates for the corresponding country and year, and match exactly in most cases where both estimates derive from the same survey. See Appendix \ref{subsec:outcome-construction} for details on outcome construction.

Selecting covariates to use as predictors of living standards requires some judgment. Conceptually, we wish to select only characteristics that are plausibly verifiable, to limit scope for strategic behavior by potential recipients \citep{grosh1995proxy,BjoerkegrenBlumenstockKnight2021} and make it easier to hold program staff accountable for enforcing eligibility criteria \citep{NiehausAttanassovaBertrandMullainathan2013targeting}. We operationalize this idea as follows. First, we identify a collection of proxy means tests actually implemented in practice in a variety of low- and middle-income countries.\footnote{These were Bangladesh \citep{KiddWylde2011}, Colombia \citep{CamachoConover2011}, India \citep{NiehausAttanassovaBertrandMullainathan2013targeting,BPLGuidelines,BPLGuidelinesUrban}, Indonesia \citep{KiddWylde2011,Alatasetal2012targeting,FernandezHadiwidjaja2018}, Malawi \citep{IE}, and Peru \citep{hanna2018universal}.} We take the union of the predictors used in these PMTs, which yields a list spanning six broad categories: household demographics, human capital, household assets, livelihood activities, geographic indicators, and community characteristics. Where appropriate we generalize from the specific variables used in past PMTs; if a PMT included a refrigerator as a predictor, for example, we would include all large household appliances on our list of eligible predictors. At the same time, we omit a few variables which, though they have been used in the past, would in our judgment typically be difficult to verify. Examples include the age of the household's primary dwelling, measures of food (in)security, indicators for use of fertilizer, or total income. We also omit indicators for the ethnicity, tribe, religion, or caste of the household or household head; while these have been used in past PMTs, they raise questions of ethical and political suitability in the given context. We prefer to err on the conservative side and omit them.\footnote{We thank Lant Pritchett for suggesting this.} Appendix \ref{subsec:predictor-selection} provides the rubric for covariate selection. We then select only variables from this list from each living standards survey for use as predictors in our analysis.


\subsection{Empirical methods}
\label{subsec:empirical-methods}

We use data from each survey to learn transfer policies via empirical minimization of the poverty rate on the survey data; empirical minimization of the poverty gap; and proxy means testing. Explicit characterizations of the empirical rate-minimizing and gap-minimizing policies and algorithms for deriving them are in Appendix \ref{sec:alg}. Here we provide a brief summary of the issues that arise and the methods we use to address them.

Unrestricted gap minimization is a convex problem, and thus tractable. Optimal transfers are (we show) functions of the conditional (on covariates) quantile functions of the distribution of living standards. We learn these conditional quantile functions for a grid of quantiles using deep learning \citep{Goodfellow-et-al-2016deeplearning}, minimizing pinball loss \citep{koenker1978regression}. When we impose the added constraint that policies must be binary, however, the problem becomes non-convex and requires more care. We express a discrete approximation of it as a nested optimization over (i) the size of the transfer, and then (ii) the eligibility rule used to determine which households received a transfer of that size. Learning the latter requires learning the expected change in loss associated with a transfer of the specified size to each household, effectively yielding a ranking of households; we do so using empirical risk minimization via deep learning. Nesting this algorithm within a grid search over values of the transfer yields the solution. The learning procedure for binary rate minimization is analogous to the binary gap minimization procedure.

Unrestricted rate minimization is also a non-convex problem. We solve a discrete approximation of it by first showing that rate-minimizing transfers must have a specific structure: conditional on a given value $x$ of the covariates, they must either be at the boundaries (i.e. $t(x) = 0$ or $t(x) = c$), or satisfy $f_{Y|X=x}(c - t) = \alpha$ for some common value $\alpha$. Intuitively, this condition states that the number of people that would be lifted out of poverty by marginally increasing the transfer amount must be the same for any values of $x$ at which the transfer amount is in the interior. We then show that, for any given value of $\alpha$, this reduces the problem to a fractional multiple-choice knapsack problem which can be solved using estimates of the covariate distribution $f_X$ (for which we use the empirical covariate distribution) and of the conditional distribution $f_{Y|X=x}$ (which we learn using an extension of Lindsey's method \citep{efron1996using}). Nesting this algorithm within a grid search over values of $\alpha$ yields the solution.

All statistical learning methods described in Appendix \ref{sec:alg} are tuned for out-of-sample prediction using a validation set \citep{hastie2009elements}. This implies that we do not need to limit the number of predictors we use to guard against overfitting up front and lets us use predictor sets that are often larger than those used in typical PMT exercises; our models use between \minDimension \ and \maxDimension \ predictors (second column of Table \ref{tab:surveys}). That said, we will also find (below) that we can use substantially smaller predictor sets without meaningful performance degradation. Finally, note that we will not be able to take full advantage of large predictor sets when learning the conditional distributions needed for unrestricted rate minimization, as the sample complexity of density estimation scales exponentially in the data dimension; we will return to this point in discussing the results below.

Implementing these learning methods requires that we specify precise training, hyperparameter selection, and evaluation procedures. To prevent iterative analysis of the same data, which could lead to ``human-in-the-loop'' overfitting, we prescribed these procedures in advance in a Data Use Plan (DUP). A complete copy of this plan is available online;\footnote{Our data use plan is available \href{https://drive.google.com/file/d/1gdIZD1Mj3rQSdzZS1Fc0mCIASry5_lEa/view?usp=sharing}{here}.} its key features are as follows.

First, we randomly partition each dataset into a 60\% training sample and a 40\% test sample. We do so maintaining geographic stratification at the level at which the survey is designed to be representative. We evaluate subsample sizes for this purpose using simple observation counts; where possible, we ensure that this also splits the sampling weights in the same proportion.

Second, we use the training set to learn hyperparameters for the policy learning algorithms. These include, for example, the topology of the neural networks used to fit nuisance parameters for gap minimization or binary rate minimization; the number of quantiles to fit during quantile regression for unrestricted gap minimization; the dimensionality of the predictor set used for unrestricted rate minimization; and various learning parameters, among others. We learn distinct hyperparameters for each policy type we wish to learn, learning them from a 67\% sub-sample of the training set and evaluating them based on the performance they induce on the complementary 33\% sub-sample (the validation set). We also considered two alternative approaches: hard-coding hyperparameters based on prior intuition, or learning them from synthetic data generated from the training set via a generative adversarial network \citep{Atheyetal2024gan}. When tested in our ``sandbox'' environment (see below), we found that all three approaches performed similarly. We therefore chose to learn hyperparameters directly from the training data as this approach is substantially simpler than using synthetic data, while providing some protection (relative to hard-coding) against ``unexpected'' data distributions we might subsequently encounter.

Third, with hyperparameters in place, we use the training set to estimate all nuisance parameters necessary for learning the transfer policies. After that, with access to the covariates from the test set, we generate the transfer schedule $\{t(X_{i})\}$ for all units in the test set, and verify that our budget constraint is enforced on the actual set of test set units we consider deploying the transfers on. This paradigm, which is often referred to as the transductive setting \citep{vapnik1998statistical}, mirrors typical practice in real-world transfer program design. Finally, we use the held-out outcomes $Y_{i}$ from the test set to evaluate the transfer schedules. This is the relevant paradigm when the policymaker expects to know the values of the covariates (though not the outcome) for the entire target population when deciding on a policy. This is the case when, for example, a government conducts a census gathering covariate information before finalizing the transfer rule for a program. 

The main exception to our otherwise rigid implementation of this plan involves the 2018--2019 Integrated Household Survey in Malawi. We used this survey as a sandbox environment in which to iteratively experiment with various approaches and guide the choices to which we subsequently committed in our Data Use Plan. Results based on this survey should therefore be interpreted with this caveat in mind. That said, we will see below that the estimated benefits of targeting in Malawi (as inversely measured by the ratio of the cost of optimized transfers to the cost of universal ones; see Figure \ref{fig:ubi_ratio}) are among the lowest of the countries in our sample, the opposite of what one might expect had our experimentation led to over-fitting. 

Any other adjustments we made to the methodology over time are noted in a changelog within the DUP itself. To date the most salient of these have been modifying the analysis weights to reflect household size as well as survey sampling weights (which we had initially neglected to specify), and removing indicators related to ethnicity, race, caste, religion, or other such demographic groups (in response to audience feedback). For purposes of inference we also conduct an additional bootstrapping step: we hold fixed the nuisance parameters learned from the training data, draw bootstrap replicates from the test sample, and recalculate for each replicate the costs of achieving some given poverty goal. This quantifies cost uncertainty due to sampling variation in the test data.

\section{Empirical Results}
\label{sec:empirical}

To build intuition we first describe learned policies and present cost results for a single country, Togo, in Section \ref{sec:togo}. We then apply our methods to the full sample and present the aggregate policy cost of achieving uniform poverty goals in Section \ref{sec:multi_country}.

\subsection{Togo Case Study}
\label{sec:togo}

\begin{figure}[tp]
\caption{Togo Transfer Distributions}
\includegraphics[width=\textwidth]{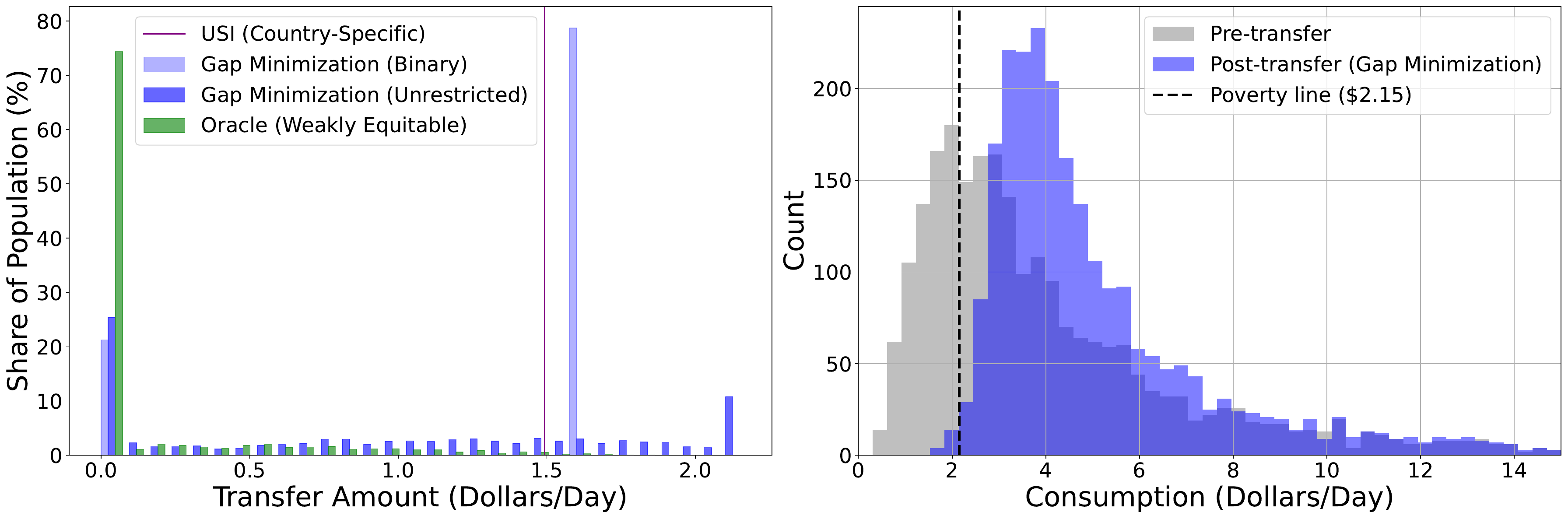}
{\footnotesize This figure reports, for Togo, the distribution of transfer sizes under various learned policies all of which achieve a 1\% post-transfer poverty rate (left), and the full pre- and post-transfer distribution of living standards under a gap-minimizing policy that achieves a 1\% post-transfer poverty rate (right).}
\label{fig:togo_distributions}
\end{figure}

We begin with illustrative results from Togo. We choose Togo because (1) it is a representative country in the sense that our results place it near the middle of the sample in terms of the ``difficulty'' of targeting, as quantified below in Figure \ref{fig:oracle_feasible_ratio}, and (2) we uniquely have data for Togo letting us examine the performance of targeting based on satellite imagery, results that we will present in Section \ref{subsubsec:sat}. Results here are based on Togo's 2018-2019 Enquête Harmonisée sur les Conditions de Vie des Ménages \citep{tgo_main_survey}, a survey of $n=\togoSampleSize$ households that contains $d=\togoCovariateDimension$ plausibly verifiable covariates.

The left panel of Figure \ref{fig:togo_distributions} visualizes the distribution of transfer sizes under several policy alternatives, all of which achieve a post-transfer poverty rate of 1\%. As expected, the (infeasible) oracle policy assigns transfer sizes of zero to many households---all those above the poverty line---and then a range of positive transfers to those below it. The binary and unrestricted gap-minimizing policies both give zero transfers to a smaller (but still meaningful) number of households; these are essentially the households that can reliably be assessed to have consumption above the poverty line based on proxies. The difference between the binary and unrestricted gap-minimization policies becomes salient for households that do receive positive transfers. The binary policy must give the same transfer amount (here the optimal choice is \$\togoBinaryGapTransferAmount) to all households receiving any transfer. In contrast, the unrestricted gap-minimizing policy can tailor transfer sizes and utilize a wide range of positive transfer values to reduce costs while still meeting the poverty reduction target. Most gap-minimizing transfers are also smaller than the amount given to households by a Universal Supplemental Income.\footnote{A few are still quite large, however---over \$2 in some cases---which is unlikely to be optimal since in the data essentially no one requires a transfer this large to exit poverty. This suggests there may be scope to further lower costs by discouraging the algorithm from learning very large transfer sizes.}

\begin{figure}[tp]
    \begin{center}
    \caption{``Inclusion error'' vs. ``exclusion error'' for Togo}
    \label{fig:togo_targeting_efficiency}
    \includegraphics[width=0.6\textwidth]{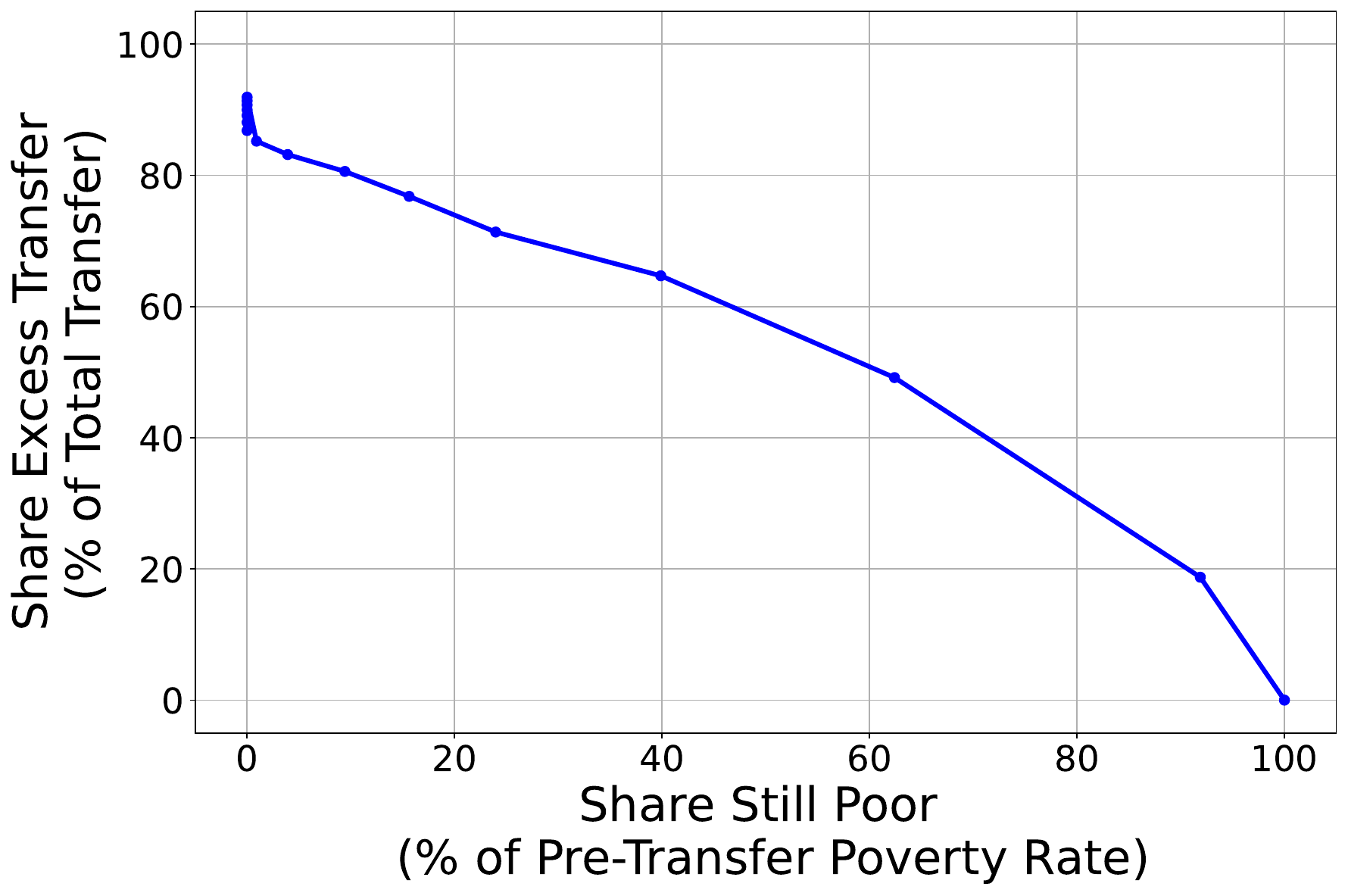}
    \end{center}
    \vspace{-1em}
    {\footnotesize This figure reports, for Togo, the relationship between the share of those initially poor who were still poor after transfers ($x$-axis), which is analogous to exclusion error in conventional analyses, and the share of transfer dollars that do not contribute to closing the poverty gap ($y$-axis), which is analogous to inclusion error. All figures refer to gap-minimizing policies.}
\end{figure}

How well do learned transfers target the poor? The right panel of Figure \ref{fig:togo_distributions} visualizes this for a gap-minimizing policy that achieves a 1\% poverty rate. It plots the pre- and post-transfer distributions of per capita consumption. The share of households below the poverty line falls substantially, but the distribution of living standards above the line also shifts rightward as some funds accrue to non-poor households. For binary policies it is conventional to summarize such tradeoffs using rates of ``inclusion error,'' i.e. of non-poor households being eligible, and of ``exclusion error,'' i.e. of poor households being ineligible. Figure \ref{fig:togo_targeting_efficiency} presents a generalization of these concepts to the case of unrestricted transfers. It considers a range of gap-minimizing policies that vary in the share of the pre-transfer poor who are still poor after transfers, indicated on the $x$-axis, analogous to exclusion error. The fraction of transferred dollars that do not contribute to closing the poverty gap, on the $y$-axis, is analogous to the inclusion error rate. As one would expect, there is a tradeoff: the least ambitious policies, which reduce the poverty rate only slightly, are able to do so very efficiently in the sense that they route 20\% or less of funds to non-poor households. More ambitious policies that lower the rate further do so at the cost of lower targeting efficiency.

\begin{figure}[tp]
	\caption{Gap- and Rate-minimizing Policy Costs, in Togo}
    \includegraphics[width=\textwidth]{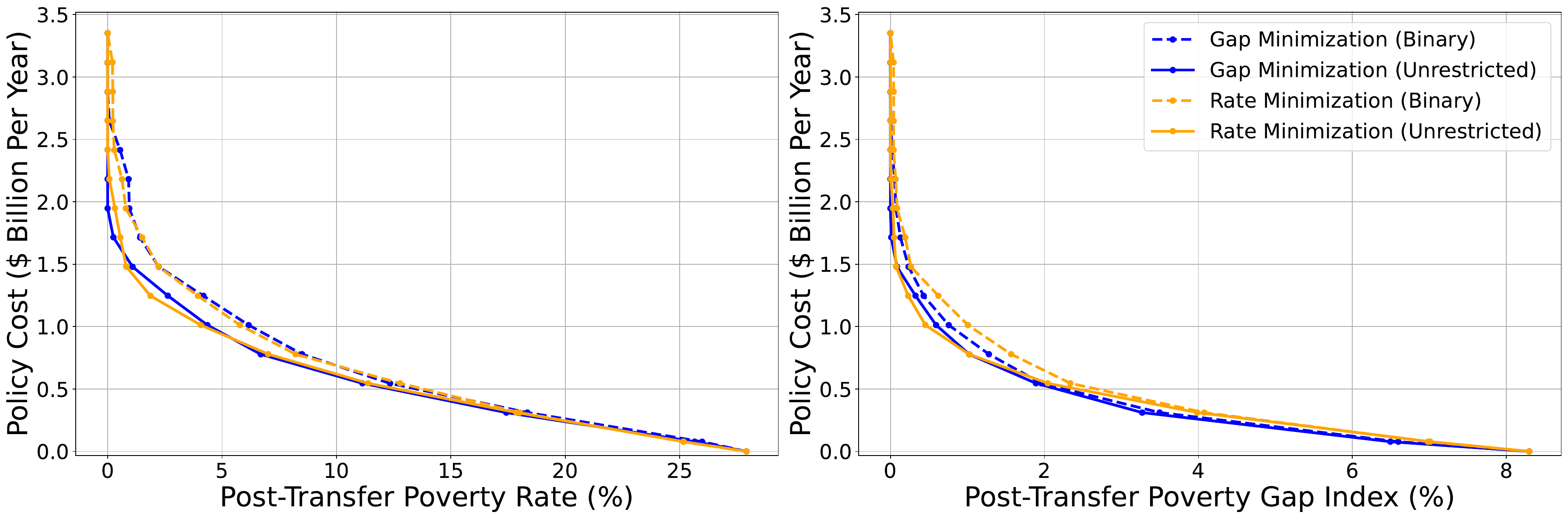}
    
    \vspace{1em}
    \footnotesize{This figure reports the costs, in billions of nominal 2023 USD annually, of the indicated optimal transfer policies learned for Togo $(n=\togoSampleSize\,,\ d=\togoCovariateDimension)$. The left-hand panel plots these against the post-transfer poverty rate, while the right-hand panel plots these against the post-transfer poverty gap index.}
    \label{fig:togo_rate_gap}
\end{figure}

Turning next to cost performance, Figure \ref{fig:togo_rate_gap} compares the performance of gap-minimizing and rate-minimizing transfer policies by plotting the total annual policy cost (in nominal 2023 USD) of achieving a particular poverty goal (either the post-transfer poverty rate or post-transfer poverty gap index). Blue lines indicate methods that minimize the poverty gap; orange lines indicate methods that minimize the poverty rate. The solid and dashed lines correspond to unrestricted and binary policies, respectively.

Several patterns emerge from these cost series, which in general will also hold in the other countries we analyze. First, as one would expect, marginal costs increase as we approach zero poverty according to either metric. Literally ending extreme poverty requires a policy effectively amounting to UBI in order to reach every last poor household, but one can come close at far lower costs. Second, unrestricted policies are less costly than binary ones (i.e., the solid lines fall below the dashed lines). This is expected, since unrestricted policies can allocate smaller transfers to households closer to the poverty line. Finally, there is no systematic benefit to rate minimization relative to gap minimization in this setting.

\subsection{Multi-Country Analysis}
\label{sec:multi_country}

We now present results aggregated across the full sample of \sampleNumCountries \ countries, with the underlying country-specific results available in Appendix \ref{sec:additional_exhibits}. We aggregate by calculating the cost of achieving a given poverty rate in \emph{every} country in our sample, and summing these. This necessarily costs (weakly) more than achieving the same poverty rate on \emph{average}, potentially by reducing poverty more in countries where this is relatively inexpensive; we do it to ensure a degree of equity between the countries in our sample. We present cost curves in their entirety, but also discuss in more detail the costs of achieving a \nationalTarget\% poverty rate in each country using policies learned via empirical poverty-gap minimization. Note that achieving this target globally would imply that the global poverty rate is \globalPovertyRateForNationalTarget \%.

\begin{figure}[tp]
	\caption{Gap minimization vs. Oracle, UBI, and PMT Benchmarks in the Full Sample}
    \includegraphics[width=\textwidth]{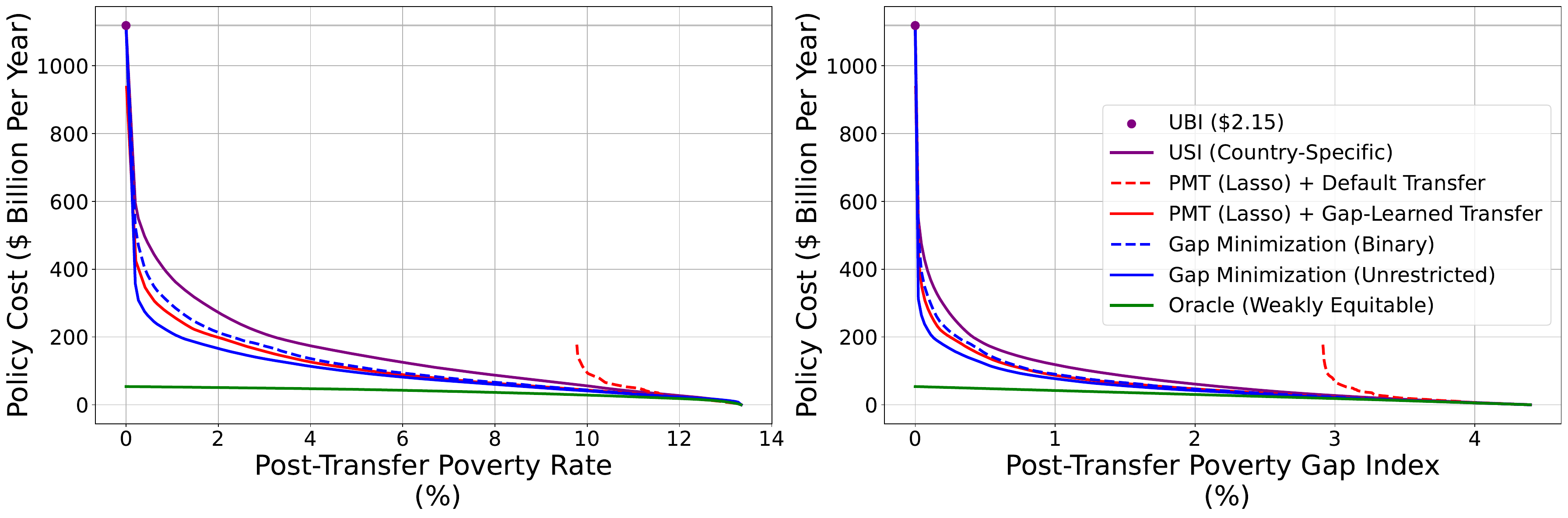}
    
    \vspace{1em}
    \footnotesize{This figure reports the costs, in billions of nominal 2023 USD annually, of the indicated optimal transfer policies learned for the full sample of countries listed in Table \ref{tab:surveys}. The left-hand panel plots these against the resulting population-weighted post-transfer poverty rate, and the right-hand panel plots these against the resulting population-weighted post-transfer poverty gap index. See Section \ref{sec:multi_country} for series definitions.}
    \label{fig:headline}
\end{figure} 

Figure \ref{fig:headline} presents the core results, plotting total annual policy cost (in nominal 2023 USD) against the resulting population-weighted headcount poverty rate (left-hand panel) or poverty gap index (right-hand panel). The primary policy of interest here is the unrestricted gap-minimizing policy, indicated in blue. The binary policy minimizes the same objective, but adds the restriction that within each country, transfers must either be zero or some common positive amount. Other series provide benchmarks. The ``UBI (\$2.15)'' series indicates the (constant) cost of giving every individual \$2.15 2017 PPP per day, without any targeting. The ``USI (Country-Specific)'' series indicates the cost of giving every individual within a given country the same transfer amount less than or equal to \$2.15, which we refer to as Universal Supplemental Income. 

The two ``PMT'' series correspond to allocating transfers with a proxy means test, fitting conditional mean consumption via lasso regression \citep{hanna2018universal} and giving a fixed amount to individuals with mean predicted consumption below a threshold (determined by the transfer size and budget). Since the transfer size is not determined algorithmically in the PMT approach, we consider two benchmarks. In the ``PMT + Default Transfer'' series, we set the transfer at 20\% of the mean pre-transfer per capita consumption among the poor, i.e. $0.2 \times \mathbb{E}[Y_i | Y_i \leq c]$. This roughly matches the sizing of two high-profile proxy means tested transfer programs (Progresa in Mexico, and the Benazir Income Support Programme in Pakistan), for example, and is more generally in the range we think of as typical for PMT-allocated transfers.\footnote{See \citet[][p. 14]{coady2003alleviating} and \citet[][p. 53]{barrientos2013social}, respectively.} In the ``PMT + Gap-Learned Transfer'' series, we use the data to learn the transfer size that leads to the largest reduction in the poverty gap for a given budget when allocated via PMT. This does \emph{not} correspond to common practice, but will help us distinguish the influence of optimal sizing of the transfer amount from that of end-to-end welfare-aware learning of the allocation rule when we compare the PMT to binary loss-minimizing policies. 

Finally, the panels include an ``Oracle'' series in green indicating the (hypothetical) cost of achieving a given poverty gap index and poverty rate with a weakly equitable policy if full information on each household's consumption were available.\footnote{The oracle policy family that we visualize in both panels is the family of policies that minimizes the poverty rate while maintaining weak equity and continuity in budget. This policy is also an optimal gap-minimizing policy in the full-information setting} This is a useful benchmark given the salience of the global poverty gap in recent discourse about the cost of ending poverty \citep[cf.][]{chandy_global_2016,KharasMcArthur2023,sumner_new_2024}, but it is not a policy that could be feasibly implemented since program administrators do not directly observe the exact amount needed by every household in the population.

Overall the results show that the lowest-cost policies cost several times more than the aggregate poverty gap, but also substantially less than benchmark universal policies. To reduce the poverty rate in all countries to \nationalTarget\%, for example, would cost \$\headlineGapNationalTarget B per year, which is \headlineGapOracleRatioNationalTarget \ times the oracle cost (\$\headlineOracleNationalTarget B), but \headlineGapUBIPercentNationalTarget \% of the cost of a \$2.15 UBI, and \headlineGapUBIVariablePercentNationalTarget\% of the cost of a country-specific Universal Supplemental Income. This figure does not appear particularly sensitive to the fact that our test set is only a sample and not the full population: a bootstrapped 95\% confidence interval for the cost is \$\headlineGapNationalTargetCILower B to \$\headlineGapNationalTargetCIUpper B. Similar ratios hold as we vary the desired post-transfer poverty rate. As the post-transfer poverty rate (or gap) approaches zero, however, the costs of all feasible policies increase at an increasing rate, as expected: requiring any of these algorithms to find the last needles in the population haystack becomes increasingly costly.

What if we restrict to binary policies? The binary gap-minimizing policy (dashed blue series) costs \headlineBinaryContPercentIncreaseNationalTarget \% more than the unrestricted gap-minimizing policy. This implies that, if politically or logistically expedient, the transfer policy can be reduced to a simple eligibility rule without massively increasing its cost. Allocating a binary transfer using a PMT, rather than via loss minimization, also yields respectable performance \emph{provided} the size of that transfer is itself learned to minimize the poverty gap (solid red series). In that case, in fact, PMT targeting outperforms binary gap minimization for the most ambitious poverty rate targets. But if instead we set the transfer size to increase mean consumption among poor recipients by 20\% (dashed red series) we cannot reduce either poverty metric very far with any budget. Such transfers are simply too small to raise more than a few poor households out of poverty.

\begin{figure}[tp]
	\caption{Gap- vs. Rate-minimizing Policy Costs in the Full Sample}
    \includegraphics[width=\textwidth]{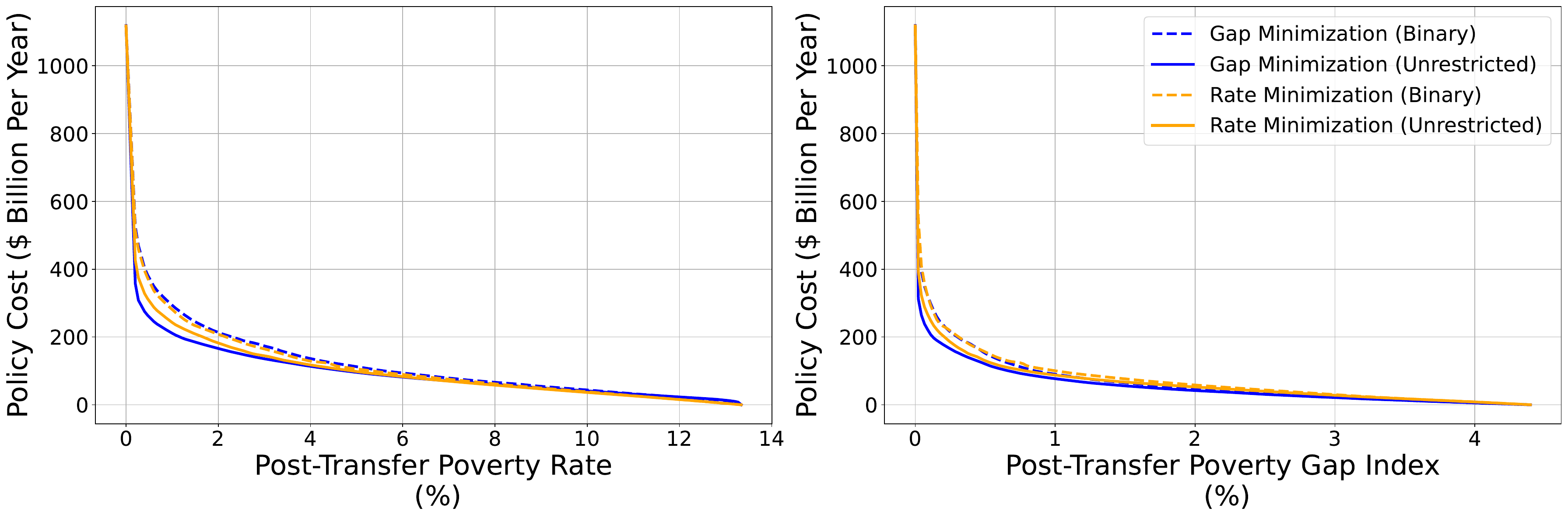}
    
    \vspace{1em}
    \footnotesize{This figure reports the costs, in billions of nominal 2023 USD annually, of the indicated optimal transfer policies learned for the full sample of countries listed in Table \ref{tab:surveys}. The left-hand panel plots these against the post-transfer poverty rate, while the right-hand panel plots these against the post-transfer poverty gap index.}
    \label{fig:rate_gap}
\end{figure}

While none of the policies examined in Figure \ref{fig:headline} are optimized to reduce the poverty rate, the left-hand panel shows that gap minimization still substantially reduces the poverty rate. This suggests that it would be possible to implement gap-minimizing policies, which are weakly equitable, while still communicating with broad audiences about progress in terms of the poverty rate, a more familiar concept. That said, Figure \ref{fig:rate_gap} directly contrasts the empirical performance of policies optimized for the rate as opposed to the gap. Axes are as in Figure \ref{fig:headline}, but here the policies include two that minimize the rate (using either unrestricted or binary-valued transfers) as well as the corresponding policies minimizing the gap. The right-hand panel shows that, as expected, policies optimized for the gap consistently reduce the gap at a lower cost than the corresponding policies optimized for the rate. The left-hand panel shows, on the other hand, that a policy optimized to reduce the gap using unrestricted transfers actually performs somewhat \emph{better} at reducing the rate than empirical rate minimization. In this sense there is little reason to prefer rate minimization in terms of performance.

The driving factor behind this somewhat surprising result is that empirical rate minimization must solve a more challenging statistical problem than empirical gap minimization, requiring that we estimate the full conditional density functions as opposed to only their quantiles. A first, immediate consequence of this challenge is that empirical rate minimization has worse sample complexity, i.e., its errors decay more slowly with sample size. A second, more subtle consequence is that conditional density estimation requires use of a bespoke algorithm (unlike conditional quantile estimation which can be implemented using off-the-shelf neural networks)---and unsurprisingly we thus find our algorithm to be less robust at handling complex and/or high dimensional data. We seek to mitigate this issue by implementing data-driven feature selection for our rate minimization algorithm;\footnote{No such feature selection is required for the quantile regression algorithm underlying our gap-minimization procedure, as off-the-shelf neural network implementations are already highly robust to receiving high-dimensional inputs.} realistically, however, it is unlikely this procedure is able to make use of as many features as gap minimization.\footnote{That said, we find that in practice gap minimization using the restricted set of features used for rate minimization performs almost identically (and, in particular, continues to reduce the rate more cost-effectively) indicating that algorithmic performance per se is the main driver (Figure \ref{fig:rate_gap_fewpredictors}).}

\begin{figure}[tp]
\caption{Cost Ratio of Gap Minimization to Universal Supplemental Income}
\begin{center}
\includegraphics[width=0.66\textwidth]{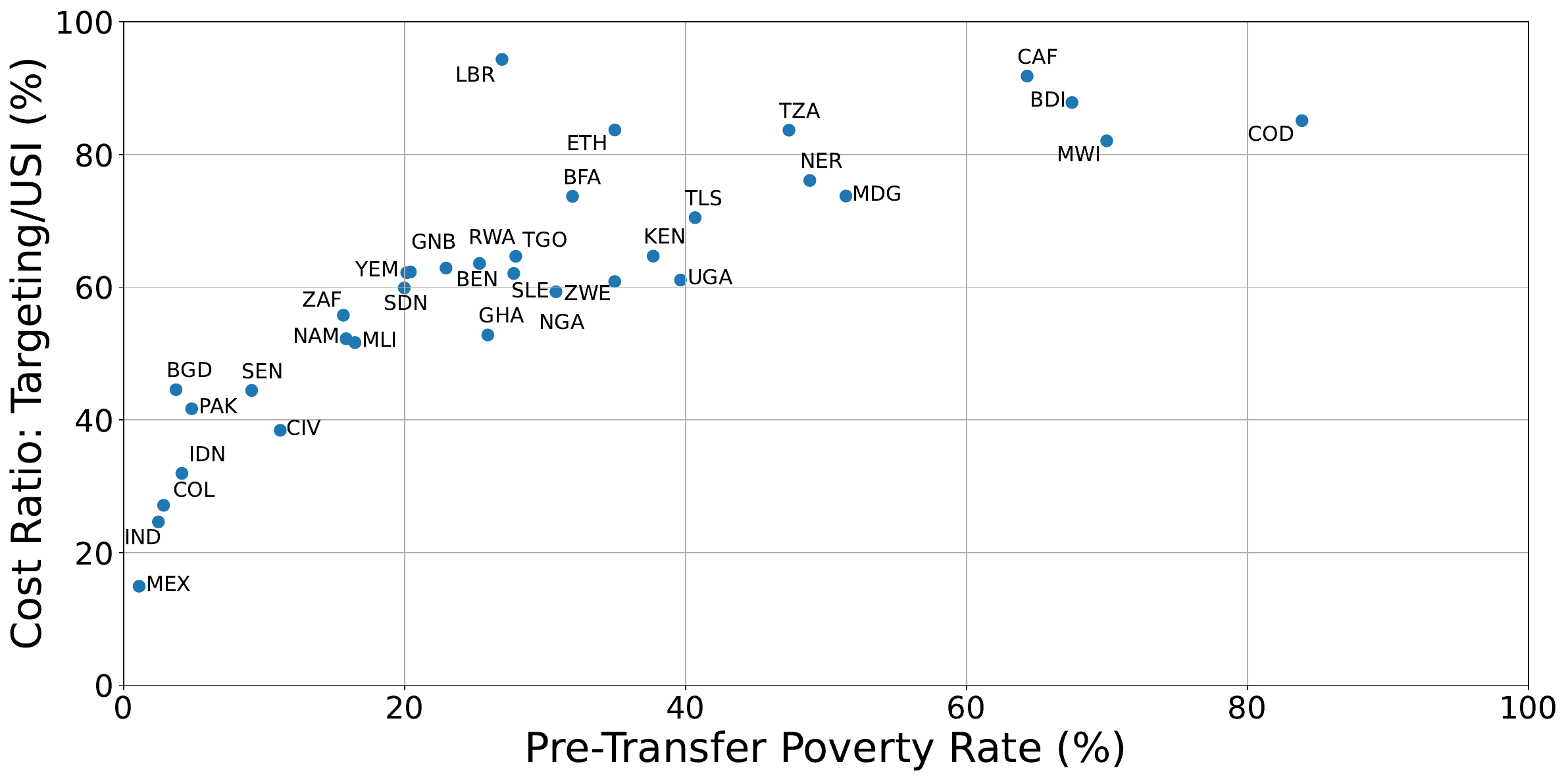}
\end{center}
\footnotesize{This figure reports, for each country in Table \ref{tab:surveys}, the country's initial poverty rate ($x$-axis), and the ratio of the cost of a policy that achieves a \nationalTarget\% post-transfer poverty rate using transfers that vary person-to-person learned via gap minimization, and the cost of a policy that does so by giving a common transfer amount to all individuals ($y$-axis).}
\label{fig:ubi_ratio}
\end{figure}

The full country-specific details underlying these results are presented in Figure \ref{fig:country1}. Generally speaking, the comparisons above for the sample of countries as a whole hold within each country as well.\footnote{For certain countries and targeting methods, we observe that the cost of learned policies can be non-monotone decreasing in the post-transfer poverty rate and gap. This behavior arises because our nuisance parameters and metrics are estimated with error in finite samples.  Binary gap and rate minimization are especially prone to exhibiting non-monotonicities because the optimal solutions for these problems can rely on different nuisance parameters even for similar budgets. In contrast, unrestricted gap minimization is less fragile in this respect because similar budgets will yield solutions that rely on similar nuisance parameters with correlated errors.} Reducing the poverty rate to \nationalTarget\% with an optimal policy costs between \oracleFeasibleRatioMin \ and \oracleFeasibleRatioMax\ times the aggregate poverty gap (Figure \ref{fig:oracle_feasible_ratio}), but still substantially less than the cost of doing so using UBI at the level of the international poverty line. In some of the poorest countries, however, the cost of achieving this goal using a Universal Supplemental Income is not much greater than the cost of doing so using variably sized transfers (Figure \ref{fig:ubi_ratio}), reflecting the fact that most households in these countries are below the poverty line.

\subsection{Variations and extensions}

This section reports comparative statics for the minimized costs of achieving poverty goals. For simplicity of exposition we focus discussion primarily on the goal of achieving a national poverty rate of \nationalTarget \% in each country, as in our summary discussion above.

\subsubsection{Alternative Poverty Line}

The \$2.15 2017 PPP line is a well-established benchmark in both public and policy discourse, derived (via periodic inflation-indexing) from the famous concept of ``dollar-a-day'' poverty introduced by the World Bank in 1990 and subsequently enshrined in the Millennium Development Goals and the Sustainable Development Goals. Yet it is also a very low bar. Lant Pritchett, among others, has argued forcefully that actors working to promote economic development should view ``dollar-a-day'' poverty as a milestone along the way, not as an end goal \citep{pritchett2024end,pritchett2025raising}, and the World Bank's 2025 revision to its global poverty line incorporated not only new information about prices, but also an upward shift in the \emph{real} standard of living used to define the threshold. This was a meaningful change, increasing by some 125 million the number of people defined as living in extreme poverty in 2022 \citep{Fosteretal2025poverty}.

\begin{figure}[tp]
\caption{Policy Costs for Poverty Reduction under \$3.00 (2021 PPP) Poverty Line}
\label{fig:headline_3_dollar}
\includegraphics[width=\textwidth]{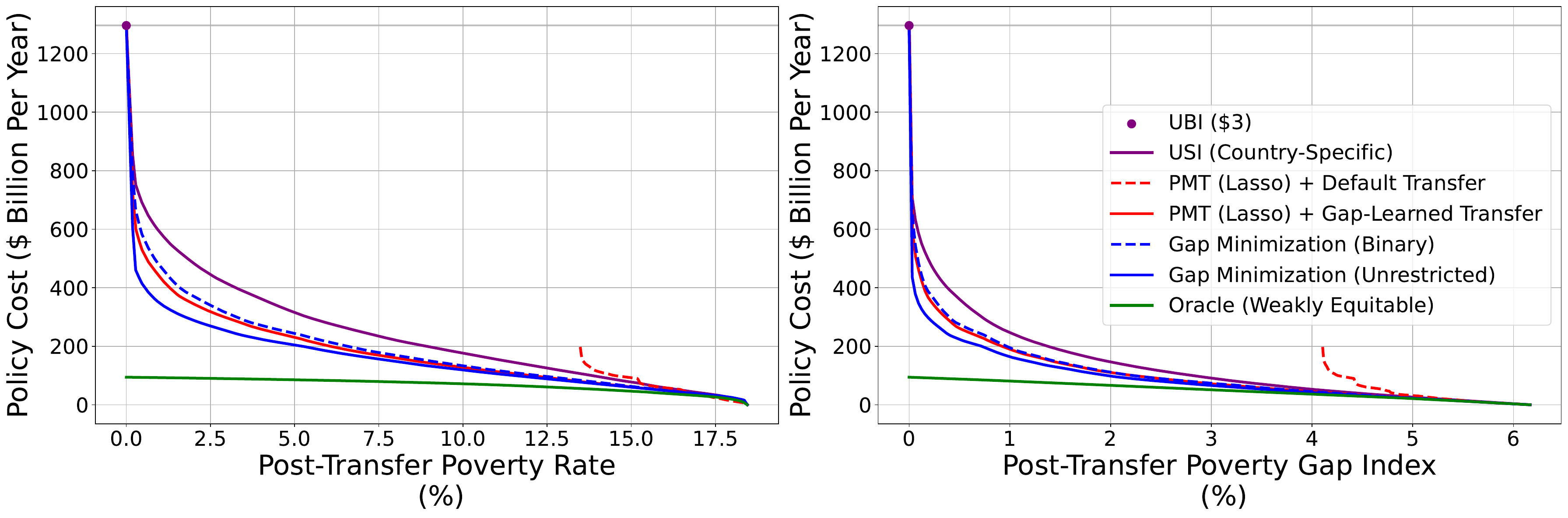}
\footnotesize{This figure reports the costs, in billions of nominal 2023 USD annually, of the indicated optimal transfer policies learned for the full sample of countries to reduce extreme poverty, as defined as living on less than \$3.00 (2021 PPP) per day. This figure is analogous to Figure \ref{fig:headline}, which reports costs to reduce extreme poverty under the \$2.15 (2017 PPP) poverty line.}
\end{figure}

Our methods and data can be used flexibly to calculate the cost of achieving poverty goals with respect to any poverty line. Here we redo our analysis using the World Bank's new \$3.00 2021 PPP line, which is of intrinsic interest as well as serving to illustrate the consequences of setting a higher bar more generally. Figure \ref{fig:headline_3_dollar} replicates Figure \ref{fig:headline}, but using this higher line to define extreme poverty. Achieving a poverty rate of \nationalTarget\% in every country in our sample using policies learned via empirical poverty-gap minimization costs \$\headlineGapNewPovertyLineNationalTarget B per year in nominal 2023 dollars. This is a \percentIncreaseContGapNewOldPovertyLineNationalTarget \% increase in the cost (again, in nominal 2023 dollars) of achieving the same rate when poverty is defined using the \$2.15 2017 PPP standard. This difference quantifies the extent to which ending extreme poverty under the new definition is indeed a substantially more ambitious goal than under the old.

\subsubsection{Poverty Minimization vs. Welfare Maximization}
\label{subsec:welfare}

\begin{figure}[tp]
    \begin{center}    
    \caption{Aggregate welfare from poverty minimization vs. welfare maximization}
    \includegraphics[width=0.5\textwidth]{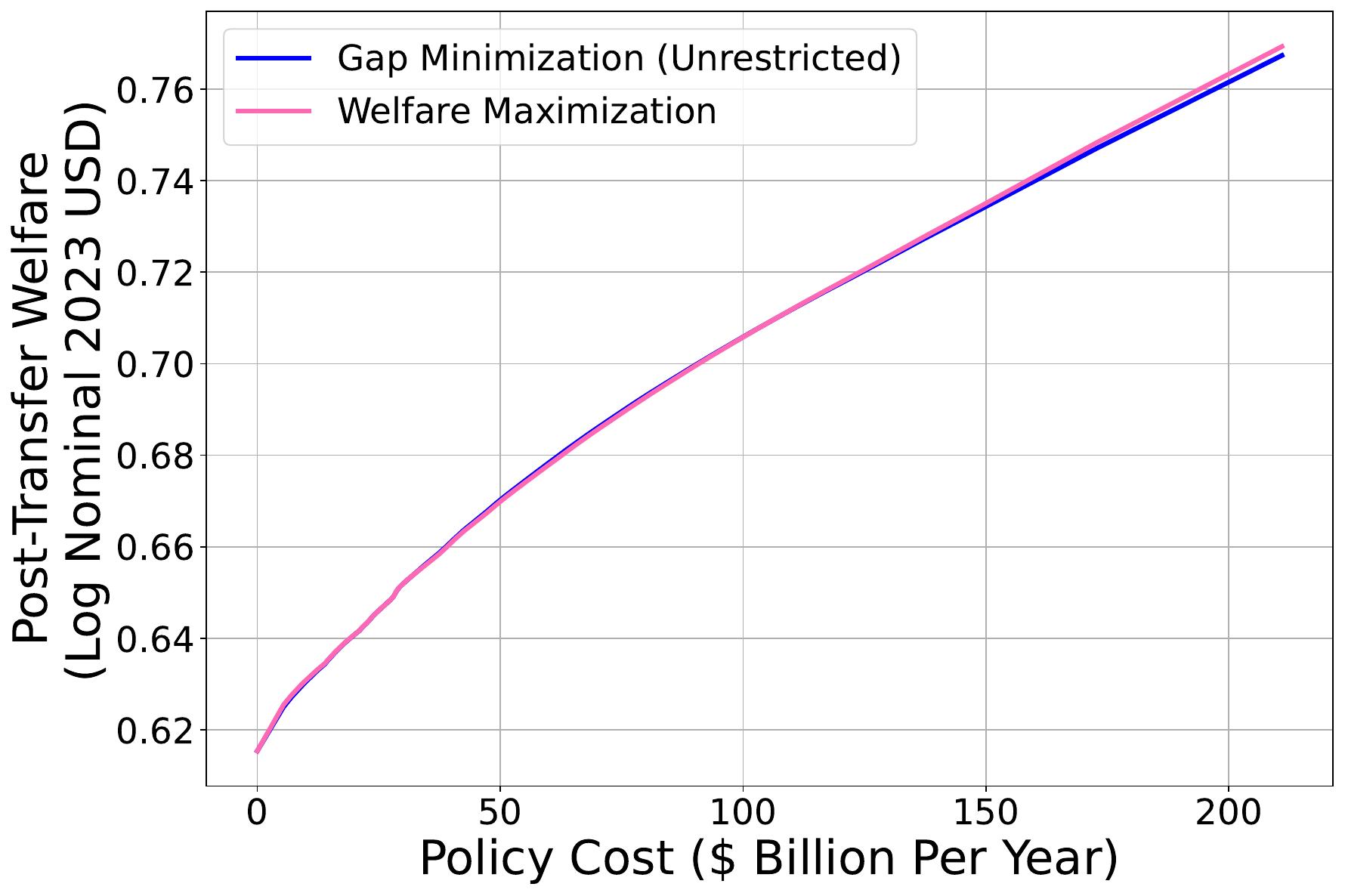}
    \label{fig:welfare}
    \end{center}
    {\footnotesize This figure reports the mean realized welfare, evaluated using log utility, i.e. $\log (Y_i + t(X_i; B))$, from policies learned to minimize the poverty gap (blue series) and to maximize welfare (magenta series), for a range of budgets $B$.}
\end{figure}

\begin{figure}[tp]
    \begin{center} 
    \caption{Realized welfare and transfer size distributions in Togo}
    \includegraphics[width=\textwidth]{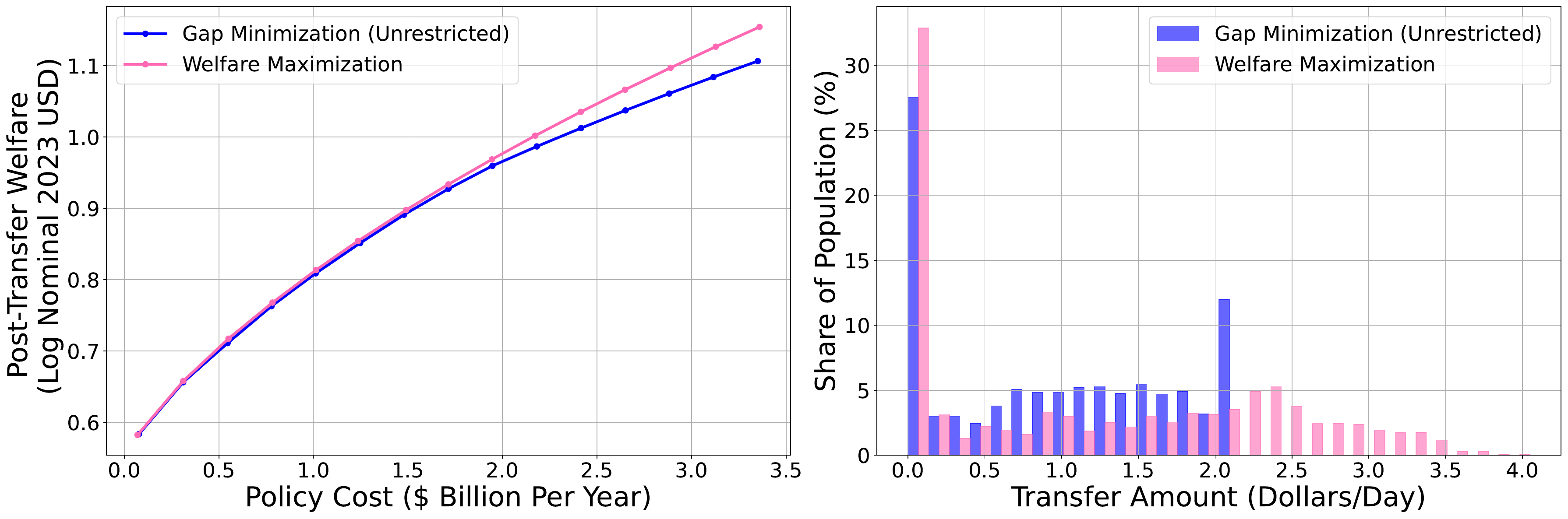}
    \label{fig:togo_welfare}
    \end{center}
    \vspace{-1em}
    {\footnotesize This figure reports, for Togo specifically, the mean realized welfare from policies as defined in Figure \ref{fig:welfare} (left-hand panel) and the corresponding transfer size distributions induced by those policies (right-hand panel).}
\end{figure}

Poverty minimization is a familiar and widely-accepted global goal with a clear end-point. But it implicitly places no value on gains for households \emph{above} the extreme poverty line. This puts it in tension with broader ethical criteria for evaluating distributive justice which---while typically very sensitive to gains among the poor---also usually value gains among the non-poor. Economic welfare analysis, in particular, typically takes as its criterion the sum of individuals' utilities. To examine this tension, we next contrast policies learned to minimize the poverty gap with those learned to maximize welfare, defined as the mean of logarithmic consumption. This imposes a degree of curvature on utility functions broadly in line with available evidence \citep{layard_marginal_2008}. We follow a process analogous to that above for poverty minimization, characterizing welfare-maximizing transfer policies analytically to the extent possible and then defining algorithms to learn solutions empirically; Appendix \ref{app:welfare} provides details.

We find that welfare maximization yields higher welfare, as expected, but only slightly so. Gap-minimizing policies that achieve a 1\% poverty rate in every sample country achieve a level of welfare that is only \percentDecreaseGaptoWelfare\% lower than that which we obtain if we maximize welfare using the same country-specific budgets (see Figure \ref{fig:welfare}). This is true despite the fact that the two approaches yield noticeably different transfer size distributions, as Figure \ref{fig:togo_welfare} illustrates for Togo. In particular, welfare maximization entails more large transfers, including some larger than \$2.15, which can never be optimal for poverty minimization.

These results are not surprising, given that the households most likely to be poor are also likely to have the highest expected marginal utility. Welfare maximization would diverge strongly from poverty minimization if the available predictors allowed us to identify households unlikely to be below the poverty line, but likely to be just above it. In practice, that does not appear to be the case.

\subsubsection{Geographic Predictors}
\label{subsec:geographic}

One alternative to targeting based on the characteristics of individual households is to target based on regional indicators \citep{coady2004targeting,elbers_poverty_2007}. One would expect such ``geographic targeting'' to be less accurate, as it uses fewer and less granular predictors, but simpler and cheaper to implement, and also less subject to gaming by potential recipients or to fraud by program personnel. We therefore examine next how loss-minimizing policies perform when we restrict the predictor sets to include \emph{only} regional indicators. These are the same geographic indicators included in the larger predictor set: indicators for the finest administrative level at which the survey documentation indicates it was designed to be representative, and indicators for urban vs. rural status where available (see Appendix \ref{subsec:predictor-selection} for details).

\begin{figure}[tp]
    \begin{center}
    \caption{Geographic targeting}
    \includegraphics[width=\textwidth]{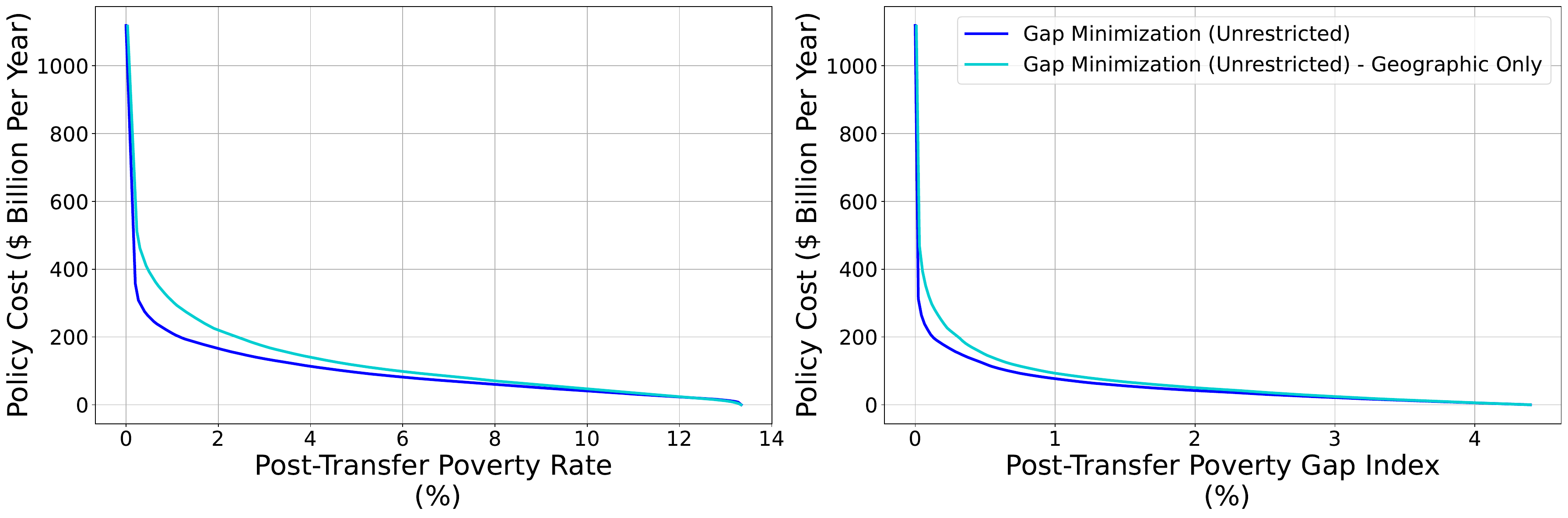}
    \label{fig:geographic}
    \end{center}
    \vspace{-1em}
    {\footnotesize This figure reports the costs, in billions of nominal 2023 USD annually, of the indicated optimal transfer policies learned for the full sample of countries to reduce extreme poverty, analogous to Figure \ref{fig:headline}. It plots values separately for policies that use the full household survey data (blue series) and policies that use only regional indicators (light blue series).}
\end{figure}

As expected, achieving poverty goals costs more when we limit to geographic predictors (Figure \ref{fig:geographic}). But the differences are not prohibitive; achieving a 1\% poverty rate using a gap-minimizing policy learned from only geographic predictors costs \percentIncreaseGeotoSurvey\% more than when using the full household survey data. Poverty reduction using a purely geographic approach appears to be a plausible alternative.

\subsubsection{Satellite-Based Predictors}
\label{subsubsec:sat}

The global availability of remote-sensing data is creating new opportunities for targeting, as remotely-sensed predictors can be obtained at lower cost and higher frequency than survey-based ones \citep{smythe_geographic_2022,yeh_using_2020}. Here, we briefly explore the possibility of determining transfer amounts based on characteristics derived from satellite imagery, rather than collected using household surveys.

We conduct this analysis using data from Togo, where we can precisely link satellite imagery to the original (un-jittered) GPS coordinates of households from a nationally-representative consumption survey \citep{tgo_main_survey,aiken_program_2023}. For this more exploratory analysis of satellite imagery, we deviate from the protocol described in Section \ref{subsec:empirical-methods} by viewing policy performance for several (7) predictor sets before selecting one.\footnote{Although this deviation introduces some risk of human-in-the-loop overfitting, we believe this concern is limited by three factors. First, the number of predictor sets we considered (7) is small relative to the total number of hyperparameters tuned following the data use plan. Second, as discussed in Section \ref{subsec:empirical-methods}, we did not see evidence of inflated performance in Malawi, which we used as a sandbox more so than we have done here in Togo. And third, selection among the satellite predictor sets we considered does not make a large difference: to reach a 1\% post-transfer rate using the best-performing set costs 87\% as much as using the worst-performing.} A full accounting of this process can be found in our \href{https://drive.google.com/file/d/1gdIZD1Mj3rQSdzZS1Fc0mCIASry5_lEa/view?usp=sharing}{Data Use Plan}.

We use the Google Satellite Embedding dataset as our predictor set \citep{brown2025alphaearth}.\footnote{Accessed at 50-meter scale using the Google EarthEngine API on January 12, 2026.} The dataset is structured as a raster, with 64 features (embeddings) per raster point. We match each household in the survey with its nearest raster point which falls within the borders of Togo. We lack location information for 16 of the 6,171 respondent households; in such cases we impute each feature using the mean among households with location data. We then proceed as in Section \ref{subsec:empirical-methods}. 

\begin{figure}[tp]
    \begin{center}
        \caption{Policy Performance using Satellite Predictors (Togo)}
        \label{fig:togo_satellite}
        \includegraphics[width=\textwidth]{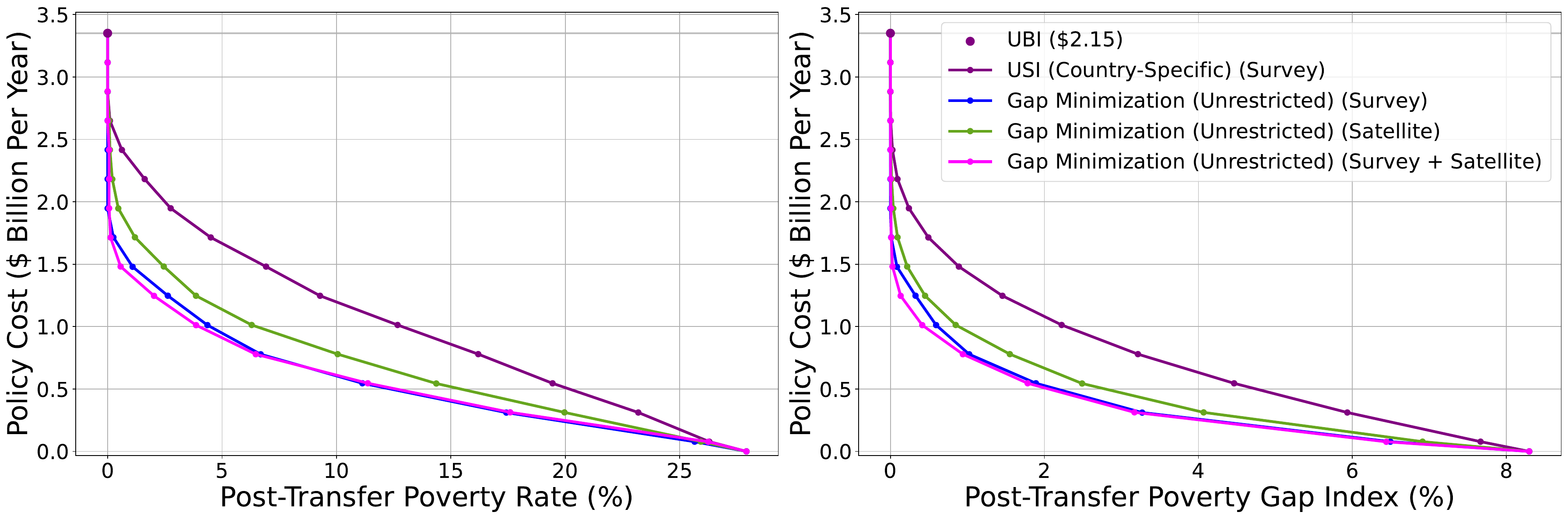}
    \end{center}
    \footnotesize{This figure reports the costs, in billions of nominal 2023 USD annually, of the indicated optimal transfer policies learned for Togo. The left-hand panel plots these against the post-transfer poverty rate, while the right-hand panel plots these against the post-transfer poverty gap index.}
\end{figure}

Figure \ref{fig:togo_satellite} presents the performance of gap-minimizing policies using satellite features only, satellite features in conjunction with survey predictors, and survey predictors only. In isolation, satellite features are useful for targeting, as they reduce the cost of achieving a \nationalTarget\% national poverty rate by \percentDecreaseSatelliteUBI \% relative to providing uniform USI. However, relative to targeting with survey data, the approach based on satellite data alone increases costs by \percentIncreaseSatelliteSurvey\%. Adding satellite features to survey predictors slightly improves performance, reducing costs by \percentDecreaseSatelliteCombined \% relative to targeting with survey features alone. Overall, satellite imagery functions as an imperfect (but less expensive) substitute for survey predictors, and may be useful in conjunction with survey predictors, in this setting.

An offsetting advantage of targeting with satellite imagery alone is the avoided cost of an eligibility questionnaire. Table \ref{tab:togo_costs} reports estimates of total program cost that take this into account, assuming costs of \$\lsmsPerHhCost\ per household surveyed to gather detailed consumption information for training and \$\screeningPerHhCost\ per household to administer an eligibility questionnaire.\footnote{The former is the mean cost of six LSMS surveys from the Sahel region of Africa reported in \citet{kilic2017costing}; the latter is the median of per-household costs in four African PMT surveys reported in \citet{schnitzer2022targeting}.} At these program scales, collecting survey predictors via a questionnaire pays off: it costs an additional \$\alphaEarthAdminSavingsMillions M, but saves \$\surveyTransferSavings M in transfer costs. This reflects the more general point that, as we discuss further below, gathering eligibility data is inexpensive relative to the cost of transfers themselves.

\subsubsection{Training Data Size}
\label{subsec:training_data_size}

In our final exercise we consider the potential benefits of varying the size of the survey dataset used to learn policies. The sample sizes for the underlying surveys on which we draw were set with objectives other than algorithmic policy learning in mind, raising the possibility that larger samples might yield better performance. Since the costs of additional surveys are third-order relative to the costs of the transfer policies we learn, even small gains in accuracy could yield very high returns on investment.

\begin{figure}
    \begin{center}
    \caption{Percent Increase in Policy Cost vs. Percent of Training Set}
    \label{fig:cost_vs_samplesize}
    \includegraphics[width=0.5\textwidth]{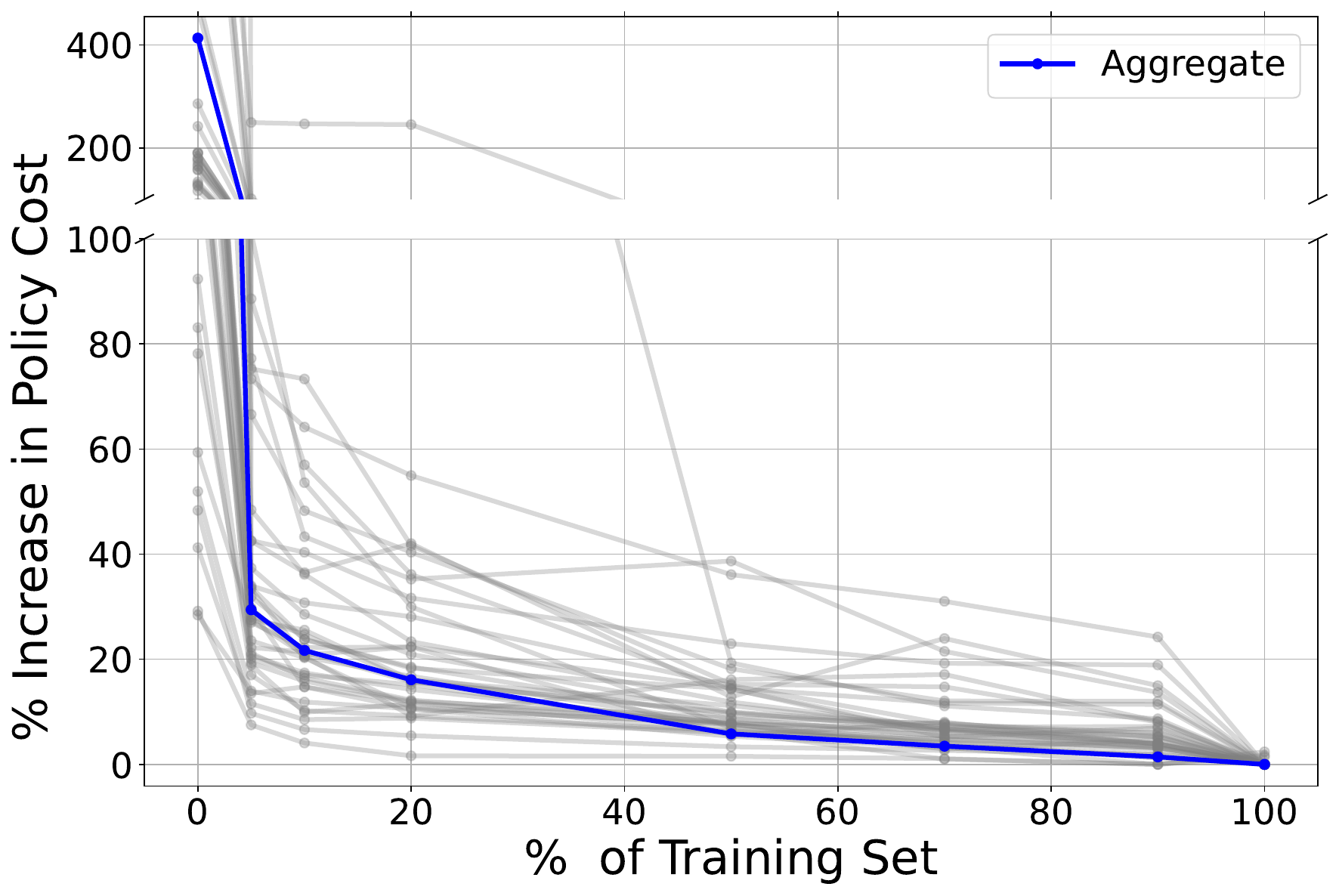}
    \end{center}
    {\footnotesize This figure reports the percentage increase in the cost ($y$-axis) of the optimal gap-minimizing policies as the fraction of the training set ($x$-axis) used for training is reduced, relative to the cost when using 100\% of the training set.}
\label{fig:sample_size}
\end{figure}

Of course, we cannot directly examine the accuracy benefits of surveys \emph{larger} than those actually conducted. Instead we quantify here the performance penalties we incur if we use smaller subsets of the data actually available. Examining how these penalties vary with the size of the dataset may support some informed guesses about the benefits of larger samples. To that end, Figure \ref{fig:cost_vs_samplesize} plots the relationship between the cost of a learned gap-minimizing policy that reduces the poverty rate to 1\%, and the share of the available training data set from each country we utilize. Costs are expressed relative to the benchmark cost when we use all available data. Note that to smooth out sampling variability we repeat this exercise 15 times for 15 subsets of the training data sampled with replacement, and report averages across those draws. 

The results suggest that the benefits of collecting additional data would be modest in relative terms. While costs spike when we use very small fractions of the data, they vary fairly little for values in the 80\%--100\% range. However, because the costs of survey data collection are typically so small relative to those of the transfers themselves, the \emph{rate} of return could still be very high. For instance, costs decrease by \percentDecreaseSampleSize \% as we increase the share of the training set used from 90\% to 100\%. If they were to decrease by a further \percentDecreaseSampleSize \% were surveys 10\% larger, the global savings (computed using the estimates in Section \ref{subsec:global_costs} below) would be on the order of \$3B, while the added data collection cost would likely run in the \$10Ms.

\section{Discussion \& Implications}
\label{sec:discussion}

Table \ref{tab:costs} summarizes the estimated cost of reducing the national poverty rate to \nationalTarget\% in each country in our sample, along with reference information on the scale of national GDP and public revenue. Note that doing so would reduce the global poverty rate to \globalPovertyRateForNationalTarget\%. We conclude in this section by discussing several implications of redistribution on the scale implied by those figures.

\begin{figure}[tp]
\caption{Policy Cost as a Share of GDP and of Government Revenue}
\includegraphics[width=\textwidth]{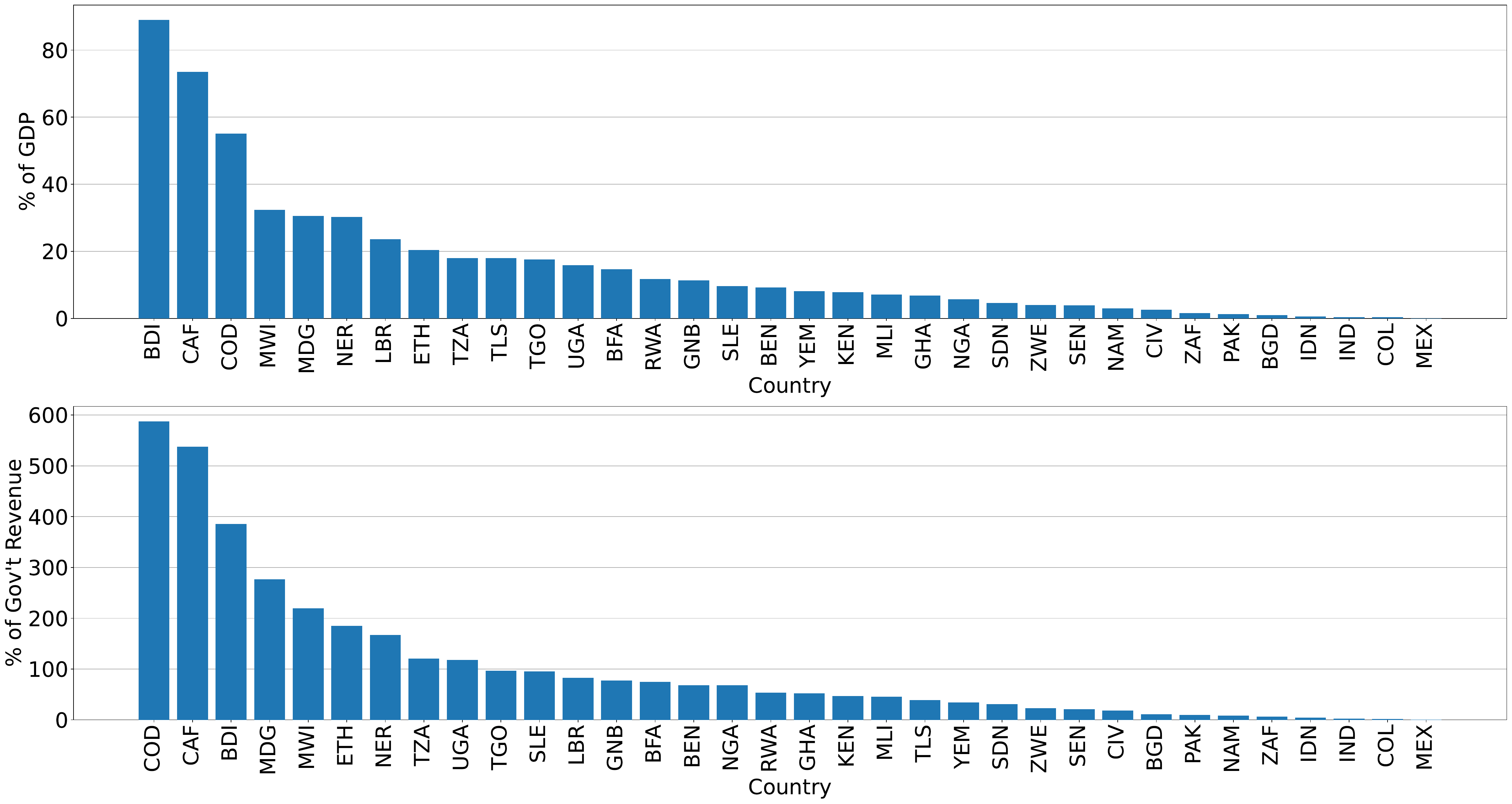}

\vspace{1em}
\footnotesize{This figure plots, for each indicated country, the ratio of the cost of a policy that achieves a \nationalTarget \% poverty rate via gap minimization to the country's GDP in the year the underlying survey was conducted (top panel) and to total government revenue in that year (bottom panel). We obtain country GDP from the World Bank (\href{https://data.worldbank.org}{data source}, accessed 13 August 2025) and government revenue percentages by country from the IMF (\href{https://www.imf.org/en/Home}{data source}, accessed 14 July 2025).}
\label{fig:gdp}
\end{figure}

The first concerns sources of funds. Figure \ref{fig:gdp} plots the amount of money spent in each country, under the learned policy that reduces the national poverty rate to \nationalTarget\%, as a proportion of that country's GDP (top panel) and total government revenue (bottom panel). Taking a simple average over the countries in our sample, implementing such a scheme would cost \sampleCostPercentGDP\% of GDP. To put this in context, this is equivalent to more than 90 years worth of the growth in tax revenue relative to GDP that \citet{BesleyPersson2014tax} document in a sample of 18 countries during the 1900s.\footnote{Figure 3 in their paper shows tax revenue growing from 8\% to 25\% of GDP over the course of 100 years; \sampleCostPercentGDP\% thus corresponds to \sampleCostPercentGDP\% / 17\% $\times$ 100 ~= 94 years worth of growth.} It thus appears plausible that many countries could contribute substantially to the cost of policies like these, but implausible that many could finance them entirely on their own. This assessment is similar to that reached by \citet{HannaOlken2025handbook}. They calculate the cost of transferring PPP \$2.15 per day to everyone below the poverty line, taking this as a rough approximation to the true cost of eliminating poverty. They note that ``in principle, one could bring everyone to the poverty line for less money than this if one could give larger transfers to those further away from the poverty line,'' but also that ``this may understate the extent of the problem, since giving transfers only to those below the poverty line assumes that one can solve targeting challenges...'' After taking both of these issues into account, the policies we learn here end up costing a similarly large share of national income.

Funding such policies using transfers from abroad, on the other hand, would raise additional questions about their macroeconomic effects. One can think of these in two parts: the effects of converting foreign currency to Local Currency Units (LCUs), and the effects of recipients spending or saving those LCUs. Both are large topics in their own right, and beyond the scope of our analysis, but we will remark briefly on related work.

With respect to currency conversion, the amounts in question are several multiples of status quo aid flows. In our sample, a country requires \sampleCostPercentGDP\% of GDP on average to reduce the national poverty rate to \nationalTarget \%, while the 2023 ODA flow to each country is only \sampleOdaPercentGDP \% of its GDP on average. This calls to mind classic questions about whether exchange rate appreciation might undermine the effects of aid more generally \citep{RajanSubramanian2011aid}. Any effects of aid inflows on exchange rates would be mitigated to the extent that transfer recipients subsequently purchased items with imported components, and credibly estimating the effects of net shocks to currency demand is an area of ongoing work.\footnote{One approach uses (instruments for) countries' open-market operations as a source of identifying variation; \citet{AdlerLisackMano2019unveiling}, for example, estimate that a purchase of 1 percentage point of GDP causes a depreciation of the real exchange rate of 1.4--1.7\%. Another approach uses shocks to currency demand caused by changes to the composition of global indices of emerging market bonds; these are arguably more clearly exogenous, but also harder to size, as passive funds' responses are mechanical and thus estimable but those of actively-managed funds are unknown. This approach yields larger elasticities. For example, \citet{BeltranHe2025fx} estimate using this approach that in a sample of countries an inflow of 0.09\% of domestic GDP led to a 1\% appreciation of the exchange rate in the days immediately following, with effects dissipating only partially over the subsequent 12 months (see Table 5.1 and Figure 4 in their paper).} That said, the broad point remains that effects on exchange rates could be large enough to matter.

Domestic spending might affect economic growth, through a demand channel (as in ``big push'' models such as that of \citet{MurphyShleiferVishny1989bigpush}) or a credit channel \citep{StiglitzWeiss1981credit}, among others. Recent evidence from large-scale experiments or natural experiments broadly supports this view: \citet{Eggeretal2022ge} estimate that the economies of villages in rural Kenya expanded by \$2.5 for every \$1 transferred into them, and \citet{GerardNaritomiSilva2024cash} similarly find substantial effects of transfers on economic activity in Brazilian municipalities.\footnote{While the transfers which \citet{GerardNaritomiSilva2024cash} study were largely domestically financed, they study the effects of gross transfers (not net of taxation), which are the relevant ones for thinking about the impacts of externally financed transfers.} Any such knock-on benefits need not accrue proportionally, however---in fact, \citet{haushofer_targeting_2022} find that the transfers in \citet{Eggeretal2022ge} had somewhat larger effects on the income and consumption of households that would otherwise have been somewhat less poor. As a result, it might be too optimistic to assume that one can simply divide the costs of the anti-poverty policies we learn here by an aggregate multiplier.

Another natural and related question is whether there is some upfront outlay that would be sufficient to achieve a given poverty goal on a lasting basis, as opposed to the flow-cost approach we have taken here. This of course inescapably requires assumptions about rates of return. One way to bound the up-front cost is to simply calculate the size of the endowment required to yield the annual flow expenditures we calculate, using any rate of return deemed plausible. But it may be possible to do better still by front-loading transfers to households themselves, as some of them may have access to much-higher-return investment opportunities than endowment managers do (see, for example, \citet{haushofer_targeting_2022} and \citet{Hussametal2022targeting}, among many others). Here the volatility of poverty would pose a challenge. Historically, households have moved back and forth across the extreme poverty line quite frequently \citep{ArmentanoNiehausVogl2025poverty,BaulchHoddinott2000mobility}, so that the amount one would ideally wish to transfer to any given household could vary substantially from year to year \citep[or even month to month; see][]{MerfeldMorduch2024poverty}. A related issue is that, as noted above, policies learned from data at a given point in time may not perform as well in subsequent years if the relationship between consumption and PMT covariates changes over time \citep{Aikenetal2025moving}. A natural extension of our framework would incorporate techniques for addressing such `covariate shift' in the statistical learning framework of Section~\ref{sec:setup} \citep[cf.][]{quinonero2008dataset,koh2021wilds}.

Finally, the gap-minimizing policies we estimate quantify the total cost of transfers themselves, but abstract from the administrative costs of delivering them at scale. In practice, real-world programs will incur costs to identify eligible households, enroll them, and disburse payments. Evidence from costing exercises for national social registries indicates that one-off, census-style registration campaigns generally cost on the order of US\$1–3 per household registered \citep{LindertEtAl2020SourcebookDeliverySystems}.\footnote{For example, Malawi's Unified Beneficiary Registry reports projected Phase~2 registration costs of approximately US\$1.27 million for 785,000 households, or  US\$1.62 per household \citep{LindertEtAl2018MalawiUBR}. For Turkey’s Integrated Social Assistance System, system development costs of roughly US\$13 million covering an estimated 10 million households imply a cost of about US\$1.30 per household \citep{LeiteEtAl2017SocialRegistries}. For Brazil’s \emph{Cadastro Único}, reported registration and re-certification costs range between US\$2.03 and US\$2.06 per household \citep{MostafaSatyro2014CadastroUnico}. For Pakistan's National Socioeconomic Registry, per-household registration costs are estimated between US\$2.30 and US\$2.70 \citep{LindertEtAl2020SourcebookDeliverySystems}.} A second component of implementation cost arises from payment delivery: governments typically compensate payment providers either through small flat per-transaction fees (fractions of a US cent to a few US cents) or through ad valorem fees of roughly 1–3\% of the amount transferred, as in Pakistan's BISP, India's Direct Benefit Transfer programs, and the Philippines' 4Ps \citep{LindertEtAl2020SourcebookDeliverySystems}.\footnote{Transaction fee structures vary across programs but typically fall within a modest range. In Pakistan’s BISP, the program pays a 3\% transaction fee \citep[][pp.~239--240]{LindertEtAl2020SourcebookDeliverySystems}. In India’s Direct Benefit Transfer system for MGNREGS payments, the Government pays a flat Re~0.50 per transaction (split across the sponsor bank, destination bank, and the National Payments Corporation of India), with an additional 1\% service fee permitted for payment service providers \citep[][p.~240]{LindertEtAl2020SourcebookDeliverySystems}. In the  Philippines' 4Ps program, the government negotiated a simplified service agreement that kept bank fees ``under~1\%'' of the payment value \citep[][p.~239]{LindertEtAl2020SourcebookDeliverySystems}.} Relative to the total value of cash transfer policies we evaluate, these fixed and recurring costs thus appear likely to be modest. In countries without foundational identification or payment infrastructure, larger one-time investments may be required to establish those systems.

\subsection{Costs at Global Scale}
\label{subsec:global_costs}

While the analysis above estimates costs for a sample of countries, as of the dates they were surveyed, it is natural to wonder what it implies for the cost of ending poverty at a global scale today. We consider in particular the cost of reducing the global poverty rate in 2023 to \globalTarget \%. To do so it is sufficient to obtain a national poverty rate of \nationalPovertyRateForGlobalTarget \% in all countries that currently have higher rates, as some other countries already have rates lower than \globalTarget \% which bring down the (population-weighted) average. 

\begin{figure}
\caption{In-sample fit and common support}
\includegraphics[width=\textwidth]{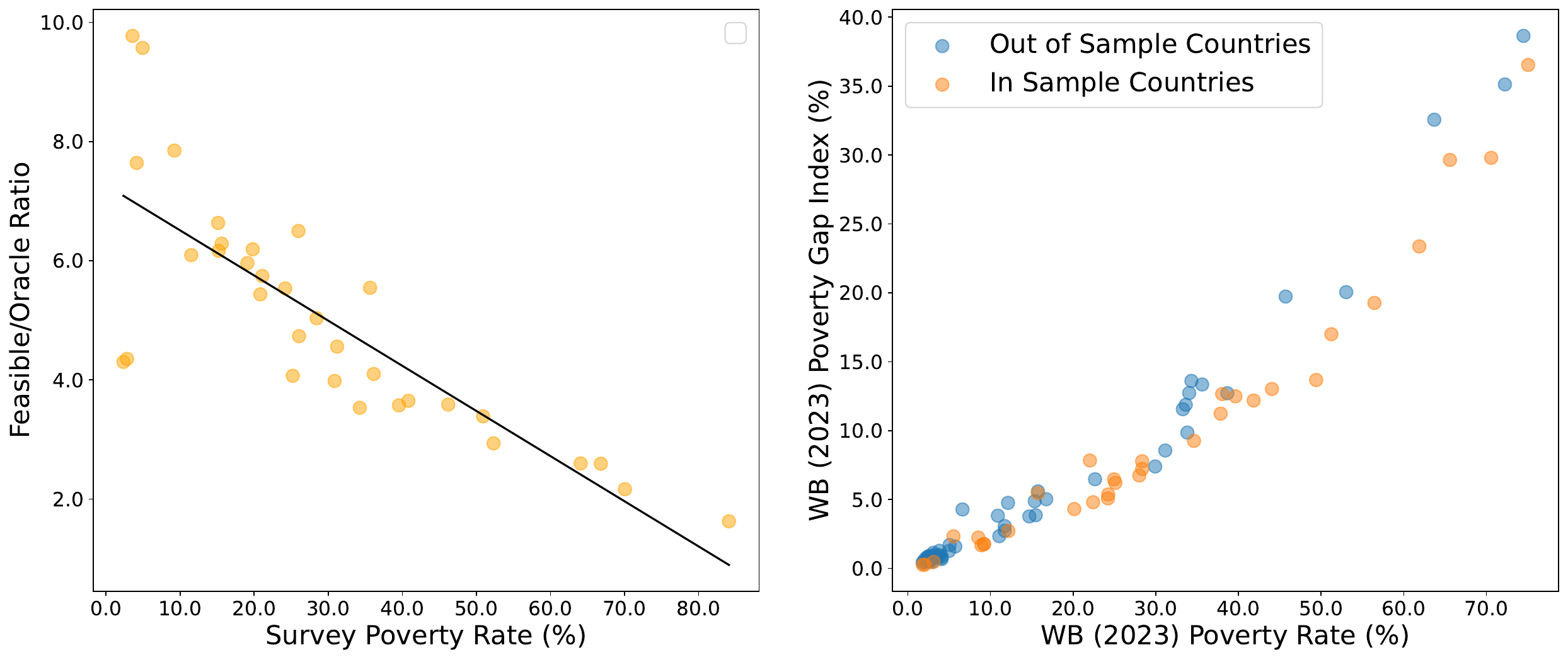}
\label{fig:scatter}
\vspace{-1em}\\
{\footnotesize The left-hand panel plots the in-sample relationship between the feasible-to-oracle policy cost ratio (given a 1.6\% poverty rate target) and the poverty rate, calculated from the various surveys we use. The right-hand panel plots each country's poverty gap index and poverty rate, using estimates as of 2023 from the World Bank, differentiating between in-sample and out-of-sample countries.}
\end{figure}

To estimate the cost of doing so, we first estimate the cost of achieving a \nationalPovertyRateForGlobalTarget \% rate in each country in our sample using unrestricted gap-minimizing transfers. We then estimate, within that sample, the relationship between (i) the ratio of the feasible policy cost to the poverty gap---which one can think of as a measure of how much one must ``overpay'' due to imperfect information---and (ii) the poverty rate. Intuitively, we expect the ratio (i) to be higher in countries with lower poverty rates, since targeting there is a harder problem. The left-hand panel of Figure \ref{fig:scatter} shows that this is the case, and that we obtain an approximately linear fit, with an $R^2$ of \extrapolationWBRSquared.\footnote{Note that we truncate fitted values so that (conservatively) they never fall below 1. We have also re-estimated results using a quadratic fit, in which case the total estimated cost changes from \extrapolationWBGlobalGDP \% to \extrapolationWBQuadraticGlobalGDP\% of global GDP. \extrapolationWBDroppedCountriesGap\  are excluded from this calculation because the corresponding conversion factors are not available from the World Bank.} We then use this estimated relationship to project the cost of achieving a poverty rate of at most \nationalPovertyRateForGlobalTarget \% in each country. We do so using estimates of poverty gaps and rates as of 2023 for all countries from the World Bank, which are based on the methodology in \citet{Mahleretal2026parsimonious}. This exercise necessarily involves extrapolation across space and across time; that said, the supports of the distributions involved overlap well (see the right-hand panel of Figure \ref{fig:scatter}).

The net result is an estimated cost of reducing the global poverty rate to \globalTarget \% of \$\extrapolationWBCost B per year, equal to \extrapolationWBGlobalGDP \% of global GDP, or \extrapolationWBOECDGDPPercent \% of OECD GDP.\footnote{Including China, the world's second-largest economy and now a major international aid donor, lowers this figure to \extrapolationWBOECDPlusChinaGDPPercent \%.} This is large in relation to some relevant benchmarks: OECD donors gave foreign aid equal to 0.21\% of global GDP in 2023,\footnote{In 2023, OECD foreign aid totaled \$223B (\url{https://www.oecd.org/en/events/2025/01/official-development-assistance-oda-2023-final-figures.html}, accessed 6 December 2025) and global GDP totalled $107T$ (\url{https://data.worldbank.org/indicator/NY.GDP.MKTP.CD}, accessed 6 December 2025).} and aid levels have since fallen sharply, particularly in the United States. In terms of sheer fiscal feasibility, however, there is no question that wealthy countries could finance a policy at this scale. Nor is the amount large relative to other, arguably less critical, global expenditures. For instance, the world spends eight times as much---2.2\% of global GDP---on alcoholic beverages.\footnote{Estimated 2024 global alcoholic beverages market size: \$2,413B as per \url{https://www.fortunebusinessinsights.com/alcoholic-beverages-market-107439}, accessed 1 September 2025. Estimated 2024 GDP: \$110,000B as per \url{https://www.imf.org/external/datamapper/NGDPD@WEO/OEMDC/ADVEC/WEOWORLD}, accessed 1 September 2025.} 

Over time the costs of such a scheme would fall if non-transfer incomes rose. This could be a result of the scheme itself if, as has typically been the case, recipients invested a share of their transfers. It could also result from economic growth unrelated to redistribution. Yet, as we noted in opening the paper, current projections are that growth in most of the countries now home to extremely poor people will slow. To see what this might imply for the cost of ending poverty, we repeat the exercise above using estimated poverty gaps and rates from the World Poverty Clock \citep{wpc}, which provides these both contemporaneously and also forecast to 2030. For 2023 these yield estimated costs of \extrapolationWPCGlobalGDP \% of global GDP, or \extrapolationWPCOECDGDPPercent \% of OECD GDP, reassuringly close to those we get using the World Bank estimates. For 2030, meanwhile, they yield estimated costs equal to \extrapolationWPCFutureCostPercentGDP \% of (current) global GDP, or \extrapolationWPCFutureCostPercentOECDGDP \% of OECD GDP. This is lower (by \extrapolationFutureCostPercentDecrease\%) than the 2023 WPC estimate, but also reinforces the point that on the current trajectory poverty will not vanish on its own. 

\begin{table}[tp]
    \centering
    \begin{threeparttable}
        \caption{Stated support for public \& private contributions}
        \label{tab:stated_support}
        \begin{tabular}{lccc}
            \toprule
             & \multicolumn{2}{c}{Support 0.5\% of GDP policy} & Would donate \\
            \cmidrule(lr){2-3}
             & To governments & Direct to people & 0.5\% of income \\
            \midrule
            OECD     & 73.5\% & 70.7\% & 59.5\% \\
            non-OECD & 90.0\% & 90.5\% & ---    \\
            \midrule
            All      & 82.4\% & 81.3\% & 59.5\% \\
            \bottomrule
        \end{tabular}

        \begin{tablenotes}[para] 
            \footnotesize
            \textit{Notes:} $n = 14,150$ respondents from 9 OECD countries (Australia, Denmark, France, Germany, Italy, Spain, Sweden, UK, USA) \& 9 non-OECD countries (Brazil, China, Ghana, India, Indonesia, Kenya, Nigeria, Pakistan, UAE). Responses collected between 15 January and 20 February 2026. Source: \citet{BanerjeeDufloGreenstone2026}.
        \end{tablenotes}
    \end{threeparttable}
\end{table}

Could redistribution at this scale garner widespread support? Recent political developments might seem to suggest otherwise. That said, Table \ref{tab:stated_support} reports related polling numbers, reproduced with permission from \citet{BanerjeeDufloGreenstone2026}, collected in 18 countries during January and February of 2026. The first column indicates that a sizeable majority (73.5\%) in OECD countries (and, less surprisingly, even more in non-OECD ones) state support for ``a policy of rich countries contributing 0.5\% of their GDP, to be given as foreign aid to governments in poor countries.''\footnote{Similarly, \citet{Fabre2025acceptance} has recently estimated that there is majority stated support in eleven high-income countries for levels of global redistribution even higher than those contemplated here.} Voters might of course be more comfortable with transfers to foreign governments than to foreign individuals, but the next column shows that support remains essentially unchanged if the public aid were ``to be given directly to people living in poor countries'' rather than to their governments. The final column then considers the scope for voluntary redistribution, not coordinated and enforced by public authorities. Respondents were asked, ``suppose that if everyone in rich countries voluntarily donated 0.5\% of their income to people living in extreme poverty, this would be enough to eliminate extreme poverty around the world. Would you personally donate 0.5\% of your income to such an effort?''--- whether, in other words, they would voluntarily ``do their part'' if there were a convincing case that everyone doing so could end poverty. Even here, a sizeable majority (59.5\%) agreed.

\clearpage
\singlespacing
\bibliographystyle{plainnat} 
\bibliography{ref.bib}

\clearpage

\appendix

\section{Algorithms}
\label{sec:alg}

In this section, we outline algorithms for learning unrestricted policies that minimize the poverty gap index, minimize the poverty rate, and maximize welfare, as well as binary policies that minimize an arbitrary loss function. We also discuss the weakly equitable oracle policy that minimizes the poverty rate.

\subsection{Poverty Gap Minimization}
We aim to solve \eqref{eq:prob} where the objective corresponds to the poverty gap index. In this case, \eqref{eq:prob} has a convex objective and is straightforward to solve using convex duality. We provide a characterization of the optimal gap-minimizing policy, a high-level overview for learning the optimal gap-minimizing policy in the population regime where we have access to the population distribution $F$, and finite-sample implementation details.

\begin{lemm}
\label{lemm:opt_solution_gap}
Suppose that Assumption \ref{assumption:cdf} holds. If $L(z) = (c - z)_{+}$ and $0 < B< c$, then the unique optimal policy that solves \eqref{eq:prob} satisfies
\begin{equation} 
\label{eq:form_gap_policy}
t^{*}_{\mathrm{gap}}(x) = (c - F_{Y \mid X=x}^{-1}(\lambda^{*}(B)))_{+}.\end{equation}
where $\lambda^{*}(B) \in [0,1]$ and $\lambda^{*}(B)$ chosen so that the budget constraint is satisfied with equality.
\end{lemm}
Lemma \ref{lemm:opt_solution_gap} implies that the family of optimal gap-minimizing policies can be parametrized as
\[ t_{\lambda}(x) = (c - F_{Y \mid X=x}^{-1}(\lambda))_{+}\]
for $\lambda \in [0, 1].$
Each policy in this family minimizes the poverty gap index for some budget $B$. Given this functional form, learning the optimal gap-minimizing policy for a particular budget $B$ boils down to finding the appropriate choice of $\lambda^{*}(B)$. We can obtain the optimal gap-minimizing policy by discretizing the interval $[0, 1]$ into a grid of quantiles $\lambda_{1}, \lambda_{2}, \dots \lambda_{J}$ and selecting $j \in [J]$ that minimizes the difference $|\EE[F]{t_{\lambda_{j}}(X)} - B|.$

\subsubsection{Finite-Data Regime}
In realistic settings, we only have access to data sampled from $F$. In this section, we assume the true data distribution $F$ is unknown but we have access to a training set $\mathcal{D}_{\text{train}} = \{(X_{i}, Y_{i})\}_{i=1}^{n_{\text{train}}}$ where $(X_{i}, Y_{i}) \sim F$ i.i.d., and we aim to optimize allocations for unlabeled samples $\mathcal{D}_{\text{test}} = \{ X_{i}\}_{i=1}^{n_{\text{test}}}$ where $X_{i} \sim F$ i.i.d. We describe the algorithm for estimating the optimal gap-minimizing policy for a budget $B$.

First, we discretize the interval $[0, 1]$ into a grid of quantiles $\lambda_{1}, \lambda_{2}, \dots \lambda_{J}$. Second, estimate each $q_{\lambda_{j}}(\cdot)$ for $j=1,2 \dots J$ with minimal distributional assumptions on $F$. Using the training set $\mathcal{D}_{\text{train}} = \{(X_{i}, Y_{i})\}$, we estimate each $\hat{q}_{\lambda_{j}}(\cdot)$ by solving a conditional quantile regression problem for quantile $\lambda_{j}$ via deep learning. In our implementation, for each $j$, we train a neural network to minimize the pinball loss parametrized by the quantile $\lambda_{j}$ \citep{koenker1978regression}.
	Recall that the pinball loss is given by
	\[ L_{\text{pinball}}(z; \lambda) = \lambda \cdot |z| \cdot \mathbb{I}( z \geq 0) + (1 - \lambda) \cdot |z| \cdot \mathbb{I}(z < 0),\]
	and we have that
	\[ q_{\lambda}(\cdot) \in \argmin_{q} \EE[F]{L_{\text{pinball}}(Y - q(X); \lambda)}.\]
	Our implementation represents $q$ with a neural network and solves the following empirical risk minimization problem
	\[ \hat{q}_{\lambda_{j}}(\cdot) \in \argmin_{q} \hEE[F]{L_{\text{pinball}}(Y - q(X); \lambda_{j})}.\]

After estimating the conditional quantiles, we use covariates from the test set $\mathcal{D}_{\text{test}} = \{X_{i}\}_{i=1}^{n_{\text{test}}}$ to estimate the optimal transfer amount for each unit in the test set.
	\[ \hat{t}_{\lambda_{j}}(X_{i}) := (c - \hat{q}_{\lambda_{j}}(X_{i}))_{+}.\]
For each $\lambda_{j}$, we can estimate the empirical policy cost. Finally, we can select the choice of $\lambda_{j}$ so that $\hEE[F]{t_{\lambda_{j}}(X_{i})} =  \frac{1}{n_{\text{test}}} \sum_{i=1}^{n_{\text{test}}} \hat{t}_{\lambda_{j}}(X_{i})$ is closest to $B$.

\subsection{Poverty Rate Minimization}
We aim to solve \eqref{eq:prob} where the objective corresponds to the poverty rate. In this case, \eqref{eq:prob} is not straightforward to solve because it has a non-convex objective. Nevertheless, we provide a characterization of the optimal deterministic rate-minimizing policy and outline an algorithm for learning a stochastic policy that is guaranteed to obtain lower or equal poverty rate than the optimal deterministic rate-minimizing policy. We also provide a high-level overview of the algorithm in the population regime where we have access to the population distribution $F$ and finite-sample implementation details.

In this section, we make the following technical assumption that the covariate space $\mathcal{X}$ is finite and discrete to permit $t$ to be finite-dimensional because the existence of minimizers of non-convex functionals over infinite-dimensional spaces is not guaranteed.
\begin{assumption}
\label{assumption:X_finite}
The covariate space $\mathcal{X}$ is finite and discrete.
\end{assumption}

We show that when $L(z) := \mathbb{I}(z < c)$, \eqref{eq:prob} has an optimal deterministic policy, and any optimal deterministic policy has a highly-structured form; it satisfies a property called $\alpha$-validity for some $\alpha > 0.$ 

\begin{defi}
For any $\alpha \geq 0$ and covariate $x \in \mathcal{X}$, the set of $\alpha$-valid transfers at $x$ is 
\begin{equation} \mathcal{T}_{\alpha}(x) := \{0, c\} \cup \{t \in (0, c) \mid f_{Y \mid X=x}(c - t) = \alpha \}.\end{equation}
\end{defi}

Using the definition of $\alpha$-valid transfers, we can now define an $\alpha$-valid transfer policy.
\begin{defi}
A transfer policy $t(\cdot)$ is $\alpha$-valid if $t(x) \in \mathcal{T}_{\alpha}(x)$ for every $x \in \mathcal{X}$. Note that $t(x)$ may make a potentially randomized choice from the set.
\end{defi}

\begin{theo}
\label{theo:alpha_valid}
Suppose that Assumptions \ref{assumption:cdf}, \ref{assumption:X_finite} hold, and $L(z):=\mathbb{I}(z < c)$ and the maximum transfer size is $c$. There exists an optimal deterministic policy that solves \eqref{eq:prob} and any optimal deterministic policy must be $\alpha$-valid for some $\alpha \geq 0.$
\end{theo}

Note that due to the non-convexity of the objective, the optimal deterministic policy is not necessarily unique. To find an optimal deterministic policy, we can recast poverty rate minimization as a two-level optimization problem, where the outer optimization sweeps over all possible $\alpha$ values and the inner optimization finds an optimal deterministic $\alpha$-valid policy among the class of deterministic $\alpha$-valid policies.
\begin{coro}
\label{coro:two_stage}
Suppose that Assumptions \ref{assumption:cdf}, \ref{assumption:X_finite} hold. If $L(z):=\mathbb{I}(z < c)$, \eqref{eq:prob} can be re-written as
\begin{equation} \label{eq:two_stage} \min_{\alpha \geq 0} \min_{t: \mathcal{X} \rightarrow \mathbb{R}_{+}} \{ \PP[F]{t(X) + Y < c} : \EE[F]{t(X)} \leq B \text{ and } t \text{ is } \alpha\text{-valid}\}.\end{equation}
\end{coro}
Note that the inner optimization problem is always feasible because $t(x)=0$ satisfies the budget constraint and is $\alpha$-valid. The inner optimization problem of \eqref{eq:two_stage} finds an optimal deterministic $\alpha$-valid policy among the class of deterministic $\alpha$-valid policies. We demonstrate that this is equivalent to solving a multiple-choice knapsack problem.

For technical convenience, we assume that for any covariate $x \in \mathcal{X}$, the set of $\alpha$-valid transfers $\mathcal{T}_{\alpha}(x)$ contains at most $K$ points for some $K < \infty$.
\begin{assumption}
\label{assumption:finite_intersections}
There exists $K < \infty$ such that $\sup_{x \in \mathcal{X}, \alpha \in \mathcal{A}} |\mathcal{T}_{\alpha}(x)| \leq K$.
\end{assumption}
This assumption requires the conditional distributions $f_{Y \mid X}$ to have a finite number of modes.

Let $\mathcal{A}$ be the space of plausible $\alpha$ values. We note that $\mathcal{A}$ is bounded because $\alpha \geq 0$ and $\alpha \leq \sup_{x \in \mathcal{X}, y \in \mathcal{Y}} f_{Y \mid X=x}(y)$. Define $z: \mathcal{X} \times \mathcal{A} \rightarrow \mathbb{R}^{K}$, where $z_{k}(x; \alpha)$ is the $k$-th smallest element of $\mathcal{T}_{\alpha}(x)$ and $z_{k}(x; \alpha) = c$ for  $|\mathcal{T}_{\alpha}(x)| < k \leq K.$ Let $p: \mathcal{X} \times \mathcal{A} \rightarrow \mathbb{R}^{K}$, where $p_{k}(x; \alpha) = \PP[F]{z_{k}(X; \alpha) + Y < c \mid X=x}.$ Note that $z$ is a vector that captures the $\alpha$-valid transfer amounts, and $p$ is a vector that captures the conditional post-transfer poverty rates at these amounts.

\begin{coro}
\label{coro:knapsack}
Fix $\alpha > 0$. Let $\pi: \mathcal{X} \rightarrow [0, 1]^{K}.$ Under Assumption \ref{assumption:finite_intersections}, the solution to the inner optimization problem in \eqref{eq:two_stage} is given by \[t(x) = \langle \pi(x), z(x; \alpha) \rangle \] where $\pi$ is the solution to a multiple-choice knapsack problem:
\begin{equation}
\label{eq:knapsack}
\begin{split}
\text{\normalfont minimize } &\EE[F]{\langle \pi(X), p(X; \alpha) \rangle}\\
\text{\normalfont subject to } &\EE[F]{\langle \pi(X), z(X; \alpha) \rangle} \leq B \\
&\langle \pi(x), \mathbf{1} \rangle = 1 \quad \forall x \in \mathcal{X}\\
&\pi_{k}(x) \in \{0, 1\} \quad \forall x \in \mathcal{X}, k \in [K].
\end{split}
\end{equation}
\end{coro}

The program in \eqref{eq:knapsack} is a multiple-choice knapsack problem. In the multiple-choice knapsack problem, the goal is to fill a knapsack up to a capacity $B$ by selecting exactly one item out of each class of items, where each item has an associated ``loss'' and ``weight.'' In our problem, each class corresponds to a unit with covariates $x$ and the items in the class are the $\alpha$-valid transfers $\mathcal{T}_{\alpha}(x)$. For clarity of exposition, if $\mathcal{T}_{\alpha}(x)$ contains less than $K$ elements, we pad the class of items with transfers $c$ in the formulation in \eqref{eq:knapsack} until each class consists of $K$ transfers. For item $k$ in the class corresponding to covariates $x$, the loss is given by the conditional post-transfer poverty rate $p_{k}(x; \alpha)$ and the weight is given by the transfer amount $z_{k}(x; \alpha)$. 

Whereas the standard multiple-choice knapsack problem is NP-hard, its fractional relaxation can be solved using a computationally efficient algorithm. We leverage this connection to develop a practical algorithm for solving \eqref{eq:knapsack}. In the fractional formulation, we permit the selection of fractional amounts of items, which permits fractional values $\pi_{k}(x) \in [0, 1]$. We emphasize that in the absence of additional assumptions, solving the fractional relaxation of \eqref{eq:knapsack} yields a stochastic policy that is guaranteed to obtain a poverty rate less than or equal to the optimal deterministic $\alpha$-valid policy. We provide conditions under which the solution to the fractional relaxation is equal to the solution to the original problem \eqref{eq:knapsack} in Appendix \ref{sec:alg_details}.

Thus, our overall algorithm for solving \eqref{eq:prob} where the objective corresponds to the poverty rate relies on the representation provided in \eqref{eq:two_stage}. We define a grid $[\alpha_{1}, \alpha_{2}, \dots, \alpha_{J}]$ over $\mathcal{A}$ and solve the fractional knapsack problem for every $\alpha_{j}$ in the grid to obtain a policy $t_{\alpha_{j}}(\cdot)$. Finally, we return the policy $t^{*}_{\text{rate}}$ that yields the lowest poverty rate over all $t_{\alpha}$ for $\alpha$ in the grid. 

\subsubsection{Algorithm}
\label{sec:alg_details}
We provide an algorithm for poverty rate minimization, which leverages a computationally efficient algorithm for solving the fractional multiple-choice knapsack problem.

\begin{algorithm}
\SetKwData{Talpha}{$t_\alpha$}
\SetKwData{RateList}{rates}
\SetKwData{RateAlpha}{rate}
\SetKwData{PolicyList}{policies}
\SetKwFunction{SolveFMCKP}{SolveFractionalMCKnapsack}
\SetKwFunction{CvxHull}{ComputeLowerConvexHull}
\SetKwFunction{AlphaValid}{ComputeAlphaValidTransfers}
\SetKwInOut{Input}{Input}\SetKwInOut{Output}{Output}
\Input{Conditional distributions $\{f_{Y \mid X=x}(\cdot) \}_{x \in \mathcal{X}}$, \\
       Covariate distribution $f_{X}(\cdot)$, \\ 
       Budget $B$, \\
       Threshold $c$, \\
       Grid size $m$}
\Output{Transfer policy $t^{*}$}
\BlankLine
\RateList $\leftarrow \emptyset$ \\
\PolicyList $\leftarrow \emptyset$ \\
Define $\alpha_{\max}$ using $\{ f_{Y \mid X=x}(\cdot)\}_{x \in \mathcal{X}}.$ \\
\For{$\alpha \in [\frac{\alpha_{\max}}{m}, \frac{2\alpha_{\max}}{m}, \dots \frac{(m -1) \cdot \alpha_{\max}}{m}, \alpha_{\max}]$}{
\For{$x \in \mathcal{X}$}{
$\mathcal{T}_{\alpha}(x) \leftarrow$ \AlphaValid{$f_{Y \mid X=x}(\cdot),\, \alpha, \, c$} \\
$\mathcal{C}_{\alpha}(x), \rho(x; \alpha) \leftarrow $ \CvxHull{$f_{Y \mid X=x}(\cdot), \, \mathcal{T}_{\alpha}(x)$}  \\
}
\emph{Apply Algorithm \ref{alg:priority}} \\
$t_{\alpha}$, \RateAlpha $\leftarrow$ \SolveFMCKP{$f_{X}(\cdot)\, \{(\mathcal{C}_{\alpha}(x), \rho(x; \alpha))\}_{x \in \mathcal{X}}, \, B $}\\
\RateList.append(\RateAlpha) \\
\PolicyList.append($t_{\alpha}$)\\
}
$t^{*} \leftarrow$ Minimum poverty rate policy in \PolicyList (based on \RateList) \\
\Return $t^{*}$
\caption{Two-level optimization procedure for poverty rate minimization.}
\label{alg:main}
\end{algorithm}

The outer loop of our algorithm is a grid search over plausible $\alpha$ values. Let $C := \sup_{x \in \mathcal{X}, y \in \mathcal{Y}} f_{Y \mid X=x}(y)$. We note that for any $\alpha > C$ the set $\mathcal{T}_{\alpha}(x)$ will only consist of $\{0, c\}$ for all $x$, meaning that the inner optimization problem is identical for these values of $\alpha$. As a result, it is sufficient to define $\alpha_{\max} = C+ \delta$ for some small $\delta > 0$ and restrict $\alpha$ to $[0, \alpha_{\max}]$. In Algorithm \ref{alg:main}, we grid this interval into $m$ values.

The inner loop of our algorithm finds the optimal $\alpha$-valid policy among the set of stochastic $\alpha$-valid policies by solving a fractional multiple-choice knapsack problem. Nevertheless, the solution is guaranteed to have equal or lower policy cost than the optimal deterministic $\alpha$-valid policy.

We solve the fractional multiple-choice knapsack problem using the computationally-efficient algorithm of \citet{zemel1980linear}. This procedure is also used by \citet{sverdrup2023qini} to solve a cost-constrained treatment allocation problem. The algorithm relies on the key observation that in the optimal solution to the fractional relaxation of \eqref{eq:knapsack}, the only transfer amounts that are active are the ones that lie on the lower convex hull of the loss-weight plane. For any $x \in \mathcal{X},$ define the lower convex hull of $(z_{k}(x;\alpha), p_{k}(x; \alpha))$ for $k=1, \dots K$ to be a set $\mathcal{C}_{\alpha}(x)$ of points with the ordering $k_{1}(x), \dots k_{|\mathcal{C}_{\alpha}(x)|}(x)$ such that 
\begin{align*}
 z_{k_{1}(x)}(x; \alpha) &<  z_{k_{2}(x)}(x; \alpha) < \dots < z_{k_{|\mathcal{C}_{\alpha}(x)|}(x)}(x; \alpha), \\
p_{k_{1}(x)}(x; \alpha) &> p_{k_{2}(x)}(x; \alpha) > \dots > p_{k_{|\mathcal{C}_{\alpha}(x)|}(x)}(x; \alpha), \\
 \rho_{k_{1}(x)}(x; \alpha) &< \rho_{k_{2}(x)}(x; \alpha) < \dots <\rho_{k_{|\mathcal{C}_{\alpha}(x)|}}(x; \alpha) < 0,
\end{align*}
where $\rho$ is the incremental cost-loss ratio for points on this convex hull as 
\begin{equation} \label{eq:cost_benefit_ratio} \rho_{k_{j}(x)}(x; \alpha) := \frac{p_{k_{j+1}(x)}(x; \alpha) - p_{k_{j}(x)}(x; \alpha)}{z_{k_{j+1}(x)}(x; \alpha) - z_{k_{j}(x)}(x; \alpha)} \quad k_{j+1}(x), k_{j}(x) \in \mathcal{C}_{\alpha}(x) \end{equation}
and let $\rho_{k}(x) := \infty$ if $k \notin \{k_{1}(x), \dots k_{|\mathcal{C}_{\alpha}(x)|}(x)\}.$

We define a thresholding rule on the incremental cost-loss ratio \eqref{eq:cost_benefit_ratio}.

\begin{defi}[Thresholding rule, \citep{sverdrup2023qini}]
\label{def:threshold}
Given cost-loss ratios $\rho$ as in \eqref{eq:cost_benefit_ratio}, a threshold $\lambda \leq 0$, and an interpolation value $\eta \in [0, 1)$, we define a thresholding rule as 
\[ T_{k_{j}(x)}(x; \rho, \lambda, \eta) = \begin{cases} 1 & \text{ if } \rho_{k_{j}(x)}(x) < \lambda < \rho_{k_{j+1}(x)}(x) \\ \eta & \text{ if } \rho_{k_{j}(x)}(x) = \lambda \\
1- \eta & \text{ if } \rho_{k_{j+1}(x)}(x) = \lambda \\
0 & \text{o.w.} \end{cases}.
\]
\end{defi}
The fractional multiple-choice knapsack algorithm can be used to obtain the optimal (stochastic) policy. For a detailed description of the algorithm, we refer readers to Algorithm \ref{alg:priority}. At the start of the procedure, all units are assigned the transfer amount $0$, which corresponds to the first point $k_{1}(x)$ on their convex hulls. Then, we initialize a priority queue. For each $x \in \mathcal{X}$, we add the pair of $x$ and the second point $k_{2}(x)$ on its convex hull $\mathcal{C}(x)$ to the queue with priority $\rho_{k_{1}}(x)$. While the post-transfer poverty rate constraint has slack and the queue is nonempty, we pop points from the queue. When a unit with covariates $x$ and point $k_{j}(x)$ on its convex hull is popped from the queue, the unit is assigned the corresponding transfer value $z_{k_{j}(x)}(x)$. If the unit has additional points on its convex hull, then the next point on its convex hull is added to the queue with priority equal to its cost-weight ratio $\rho_{k_{j+1}(x)}$.  The sequence of updates is dictated by the priority queue ordered by $\rho$. When solving the problem in the population case, the time-complexity of the algorithm is log-linear in $|\mathcal{X}| \cdot K.$ While this time-complexity may seem prohibitively large when the covariate space is continuous, in practice, this algorithm is only applied to units in the test set, so the time-complexity is at most $n_{\text{test}} \cdot K.$

The time complexity of this algorithm is $|\mathcal{X}| \cdot K \log |\mathcal{X}| \cdot K$. The worst-case run time arises when $ B > \EE[F]{z_{k_{|\mathcal{C}_{\alpha}(X)|}}(X)},$ so all points on the convex hull for all units will be added and removed from the priority queue. Since there are at most $|\mathcal{X}| \cdot K$ points added and removed this yields a complexity of $|\mathcal{X}| \cdot K \log |\mathcal{X}| \cdot K$.

\begin{algorithm}
\SetKwInOut{Input}{Input}\SetKwInOut{Output}{Output}
\SetKwData{Cost}{cost}
\SetKwData{Loss}{loss}
\SetKwData{Queue}{queue}
\SetKw{Break}{break}
\SetKw{Return}{return}
\SetKwFunction{PolicyCost}{ComputePolicyCost}
\Input{Convex hulls and incremental cost-loss ratios $\{ (\mathcal{C}(x), \rho(x))\}_{x \in \mathcal{X}},$ \\
       Covariate distribution $f_{X}(\cdot)$, \\ 
       Budget $B$}
\Output{Transfer policy $t$, Poverty rate \Loss}
\BlankLine
\emph{Initialize priority queue.} \\
\Cost $\leftarrow 0$, \quad \Loss $\leftarrow 0$ \\
\Queue $\leftarrow \emptyset$ \\
\For{$x \in \mathcal{X}$}{
 \emph{Assign unit to first point on convex hull.} \\
 $k_{1}(x) \leftarrow $ First point on convex hull $\mathcal{C}(x)$ \\
 \Cost += $f_{X}(x) \cdot z_{k_{1}(x)}(x)$, \quad \Loss += $f_{X}(x) \cdot p_{k_{1}(x)}(x)$ \\ 
 \emph{Enqueue next point on hull.} \\
 $k_{2}(x) \leftarrow $ Second point on convex hull $\mathcal{C}(x)$ \\
 \Queue.add($(x, k_{2}(x))$ with priority $\rho_{k_{1}(x))}$)
}
\While{\Cost $< B$ and \Queue.size() $> 0$ }{
	$(x, k_{j}(x)) \leftarrow $ \Queue.pop() \\
	\emph{Subtract current transfer to unit $x$ from \Cost and \Loss.} \\
	$\Cost -= f_{X}(x) \cdot z_{k_{j-1}(x)}(x)$, \quad $\Loss -= f_{X}(x) \cdot p_{k_{j-1}(x)}(x)$ \\
	\emph{Allocate transfer $k_{j}(x)$ to unit $x$, record new \Cost and \Loss.} \\
	$t(x) \leftarrow z_{k_{j}(x)}(x)$ \\
	$\Cost += f_{X}(x) \cdot z_{k_{j}(x)}(x)$, \quad $\Loss += f_{X}(x) \cdot p_{k_{j}(x)}(x)$ \\
	\If{ $\Cost  > B$}{
	   Perform fractional adjustment for unit $x$-- updating $t$, \Cost, \Loss. \\
	   \Break
	}
 \If{there remain points on convex hull for unit $x$}{
    $k_{j+1}(x) \leftarrow $Next point on $\mathcal{C}(x)$ \\
    \Queue.add($(x, k_{j+1}(x))$ with priority $\rho_{k_{j}(x)}$)
 }
}
\Return $t$, \Loss.
\caption{Solve fractional multi-choice knapsack.}
\label{alg:priority}
\end{algorithm}

\subsubsection{Finite-Data Regime}
Thus far, we have given a characterization of the optimal deterministic policy that minimizes the poverty rate and a practical algorithm for poverty rate minimization. However, in realistic settings, we only have access to data sampled from $F$. In this section, we assume the true data distribution $F$ is unknown but we have access to a training set $\mathcal{D}_{\text{train}} = \{(X_{i}, Y_{i})\}_{i=1}^{n_{\text{train}}}$ where $(X_{i}, Y_{i}) \sim F$ i.i.d., and we aim to optimize allocations for unlabeled samples $\mathcal{D}_{\text{test}} = \{ X_{i}\}_{i=1}^{n_{\text{test}}}$ where $X_{i} \sim F$ i.i.d.

Algorithm \ref{alg:main} relies on the conditional density function $f_{Y \mid X}$ to compute the optimal transfer policy. In finite samples, we consider the plug-in estimator for the optimal transfer policy, which is obtained by running Algorithm \ref{alg:main} with an estimator of the conditional density $\hat{f}_{Y \mid X}$ and the empirical covariate distribution $\hat{f}_{X}(x) = \frac{1}{n_{\text{test}}} \cdot \mathbb{I}(X_{i} \in \mathcal{D}_{\text{test}})$ instead of the true conditional and covariate distributions.


To estimate $\hat{f}_{Y \mid X}$, we apply an extension of Lindsey's method \citep{efron1996using} to the training data $\mathcal{D}_{\text{train}}$. Lindsey's method is a popular technique for marginal density estimation, and our approach is a straightforward extension of this method for conditional density estimation. To model the conditional densities, we consider the following exponential family of densities on $\mathcal{Y}$
\begin{equation} \label{eq:special_exp_family} h(y \mid x) = h_{0}(y) \cdot \exp(s(y)^{T} \Theta x - \psi(\Theta x)). \end{equation}
Here, $h_{0}(y)$ represents a carrier density, $s(y)$ is a $J$-dimensional vector of sufficient statistics, $\Theta$ is a $J \times d$ matrix of parameters, and $\psi$ is the log partition function, the normalizing function that ensures $h(y \mid x)$ integrates to $1$ over $\mathcal{Y}$. Many choices of carrier density and sufficient statistics are possible. Following Lindsey's method, we set the carrier density to be a nonparametric estimate of $f_{Y}$, the marginal distribution of $Y$. In addition, we consider a basis of B-Spline functions to form the sufficient statistics.

We compute the conditional density estimator as follows.
\begin{enumerate}
	\item We estimate $f_{Y}$, the marginal density of $Y$, by computing a kernel density estimate using $\{ Y_{i}\}_{i=1}^{n_{\text{train}}}$. We denote this estimate as $\hat{f}_{Y}.$
	\item Setting $\hat{f}_{Y}$ as the carrier density in \eqref{eq:special_exp_family}, we estimate $\Theta$ via maximum likelihood. To estimate the log-likelihood of \eqref{eq:special_exp_family}, we first approximate the log-partition function by discretizing $\mathcal{Y}$ into $M$ bins with midpoints $[y_{1}, \dots y_{M}]$ and corresponding widths $[\Delta_{1}, \Delta_{2}, \dots \Delta_{M}]$. We note that the log partition function can be approximated by
	\begin{align*}
	\psi(z) &= \log( \int \hat{f}_{Y}(y) \cdot \exp(s(y)^{T}z)) \approx \log(\sum_{j=1}^{M} \hat{f}_{Y}(y_{j}) \cdot \exp(s(y_{j})^{T}z) \cdot \Delta_{j}).
	\end{align*}
	Thus, the log-likelihood of \eqref{eq:special_exp_family} can be written as 
	\begin{equation}
	\label{eq:log_likelihood}
	\log h(y_{i} \mid x_{i}; \Theta) =  s(y_{i})^{T}\Theta x_{i} - \log\Big(\sum_{j=1}^{M} \hat{f}_{Y}(y_{j}) \cdot \exp(s(y_{j})^{T}\Theta x_{i})  \cdot \Delta_{j} \Big) + c,
	\end{equation}
	where $c$ is a constant factor that does not depend on $\Theta$. Notably, the log-likelihood is concave in $\Theta$, so standard optimization tools can be applied to obtain an estimate of $\Theta$. In our work, we optimize $\Theta$ via stochastic gradient descent.
\end{enumerate}
Like Lindsey's method, this is a hybrid approach between a parametric and nonparametric method because it models the relationship between high-dimensional covariates and the density value with a generalized linear model but allows the density function to take on a flexible form by fitting the carrier density nonparametrically.

\subsection{Welfare Maximization}
\label{app:welfare}
We formally define the problem of welfare maximization. Let $w : \mathbb{R}_{+} \rightarrow \mathbb{R}$ be an increasing, strictly concave, differentiable welfare function that is bounded above by a constant $C < \infty$. Welfare maximization is defined as the following problem
\begin{equation}
\label{eq:welfare_maximization}
\max_{t: \mathcal{X} \rightarrow \mathbb{R}_{+}} \{\EE[F]{w(Y + t(X))} : \EE[F]{t(X)} \leq B \}.
\end{equation}

Note that welfare maximization is analogous to loss minimization, so we can apply the same techniques as in Theorem \ref{theo:simple_equity} to obtain the optimal policy for welfare maximization.

\subsubsection{Algorithm}
We compute the solution to a discrete approximation of \eqref{eq:welfare_maximization}. In our approximation, we consider policies that only allocate transfer amounts that lie in a discrete grid. In particular, let $\mathcal{T} := [t_{1}, t_{2}, \dots t_{J}]$ denote a grid of transfer amounts on $[0, \bar{t}]$, where $\bar{t}$ is the maximum transfer size.

First, we define the function $G_{x}(t) := \EE[F]{w(Y) + t) \mid X=x}.$ We obtain a discrete approximation to the inverse function $(G_{x}')^{-1}: \mathbb{R}_{+} \rightarrow \mathbb{R}_{+}.$ Let the function $H_{x}: \mathbb{R}_{+} \rightarrow \mathcal{T}$ denote the discrete approximation. To construct $H_{x},$ obtain $G_{x}'(t): \mathcal{X} \mapsto \mathbb{R}_{+}$ for $t \in \mathcal{T}.$ Next, use the functions $\{G_{x}'(t)\}_{t \in \mathcal{T}}$ to define
	\[ H_{x}(\lambda) = t_{j^{*}},\, \where j^{*} = \sup \{ j \in [J] \mid G_{x}'(t_{j}) \geq \lambda \}. \]
Thus, $H_{x}(\cdot)$ is a discrete approximation of $(G_{x}')^{-1}(\cdot).$

Second, use $H_{x}(\cdot)$ to find the dual parameter $\lambda$ corresponding to budget $B$. Define bounds $\lambda_{\min} := \inf_{x \in \mathcal{X}, j \in [J]} G_{x}'(t_{j})$ and $\lambda_{\max} := \sup_{x \in \mathcal{X}, j \in [J]} G_{x}'(t_{j}).$ Consider a grid $\{\lambda_{1}, \lambda_{2}, \dots \lambda_{J}\}$ on $[\lambda_{\min}, \lambda_{\max}]$. We return the policy $t_{j}(x) = (H_{x}(\lambda_{j}))_{+}$ that minimizes $|\EE[F]{(H_{X}(\lambda_{j}))_{+}} - B|$ over the grid.

\subsubsection{Finite-Data Regime}
We consider the setting where we only have access to data sampled from $F$. In this section, we assume access to a training set $\mathcal{D}_{\text{train}} = \{(X_i, Y_i)\}_{i=1}^{n_{\text{train}}}$, where $(X_{i}, Y_{i}) \sim F$ i.i.d., and we aim to optimize allocations for unlabeled samples $\mathcal{D}_{\text{test}} = \{ X_{i} \}_{i=1}^{n_{\text{test}}}$ where $X_{i} \sim F$ i.i.d. We use the following procedure to estimate the optimal policy for a budget $B.$

In finite samples, we must estimate the functional parameters that appear in the above algorithm. In particular, we must estimate $G_{x}'(t)$ for $t \in \mathcal{T}$ from the training data. To do so, we can apply any nonparametric regression technique to solve
			\[ \min_{h: \mathcal{X} \rightarrow \mathbb{R}_{+}} \EE[F]{(w'(t + Y) - h(X))^{2}}.\]


\subsection{Binary Policies}
\label{sec:binary}
In this section, we solve \eqref{eq:prob} with the additional constraint that the policy must be binary-valued. In contrast to proxy means testing as in Definition \ref{defi:pmt}, we minimize a particular poverty measure and optimize not only who receives the transfer but also the optimal transfer size.

Learning the optimal binary transfer policy that minimizes a particular loss function $L$ among all possible binary transfer policy can be framed as the following optimization problem
\begin{equation}
\label{eq:binary_opt}
\begin{split}
\text{minimize } &\EE[F]{L(t(X) + Y)} \\
\text{subject to } &\EE[F]{t(X)} \leq B \\
&t(x) \in \{0, \bar{t} \} \quad \forall x \in \mathcal{X} \\
&\bar{t} \in [0, c].
\end{split}
\end{equation}
However, the above optimization problem is non-convex because it involves jointly optimizing over the maximum transfer value $\bar{t}$ and the transfer policy $t$.

We can rewrite \eqref{eq:binary_opt} as follows
\begin{equation}
\label{eq:binary_opt_decomp}
\min_{\bar{t} \in [0, c]} \Big\{ \min_{\pi: \mathcal{X} \rightarrow \{0, 1\}} \EE[F]{L(\bar{t} + Y) \cdot \pi(X) + L(Y) \cdot (1 - \pi(X))} \text{ subject to } \EE[F]{\pi(X)} \leq B/\bar{t} \Big\}.
\end{equation}

Solving the inner minimization of \eqref{eq:binary_opt_decomp} allows us to compute the optimal binary transfer policy among binary transfer policies where the maximum transfer value is a fixed value $\bar{t}$. The inner minimization of \eqref{eq:binary_opt_decomp} can be viewed as a capacity-constrained classification problem. Let $\pi_{\bar{t}}(\cdot; B)$ be the solution to the inner minimization of \eqref{eq:binary_opt_decomp}. The corresponding binary transfer policy can be defined as $t_{\bar{t}}(\cdot; B) := \bar{t} \cdot \pi_{\bar{t}}(x; B).$ 

The function $\pi_{\bar{t}}(\cdot; B)$ identifies the units who will benefit most (in terms of conditional loss reduction) from the $\bar{t}$-valued transfer. We can define the conditional loss reduction function for a $\bar{t}$-valued transfer
\begin{equation}
\label{eq:conditional_loss_reduction_function}
\rho_{\bar{t}}(x) := \EE[F]{L(\bar{t} + Y) - L(Y) \mid X=x}.
\end{equation}
For poverty measures, $L$ is decreasing so $\rho_{\bar{t}}(x) \leq 0$ for any $\bar{t} \in \mathbb{R}_{+}$.

Following from Theorem 1 from \citet{sun2021treatment}, there are constants $\rho_{\bar{t}, B}^{*} \in \mathbb{R}, a_{\bar{t}, B} \in [0, 1]$, such that the optimal solution to the inner minimization of \eqref{eq:binary_opt_decomp} has the form
\begin{equation}
\pi_{\bar{t}}(x; B) = \begin{cases} 1 & \rho_{\bar{t}}(x) < \rho_{\bar{t}, B}^{*}  \\ a_{\bar{t}, B} & \rho_{\bar{t}}(x) = \rho_{\bar{t}, B}^{*} \\ 0 & \rho_{\bar{t}}(x) > \rho_{\bar{t}, B}^{*},
\end{cases}
\end{equation}
where either $\rho_{\bar{t}, B}^{*} = a_{\bar{t}, B} = 0$ (i.e., we have sufficient budget to treat all units) or $\rho_{\bar{t}, B}^{*} < 0$ and the pair $(\rho_{\bar{t}, B}^{*}, a_{\bar{t}, B})$ is the unique pair for which the policy cost is $B$. If $X_{i}$ has continuous density, $\PP[F]{\rho_{\bar{t}}(X) = \rho_{\bar{t}, B}^{*}} = 0$ and the policy $\pi_{\bar{t}}(\cdot; B)$, and so the optimal policy will almost surely be integer-valued.

We operationalize this result to provide an algorithm for optimal binary policies. Our overall algorithm first discretizes the interval $[0, c]$ into a grid of possible maximum transfer amounts $[\bar{t}_{1}, \bar{t}_{2}, \dots \bar{t}_{J}]$. For each $j=1, 2, \dots J$, we compute $\pi_{\bar{t}_{j}}(\cdot; B)$ by ranking units by $\rho_{\bar{t}_{j}}(\cdot)$ in increasing order and allocating transfers $\bar{t}_{j}$ sequentially until the budget $B$ is exhausted. We set the optimal binary policy $t_{\text{binary}}^{*}$ under loss function $L$ to be the policy that yields the lowest expected loss over the grid, i.e.
\[ t_{\text{binary}}^{*}(x; B) = \bar{t}_{j^{*}} \cdot \pi_{\bar{t}_{j^{*}}}(x; B),\, \where j^{*} = \argmin_{j \in [J]} \EE[F]{L(t_{\bar{t}_{j}}(X_{i}; B) +  Y_{i})}.\]
We note that this procedure will yield the optimal binary-valued policy under any decreasing poverty measure $L$ and does not require the loss function $L$ to be convex. As a result, we will use this approach to learn optimal binary-valued policies under both the poverty rate and the poverty gap index.

\subsubsection{Finite-Data Regime}
Again, we consider the setting where we only have access to data sampled from $F$. In this section, we assume the true data distribution $F$ is unknown but we have access to a training set $\mathcal{D}_{\text{train}} = \{(X_{i}, Y_{i})\}_{i=1}^{n_{\text{train}}}$ where $(X_{i}, Y_{i}) \sim F$ i.i.d., and we aim to optimize allocations for unlabeled samples $\mathcal{D}_{\text{test}} = \{ X_{i}\}_{i=1}^{n_{\text{test}}}$ where $X_{i} \sim F$ i.i.d. We use the following procedure to estimate the optimal binary policy for a budget $B$. 

First, we discretize the interval $[0, c]$ into a grid of possible maximum transfer amounts $\bar{t}_{1}, \bar{t}_{2}, \dots \bar{t}_{J}.$ Second, we estimate nuisance parameters $\rho_{\bar{t}_{j}}(\cdot)$ for $j=1, 2, \dots J$. Using the training set $\mathcal{D}_{\text{train}} = \{(X_{i}, Y_{i})\}_{i=1}^{n_{\text{train}}}$, we define pseudo-labels 
		\[Z_{i} = L(\bar{t}+ Y_{i}) - L(Y_{i})\] for $i=1,2, \dots, n_{\text{train}}.$
Estimate $\hat{\rho}_{\bar{t}_{j}}(\cdot)$ via deep learning using $\mathcal{D}_{\text{train}}$ and pseudolabels. Observe that 
		\[ \rho_{\bar{t}} \in \argmin_{\rho} \EE[F]{(Z_{i} - \rho(X_{i}))^{2}}.\]
Our implementation represents $\rho_{\bar{t}}$ with a neural network and solves the following empirical risk minimization problem
	\[ \hat{\rho}_{\bar{t}_{j}}(\cdot) \in \argmin_{\rho} \hEE[F]{(Z_{i} - \rho(X_{i}))^{2}}.\]
Now, we must learn the optimal transfer size $\bar{t}_{j}$. For each $j=1,2, \dots J$, create a ranked list of units by sorting $\hat{\rho}_{\bar{t}_{j}}(X_{i})$ in increasing order for $X_{i}$ in the training set. We can estimate the threshold $\rho^{*}_{\bar{t}_{j}, B}$ by allocating transfers to the $B/\bar{t}_{j}$-fraction of units at the top of the ranked list for each $j=1, 2, \dots J$. After that, we can form $\hat{t}_{\bar{t}_{j}}(\cdot; B)$ using the estimated conditional loss reduction function $\hat{\rho}_{\bar{t}_{j}}(\cdot)$ and estimated threshold $\hat{\rho}^{*}_{\bar{t}_{j}, B}.$ Then, we can select 
\[ \hat{j}^{*} \in \argmin_{j \in [J]} \hEE[F]{L(\hat{t}_{\bar{t}_{j}}(X_{i}; B) + Y_{i})}.\]

We assign transfers to units in the test set $\mathcal{D}_{\text{test}}$ by ranking units in the test set by $\hat{\rho}_{j^{*}}$ in increasing order and allocating $\bar{t}_{\hat{j}^{*}}$-valued transfers to units starting at the top of the ranked list until we hit the budget constraint $B$, which ensures that the budget constraint is exactly satisfied on the test set.


\subsection{Weakly Equitable Oracle Policy}
\label{app:oracle}
This subsection focuses on oracle policies, which can allocate transfers on the basis of individual-level consumption. We characterize the policy family that minimizes the poverty rate subject to the weak equity constraint and the constraint that the policy family is continuous in budget. To characterize this policy, we first introduce a definition of weak rank preservation and show that weak equity implies this property for policy families that are continuous in budget. Finally, we use the property of weak rank preservation to characterize the optimal policy.

\begin{defi}[Weak Rank Preservation]
A full information policy family $t(y; B)$ weakly preserves ranks if for $y_{1}, y_{2} \in \mathcal{Y}$
\[ y_{1} + t(y_{1}; B') \leq y_{2} + t(y_{2}; B') \implies y_{1} + t(y_{1}; B) \leq y_{2} + t(y_{2}; B)\]
for all $0 \leq B' \leq B.$
\end{defi}
Transfer policies that weakly preserve ranks do not permit a poorer unit's post-transfer consumption to exceed that of a unit who is initially richer. This rules out the scenario where a unit who is initially poorer than another receives a transfer $t(y_{1}; B)$ that allows them to ``leapfrog'' the other unit in terms of post-transfer consumption, i.e. $y_{1} + t(y_{1}; B) > y_{2} + t(y_{2}; B)$. We can consider such policies as yielding post-transfer consumption that is monotone increasing in pre-transfer consumption.

We observe that weak equity (Definition \ref{defi:weakly_equitable}) and continuity in budget imply this property.

\begin{lemm}
\label{lemm:weak_equity_ranks}
If $t(y; B)$ is a full-information weakly equitable policy family and $t(y; B)$ is continuous in $B$, then $t(y; B)$ weakly preserves ranks.
\end{lemm}

Let $\mathcal{W}$ denote the class of policy families that satisfy weak equity and are continuous in budget. The following lemma characterizes the optimal rate-minimizing policy family among policy families that are weakly equitable and continuous in budget.
\begin{lemm}
\label{lemm:oracle_rate_we}
Suppose $F$ is a population distribution over $Y_{i}$ that has positive density on $(0, c)$. Consider minimizing the post-transfer poverty rate subject to a budget constraint and a weak equity constraint in the full-information setting:
\begin{equation}
\label{eq:oracle_prob}
\min_{t(\cdot; \cdot): \mathcal{Y} \times \mathbb{R}_{+} \rightarrow \mathbb{R}_{+}} \{\PP[F]{Y + t(Y; B) < c} : \EE[F]{t(Y; B)} \leq B,\, t \in \mathcal{W} \}.
\end{equation}
The optimal policy that solves this optimization problem is
\begin{equation}
\label{eq:optimal_oracle}
t(y; B) = \begin{cases} c - \lambda(B) & 0 \leq y \leq \lambda(B) \\ c - y & \lambda(B) < y \leq c \\
0 & y > c.
 \end{cases}.
\end{equation}
\end{lemm}

\section{Proofs}
\label{sec:proofs}

\subsection{Additional Lemmas}
We define new notation for the following lemmas. Define 
\begin{equation}
\begin{split}
G_{x}(t) &= -\EE[F]{L(Y + t) \mid X=x}\\
G'_{x}(t) &= -\frac{d}{dt}\EE[F]{L(Y + t) \mid X=x}
.
\end{split}
\end{equation}

\begin{lemm}
\label{lemm:regularity}
Suppose Assumption \ref{assumption:cdf} holds. Let $L$ be decreasing, strictly convex, differentiable, bounded below by $C > - \infty$. Then, $G_{x}(t)$ is differentiable on $\mathbb{R}_{+}$ and strictly concave.
\end{lemm}

\begin{lemm}
\label{lemm:fgt_regularity}
Suppose Assumption \ref{assumption:cdf} holds. Let $L$ denote an FGT index with $\alpha \geq 1$. Then, $G_{x}(t)$ is differentiable on $\mathbb{R}_{+}$ and strictly concave on $[0, c]$.
\end{lemm}

\subsection{Proof of Theorem \ref{theo:inequity}}
We consider two cases. We first focus on the case where $L$ is nonconvex in a region where it is continuously differentiable. Then, we will consider the case that $L$ is nonconvex in a region where it is not continuously differentiable.

Our proof relies on the truncated Gaussian kernel to construct distributions under which weak equity violations occur. The truncated Gaussian kernel is given by
\begin{equation} 
\label{eq:Gaussian_kernel}dK_{h}(y; m) = \frac{\phi((y - m)/ h)}{h \cdot (1 - \Phi(-m/h))} \quad y \geq 0\end{equation}
In addition, as $h\rightarrow 0$, $dK_{h}(y; m)$ converges weakly to $\delta_{m}$, a point mass at $m$. We also note that if $m_{1} < m_{2}$, then $K_{h}(\cdot; m_{1}) \preceq_{\text{SD}} K_{h}(\cdot; m_{2})$.

The second half of our proof also uses the Gamma kernel to smooth loss functions that are not continuously differentiable. The Gamma kernel is given by
\begin{equation}
\label{eq:Gamma_kernel}
dQ_{b}(z) = \frac{z}{b^{2}} e^{-z/b} \quad z \geq 0.
\end{equation}

\subsubsection{Continuously Differentiable Case}
\label{sec:continuous_case}
In the case where $L$ is nonconvex in a region where it is continuously differentiable, we show that it is possible to construct a population distribution $F_{0}$ with the following properties
\begin{enumerate}
\item $F_{0}$ is a mixture distribution over two types, i.e. $F_{X} = \frac{1}{2} \cdot \mathbb{I}(X =x_{1}) + \frac{1}{2} \cdot \mathbb{I}(X=x_{2})$.
\item The conditional distribution $F_{0, Y\mid X=x_{i}}$ is a point mass, i.e. $F_{0, Y \mid X=x_{i}} = \mathbb{I}(y=y_{i})$. In addition, $0 <y_{1} < y_{2}$, so $F_{Y \mid X=x_{1}} \preceq_{\text{SD}} F_{Y \mid X=x_{2}}$. 
\item There exists a strictly unique optimal transfer policy under $F_{0}$ and budget $B > 0$ called $t^{*}_{0}(\cdot; B)$. The optimal transfer policy $t_{0}^{*}(\cdot, B)$ is not weakly equitable under $F_{0}$. In particular, $t_{0}^{*}(x_{1}; B) - t_{0}^{*}(x_{2}; B) < 0.$
\end{enumerate}
We will use $F_{0}$ to construct a population distribution that satisfies Assumption \ref{assumption:cdf} under which the optimal policy that solves \eqref{eq:prob} is not weakly equitable. Given that $F_{0}$ has conditional distribution $F_{0, Y \mid X=x_{i}} = \mathbb{I}(y = y_{i})$, we define $F_{h}$ be a distribution over $(X, Y)$ where $F_{h, X} = F_{0, X}$ and $F_{h, Y \mid X} = K_{h}(y; y_{i}),$ where we recall that $K_{h}$ is the truncated Gaussian kernel defined in \eqref{eq:Gaussian_kernel}.

\begin{paragraph}{Construction of $F_{0}$ and Budget $B$.}
\label{sec:construction}
Since $L$ is non-convex on a continuously differentiable region, there exists two points $y_{1} < y_{2}$ with $L'(y_{1}) > L'(y_{2})$. Since $L'$ is continuous, there exists $\delta_{1}(\epsilon_{1}) > 0$, so that $|L'(y_{1}) - L'(z)| < \epsilon_{1}$ for $|y_{1} - z|< \delta_{1}(\epsilon_{1}).$ Again, by the continuity of $L'$, for any $\epsilon_{2}$, there exists $\delta_{2}(\epsilon_{2}) > 0$, so that $|L'(y_{2}) - L'(z)| < \epsilon$ for $|y_{2} - z| < \delta_{2}(\epsilon_{2}).$ We can choose $\epsilon_{1}, \epsilon_{2}$ sufficiently small so that $ L'(y_{2}) +\epsilon_{2} < L'(y_{1}) - \epsilon_{1}.$ Let $\epsilon = \min_{i \in \{1, 2\}} \epsilon_{i}$. It is straightforward to see that 
\begin{equation} 
\label{eq:key_construction} L'(y_{2}) + \epsilon < L'(y_{1}) - \epsilon. 
\end{equation}
Let $\Delta = \min(\delta_{1}(\epsilon), \delta_{2}(\epsilon))/2.$ For $z_{i} \in B_{\Delta}(y_{i})$, we have that 
\[L'(z_{2}) < L'(y_{2}) +\epsilon < L'(y_{1}) - \epsilon < L'(z_{1}).\] This implies that for $t \in [0, \Delta]$, we have that
\begin{equation} \label{eq:construction1} L'(y_{1} + t) \geq L'(y_{1}) - \epsilon \end{equation}
and
\begin{equation} \label{eq:construction2} L'(y_{2} + t)  \leq L'(y_{2}) + \epsilon.\end{equation}
Combining \eqref{eq:construction1}, \eqref{eq:construction2} and \eqref{eq:key_construction}, implies that for all $t_{1}, t_{2} \in [0, \Delta]$
\[ L'(y_{2} + t_{2}) < L'(y_{1} + t_{1}).\]

Set $F_{0, Y \mid X=x_{i}} = \mathbb{I}(y = y_{i})$ for $i \in \{1, 2\}.$ Also, note that we have selected $\Delta$ so that $L$ is continuous on open neighborhoods $O_{1}, O_{2}$ that contain $[y_{1}, y_{1} + \Delta]$ and $[y_{2}, y_{2} + \Delta]$, respectively.

\paragraph{Optimal Policy under $F_{0}$ and Budget $B=\Delta/2$.}
For this choice of population distribution $F_{0}$ and $B = \Delta/2$, the optimal policy that solves \eqref{eq:prob} is $t^{*}(x) =0 \cdot \mathbb{I}(X = x_{1}) + \Delta \cdot \mathbb{I}(X= x_{2})$. Note that this policy is non-monotone increasing in consumption because $F_{0, Y \mid X=x_{1}} \preceq_{\text{SD}} F_{0, Y \mid X=x_{2}} $ but $t^{*}(x_{2}) > t^{*}(x_{1})$ and is thus, not weakly equitable.

We consider the following optimization problem to characterize the optimal policy. Denote the objective as $J(t_{1}, t_{2}):= \frac{1}{2} (L(y_{1} + t_{1}) + L(y_{2} + t_{2}))$.

\begin{equation}
\label{eq:original_opt}
\min_{t_{1}, t_{2} \in [0, \Delta]} \Big\{ J(t_{1}, t_{2})  : t_{1} + t_{2} \leq \Delta\Big\}. 
\end{equation}

 First, we show that the budget constraint binds. For the sake of contradiction, suppose that $(t_{1}^{*}, t_{2}^{*})$ is the optimal policy but $t_{1}^{*} + t_{2}^{*} < \Delta$. Then, we can consider the policy $(t_{1}^{*}, t_{2}^{*} + \epsilon)$. Then we have that
\begin{align*}
J(t_{1}^{*}, t_{2}^{*} + \epsilon) - J(t_{1}^{*}, t_{2}^{*}) &= \frac{1}{2}(L(y_{2} + t_{2}^{*} + \epsilon ) - L(y_{2} + t_{2}^{*})) \\
&= \frac{1}{2} \cdot L'(y_{2} + t) \cdot \epsilon
\end{align*}
for some $t \in [t_{2}^{*}, t_{2}^{*} + \epsilon]$ by the Mean Value Theorem. Since $L'(y_{2} + t_{2}) < L'(y_{1} + t_{1}) \leq 0$ where $t_{1}, t_{2} \in [0, \Delta]$, we have that $J(t_{1}^{*}, t_{2}^{*} + \epsilon) - J(t_{1}^{*}, t_{2}^{*}) < 0$. This is a contradiction because we assumed that $(t_{1}^{*}, t_{2}^{*})$ was the optimal policy. Thus, the budget constraint must bind.

Since the budget constraint must bind, we can restrict to solutions that satisfy $t_{1} + t_{2} = \Delta.$ Suppose the optimal policy has $t_{1}^{*} > 0$ (and satisfies the budget constraint). We consider the policy given by $(t_{1}^{*} - \epsilon, t_{2}^{*} + \epsilon)$, which still satisfies the budget constraint. Then we note that

\begin{align*}
J(t_{1}^{*} - \epsilon, t_{2}^{*} + \epsilon) - J(t_{1}^{*}, t_{2}^{*}) &=\frac{1}{2} \Big(L(y_{1} + t_{1}^{*} - \epsilon) - L(y_{1} + t_{1}^{*}) +  L(y_{2} + t_{2}^{*} + \epsilon ) - L(y_{2} + t_{2}^{*})\Big) \\
&= \frac{1}{2} \cdot \epsilon \cdot (L'(y_{2} + t_{2}) - L'(y_{1} + t_{1})),
\end{align*}
where $t_{1} \in [t_{1}^{*} - \epsilon, t_{1}^{*}], t_{2} \in [t_{2}^{*}, t_{2}^{*} + \epsilon].$ We note that we must have that $J(t_{1}^{*} - \epsilon, t_{2}^{*} + \epsilon) - J(t_{1}^{*}, t_{2}^{*}) < 0$ because $L'(y_{2} + t_{2}) < L'(y_{1} + t_{1})$ and $L' \leq 0$ for $t_{1}, t_{2} \in [0, \Delta].$ This is a contradiction, so the optimal policy must provide the entire  budget to $x_{2}$.
\end{paragraph}
Thus, the optimal policy under $F_{0}$ and $B= \Delta/2$ is not weakly equitable.

\paragraph{Weak Equity Violation under $F_{h}$} Recall that $F_{h, Y|X} = K_{h}(y; y_{i})$ for $i=1, 2$. Define $J(t_{1}, t_{2}; h) := \frac{1}{2} \sum_{i \in \{1, 2\}} \EE[F_{h}]{ L(Y + t_{i}) \mid X=x_{i}}.$ Note that $J(t_{1}, t_{2}; h) = \EE[F_{h}]{L(Y + t(X))},$ where $t(x_{i}) = t_{i}$ for $i=1, 2.$ We consider solving
\begin{equation} 
\label{eq:kernel_prob}
\min_{t_{1}, t_{2} \in [0, \Delta]} \{ J(t_{1}, t_{2}; h) : t_{1} + t_{2} \leq \Delta\}.\end{equation}
Let $\Theta^{*}(h)$ denote the set of minimizers $(t_{1}, t_{2})$ of \eqref{eq:kernel_prob} at kernel bandwidth $h$.  We show that there exists $h > 0 $ such that $\Theta^{*}(h)$ is nonempty and all $(t_{1}, t_{2}) \in \Theta^{*}(h)$ violate weak equity for $(B, B')$ where $B = \Delta/2$ and $B' = 0.$ In other words, there exists $h > 0$ such that all $(t_{1}, t_{2}) \in \Theta^{*}(h)$ have the property that $t_{2} > t_{1}.$

We establish that $\Theta^{*}(h)$ is nonempty. Since $L$ is monotone, it has countably many discontinuities. The set of discontinuities of $L(t + y)$ have measure zero under $dK_{h}(\cdot; y_{i})$. Since $L$ is bounded, the dominated convergence theorem yields that $J$ is continuous in $t_{1}, t_{2}$. A continuous function has a minimizer on a compact set, so $\Theta^{*}(h)$ is nonempty for $h > 0.$

We first establish that uniform convergence on the feasible set: we show that
\begin{equation} \label{eq:uniform_convergence} \sup_{t_{1}, t_{2} \in [0, \Delta]} | J(t_{1}, t_{2}; h) - J(t_{1}, t_{2}) | \rightarrow 0. \end{equation}

Recall that $L$ is continuous on $O_{i}$ that contains $[y_{i}, y_{i} + \Delta]$. Let $C_{i}$ be a closed neighborhood that contains $[y_{i}, y_{i}+ \Delta].$ Note that $L$ is uniformly continuous on $C_{i}$. This means that for $\epsilon > 0$, there exists $\delta > 0$ so that $|y- y_{i}| < \delta$ implies $|L(t_{i} + y) - L(t_{i} + y_{i})| < \epsilon$ (for all $t_{i} \in [0, \Delta]$). As a result, we have that
\begin{align*}
&\sup_{t_{i} \in [0, \Delta]} \Big| \EE[F_{h}]{L(t_{i} + Y) \mid X=x_{i}} - L(t_{i} + y_{i}) \Big| \\
&\leq \sup_{t_{i} \in [0, \Delta]} \int |L(t_{i} + y) - L(t_{i} + y_{i}) | dK_{h}(y; y_{i}) \\
&\leq \sup_{t_{i} \in [0, \Delta]} \int_{y: |y - y_{i}| < \delta } |L(t_{i} + y) - L(t_{i} + y_{i})| dK_{h}(y; y_{i})  \\
&\indent+ \sup_{t_{i} \in [0, \Delta]} \int_{y: | y - y_{i}| > \delta} |L(t_{i} + y) - L(t_{i} + y_{i})| dK_{h}(y; y_{i}) \\
&\leq \sup_{t_{i} \in [0, \Delta]} \int_{y: |y - y_{i}| < \delta } |L(t_{i} + y) - L(t_{i} + y_{i})| dK_{h}(y; y_{i})  \\
&\indent+ \sup_{t_{i} \in [0, \Delta]} \int_{y: | y - y_{i}| > \delta} (|L(t_{i} + y)| +  |L(t_{i} + y_{i})|) dK_{h}(y; y_{i}) \\
&= \epsilon + 2||L||_{\infty} \cdot \PP[F_{h}]{|Y - y_{i}| > \delta \mid X=x_{i}}.
\end{align*}
We note that $\PP[F_{h}]{|Y - y_{i}| > \delta \mid X=x_{i}} \rightarrow 0$ because $Y \mid X=x_{i} \xrightarrow{p} y_{i}.$ Since $\epsilon$ is arbitrary, we have that \[\sup_{t_{i} \in [0, \Delta]} |\EE[F_{h}]{L(t_{i} + Y) \mid X=x_{i}} - L(t_{i} + y_{i})| \rightarrow 0,\] and \eqref{eq:uniform_convergence} must hold.

Now, we prove the desired claim. Denote the minimizer of \eqref{eq:original_opt} by $(t_{1}^{*}, t_{2}^{*}).$ Let $r = \Delta/4$. Consider a ball about the minimizer in the feasible set $B_{r}((t_{1}^{*}, t_{2}^{*})).$ Note that all points in this ball violate weak equity because $(t_{1}^{*}, t_{2}^{*}) = (0, \Delta)$ and perturbing each coordinate by at most $r= \Delta/4$ still yields a point $(t_{1}, t_{2})$ where $t_{2} > t_{1}$. Let $\mathcal{T} = \{ (t_{1}, t_{2}) \in [0, \Delta]^{2} \mid t_{1} + t_{2} \leq \Delta \}.$ i.e. $K:= \mathcal{T} \setminus B_{r}((t_{1}^{*}, t_{2}^{*}))$. Note that $K$ must be compact. Since $(t_{1}^{*}, t_{2}^{*})$ is the unique minimizer of \eqref{eq:original_opt}, then there exists $\delta > 0$ such that
\[ J(t_{1}, t_{2}; 0) > J(t_{1}^{*}, t_{2}^{*};0) + \delta\]
for $(t_{1}, t_{2}) \in K$. This follows from the fact that $J$ is continuous on a compact set $K$ so it must attain a minimum on $K$, and this minimum value must be strictly larger than the objective value attained by $(t_{1}^{*}, t_{2}^{*})$ by the uniqueness of the solution to \eqref{eq:original_opt}. By uniform convergence, there exists $b > 0$ so that $|h| < b$ implies that 
\begin{equation} \label{eq:uniform_convergence_2} |J(t_{1}, t_{2}; h) - J(t_{1}, t_{2}; 0)| \leq \frac{\delta}{3} \end{equation}
for all $(t_{1}, t_{2}) \in \mathcal{T}.$ For $|h| < b$ and $(t_{1}, t_{2}) \in K$, we have that 
\begin{align}
J(t_{1}, t_{2}; h) &\geq J(t_{1}, t_{2}; 0) - \frac{\delta}{3} \\
&\geq J(t_{1}^{*}, t_{2}^{*}; 0) + \frac{2\delta}{3}. \label{eq:ineq1}
\end{align}
If $|h| \leq b$, we also have that
\begin{equation}
\label{eq:ineq2}
J(t_{1}^{*}, t_{2}^{*}; h) \leq J(t_{1}^{*}, t_{2}^{*}; 0) + \frac{\delta}{3}. 
\end{equation}
Combining \eqref{eq:ineq1} and \eqref{eq:ineq2} imply that
\[J(t_{1}, t_{2}; h) \geq J(t_{1}^{*}, t_{2}^{*}; h) + \frac{\delta}{3}.\]
This implies that any point in $K$ incurs strictly higher objective value than $(t_{1}^{*}, t_{2}^{*})$ under objective function $J(\cdot; h)$ for $h$ sufficiently small. In this case, any minimizer of \eqref{eq:kernel_prob} cannot lie in $K$ and must instead lie in $B_{r}((t_{1}^{*}, t_{2}^{*})).$ This implies that there exists $h > 0$ such that any optimal policy that solves \eqref{eq:kernel_prob} for this $h$ must violate weak equity for $(B, B') = (\Delta/2, 0).$

\subsubsection{Non-Continuously Differentiable Case}
Now, we consider the case where $L$ is not continuously differentiable in the region where it is non-convex. In this case, we will be able to show that it is possible to construct a smoothed loss function $\tilde{L}$ that preserves the non-convexity of $L$ but is also continuously differentiable by using the Gamma kernel $Q_{b}$ for smoothing. For this smoothed loss function $\tilde{L}$, we can construct a distribution $F_{h}$ that satisfies Assumption \ref{assumption:cdf} using the above proof technique under which the optimal policy that solves \eqref{eq:prob} with population distribution $F_{h}$ and loss function $\tilde{L}$. Using the smoothing kernel $Q_{b}$ and the distribution $F_{h}$, we can show that there is a distribution $F$ under which the optimal policy that solves \eqref{eq:prob} with the original loss function $L$ is not weakly equitable.

We prove the following technical lemma.
\begin{lemm}
\label{lemm:nonconvex_continuity}
If $L: \mathbb{R}_{+} \rightarrow \mathbb{R}$ is a non-convex and monotone decreasing function, then there exist continuity points $t_{0}, t_{1}, t_{2}$ of $L$ such that $t_{0} = \lambda t_{1} + (1- \lambda)t_{2}$ for some $\lambda \in (0, 1)$ and
\begin{equation} \label{eq:strict_nonconvexity} L(t_{0}) - \lambda L(t_{1}) - (1- \lambda) L(t_{2}) > 0. \end{equation}
\end{lemm}
We apply Lemma \ref{lemm:nonconvex_continuity} and obtain continuity points $t_{0}, t_{1}, t_{2}$ of $L$ that yield  \eqref{eq:strict_nonconvexity}. Define 
\[ \epsilon := L(t_{0}) - \lambda L(t_{1}) - (1- \lambda) L(t_{2}).\]

We define the smoothed loss function. 
\begin{align*}
\tilde{L}_{b}(t) &:= \int_{0}^{\infty} L(t + z) dQ_{b}(z),
\end{align*}
where $Q_{b}$ is the Gamma kernel. As $b \rightarrow 0$, the Gamma kernel converges in distribution to the Dirac measure at $0$, i.e. $dQ_{b} \xrightarrow{d} \delta_{0}$. First, we verify that $\tilde{L}_{b}(t)$ is continuously differentiable on $\mathbb{R}_{+}$ for any $b > 0$ in the following lemma.
\begin{lemm}
\label{lemm:smoothed_loss_diff}
Let $L$ be a bounded and Lebesgue integrable function on $\mathbb{R}_{+}.$ Then, $\tilde{L}_{b}$ is continuously differentiable.
\end{lemm} 

Second, we show that the non-convexity exhibited by $L$ is preserved by $\tilde{L}_{b}.$ By weak convergence, $b \rightarrow 0$, $\tilde{L}_{b}(t) \rightarrow L(t)$ for continuity points $t$ of $L$. This implies that there exists $b$ sufficiently small so that $|\tilde{L}_{b}(t_{1}) - L(t_{1})| < \frac{\epsilon}{4},$ $|\tilde{L}_{b}(t_{2}) - L(t_{2})| < \frac{\epsilon}{4}$, and lastly $|\tilde{L}_{b}(t_{0}) - L(t_{0})| < \frac{\epsilon}{4}.$ This implies that
\begin{equation} 
\label{eq:non_convexity}
\tilde{L}_{b}(t_{0}) - \lambda \tilde{L}_{b}(t_{1}) - (1- \lambda) \tilde{L}_{b}(t_{2}) \geq \frac{\epsilon}{2}.
\end{equation}
Thus, $\tilde{L}_{b}$ is nonconvex in a region where it is continuously differentiable.

Since $\tilde{L}_{b}$ is nonconvex in a region where it is continuously differentiable, we can apply the proof in Section \ref{sec:continuous_case} to show that there is a distribution $F_{h}$ under which the optimal policy under $\tilde{L}$ is not weakly equitable. This distribution $F_{h}$ has covariate distribution $F_{h, X}(x) = \sum_{i \in \{1, 2\}} \frac{1}{2} \mathbb{I}(x=x_{i})$ and $F_{h, Y \mid X=x_{i}} = K_{h}(\cdot; y_{i})$. Define a population distribution $F$ over $(X, Y)$ where $F_{X} = \sum_{i \in \{1, 2\}} \frac{1}{2} \mathbb{I}(x=x_{i})$ and $F_{Y \mid X=x_{i}}$ is the distribution over $Z_{1} + Z_{2, i}$, where $Z_{1} \sim Q_{b}(\cdot; 0)$ and $Z_{2, i} \sim K_{h}(\cdot; y_{i})$. We note that $F_{Y \mid X=x_{1}} \preceq_{\text{SD}} F_{Y \mid X=x_{2}}$ in this case. We observe that
\begin{align*}
\EE[F]{L(t + Y) \mid X=x_{i}} &= \EE[]{L(t + Z_{1} + Z_{2, i}) \mid X=x_{i}} \\
&= \int_{0}^{\infty} \int_{0}^{\infty} L(t + z_{1} + z_{2, i})  dQ_{b}(z_{1}) \cdot dF_{h, Y \mid X}(z_{2, i}) \\
&= \int_{0}^{\infty} \tilde{L}(t + z_{2, i}) dF_{h, Y \mid X}(z_{2, i}) \\
&= \EE[F_{h, Y \mid X}]{\tilde{L}(t + Y) \mid X=x_{i}}.
\end{align*}
Thus, any optimal policy $t^{*}$ that solves \eqref{eq:prob} with loss function $\tilde{L}$ and population distribution $F_{h}$ is also the optimal policy that solves \eqref{eq:prob} with loss function $L$ and population $F$ described above. Since $t^{*}$ is not weakly equitable under $F_{h}$, $t^{*}$ is also not weakly equitable under $F.$

\subsection{Proof of Theorem \ref{theo:simple_equity}}

\subsubsection{Existence}
To establish existence of a minimizer, we show the following. Suppose $L, B$ satisfy conditions (a) or (b). If a policy $t^{*}$ solves the pointwise optimization problem 
\begin{equation} 
\label{eq:pointwise_opt}
\min_{t(x) \in \mathbb{R} } \{\EE[F]{L(t(x) + Y) \mid X=x} + \lambda \cdot t(x): t(x) \geq 0\} \end{equation}
for every $x \in \mathcal{X}$ and a fixed $\lambda > 0$ and satisfies $\EE[F]{t^{*}(X)} = B$, then $t^{*}$ must be a solution to \eqref{eq:prob}. Then, we show that a policy that solves the pointwise optimization problem \eqref{eq:pointwise_opt} for $\lambda > 0$ and satisfies the budget constraint exists, concluding the proof.

We show the first part holds. Let $t^{*}(X)$ be a policy that solves \eqref{eq:pointwise_opt} and satisfies the budget constraint. Then for any policy $t$ in the feasible set of \eqref{eq:prob}, we have that
\[ \EE[F]{L(t^{*}(x) + Y) \mid X=x} + \lambda \cdot t^{*}(x) \leq \EE[F]{L(t(x) + Y) \mid X=x} + \lambda \cdot t(x)\]
for all $x \in \mathcal{X}$. Taking expectations of both sides, we have that
\[ \EE[F]{L(t^{*}(X) + Y)} + \lambda \EE[F]{t^{*}(X)} \leq \EE[F]{L(t(X) + Y)} + \lambda \EE[F]{t(X)}.\]
We note that $\EE[F]{t^{*}(X)} = B$ by assumption and $\EE[F]{t(X)} \leq B$ by feasibility. As a result, we have that
\[ \EE[F]{L(t^{*}(X) + Y)} + \lambda \cdot B \leq \EE[F]{L(t(X) + Y)} + \lambda \EE[F]{t(X)} \leq \EE[F]{L(t(X) + Y)} + \lambda \cdot B.\]
Then, we have that $\EE[F]{L(t^{*}(X) + Y)} \leq \EE[F]{L(t(X) + Y)}$, so $t^{*}$ must be a minimizer of \eqref{eq:prob}.

Now, we show that a solution $t^{*}$ that solves \eqref{eq:pointwise_opt} and satisfies the budget constraint exist. We start by proving the existence of the solution to \eqref{eq:pointwise_opt} for any $\lambda > 0$. We note that \eqref{eq:pointwise_opt} can be re-written as
\begin{equation}
\label{eq:solve_dual} 
\min_{t(x) \in \mathbb{R}} \{ -G_{x}(t(x)) + \lambda t(x) : t(x) \geq 0 \}.
\end{equation}
Suppose that the first set of conditions holds, i.e., $L$ is decreasing, strictly convex, differentiable, bounded below by a constant $C > - \infty$ and $B > 0$. In this case, we can apply Lemma \ref{lemm:regularity} to see that $G_{x}(t)$ is differentiable. As a result, the objective of \eqref{eq:solve_dual} is continuous. In addition, the objective of \eqref{eq:solve_dual} is also coercive for $\lambda > 0$ because $L$ is bounded below, so $-G_{x}(t(x)) + \lambda t(x) \rightarrow \infty$ as $t(x) \rightarrow \infty$. Since a continuous, coercive function must have at least one global minimizer on $[0, \infty)$, then \eqref{eq:solve_dual} has at least one global minimizer. Lemma \ref{lemm:regularity} also shows that $G_{x}(t)$ is strictly concave, so the objective of \eqref{eq:solve_dual} is strictly convex. This implies that \eqref{eq:solve_dual} has at most one minimizer on $[0, \infty).$ The unique solution that solves \eqref{eq:solve_dual} must satisfy the KKT conditions that $G_{x}'(t(x)) = \lambda$ or take value at the boundary $t(x) = 0$. We also note that $G_{x}'$ is strictly decreasing so is invertible on its image, i.e. $(0, G_{x}'(0)]$. The solution must be given by
\begin{equation}
\label{eq:t_lambda}
t_{\lambda}(x) = \begin{cases} (G_{x}')^{-1}(\lambda) & 0 < \lambda < G_{x}'(0) \\
 0  & \lambda  \geq G_{x}'(0).
 \end{cases}
\end{equation}

Suppose that the second set of conditions holds, i.e. $L$ is an FGT index with $\alpha \geq 1$ and $0 < B <c$. In this case, we can apply Lemma \ref{lemm:fgt_regularity} to see that $G_{x}(t)$ is differentiable on $(0, \infty)$. Since $L$ is an FGT index, the solution to \eqref{eq:solve_dual} is equivalent to the solution to
\begin{equation}
\label{eq:solve_dual_bounded} 
\min_{t(x) \in \mathbb{R}_{+}} \{ -G_{x}(t(x)) + \lambda t(x) : 0 \leq t(x) \leq c \}
\end{equation}
because $L(z) = 0$ for $z \geq c$. Since the objective of \eqref{eq:solve_dual_bounded} is continuous and the feasible set is compact, \eqref{eq:solve_dual_bounded} must have at least one minimizer on $[0, c].$ By Lemma \ref{lemm:fgt_regularity}, we also have that $G_{x}(t)$ is strictly concave on $[0, c]$ for every $x \in \mathcal{X}$. Thus, \eqref{eq:solve_dual_bounded} has at most one solution on $[0, c]$, so the minimizer must be unique. The unique solution that solves \eqref{eq:solve_dual_bounded} must satisfy the KKT conditions that $G_{x}'(t) = \lambda$ or take value on the boundary $t(x) = 0$. We also note that $G_{x}'$ is strictly decreasing on $[0, c]$ so is invertible on this region. Thus, the optimal solution to \eqref{eq:solve_dual} is also given by \eqref{eq:t_lambda}.

We argue that there exists $\lambda \geq 0$ so that $\EE[F]{t_{\lambda}(X)} = B$. To show this, we note that $t_{\lambda}(x)$ is continuous in $\lambda$, as is its expectation. Note that $\EE[F]{t_{\lambda}(X)} \rightarrow 0$ as $\lambda \rightarrow \infty$: To see this, note that
\[  -G_{x}(t_{\lambda}(x)) + \lambda t_{\lambda} (x) \leq - G_{x}(0)\]
because $t_{\lambda}(x)$ is the minimizer of \eqref{eq:solve_dual}. In addition, because $L$ is bounded below
\[  C + \lambda t_{\lambda} (x) \leq -G_{x}(t_{\lambda}(x)) + \lambda t_{\lambda} (x).\]
Combining the two inequalities yields
\begin{align*} 
t_{\lambda}(x) &\leq \frac{-C - G_{x}(0)}{\lambda}.
\end{align*}
We note that $- G_{x}(0) \geq C$, so as $\lambda \rightarrow \infty$, we have that $t_{\lambda}(x) \rightarrow 0.$ Taking expectations yields the desired result.

Under conditions (a), we can also show that as $\lambda \rightarrow 0$, we have that $t_{\lambda}(x) \rightarrow \infty$. To see this, we note that $L$ is strictly convex, decreasing, and bounded below, so $- G'_{x}(t) \rightarrow 0$ as $t \rightarrow \infty.$ This implies that $(G_{x}')^{-1}(\lambda) \rightarrow \infty$ as $\lambda \rightarrow 0.$ Combined with the above result, we can use the Intermediate Value Theorem to show that there exists $\lambda \geq 0$ that yields $\EE[F]{t_{\lambda}(X)} = B.$ Similarly, under conditions (b), we have that $(G_{x}')^{-1}(0) = c$. Again, we can use the Intermediate Value Theorem to show that there exists $\lambda \geq 0$ that yields $\EE[F]{t_{\lambda}(X)} = B$ for $0 < B < c.$

\subsubsection{Uniqueness}
Under conditions (a), the objective of \eqref{eq:prob} is guaranteed to be strictly convex in $t$. The feasible set is convex, so \eqref{eq:prob} must have at most one global minimizer.

Under conditions (b), \eqref{eq:prob} can be rewritten as 
\[ \min_{t: \mathcal{X} \rightarrow \mathbb{R}} \{ \EE[F]{L(t(X) + Y) } : \EE[F]{t(X)} \leq B, 0 \leq t(x) \leq c\}\]
because $L$ is an FGT index, and no optimal policy will allocate a transfer larger than $c$. Applying Lemma \ref{lemm:fgt_regularity} yields that the objective of this optimization problem is strictly convex on the feasible set.

\subsubsection{Weak Equity}
The existence and uniqueness results imply that the optimal policy has the form given in \eqref{eq:t_lambda} for some $\lambda \geq 0.$ We use this structure to verify weak equity.

\paragraph{Incremental transfer is monotone decreasing in post-transfer consumption.}

First, we show that the simpler property in \eqref{eq:simple_version_consumption} holds. Second, we show that the optimal policy that is obtained by solving \eqref{eq:prob} with budget $B$ is the same as the cumulative optimal policy that is obtained by summing the optimal policy $t(x; B')$ that solves \eqref{eq:prob} with a small budget $B'$ and the policy $\Delta t(x; B - B')$ that solves \eqref{eq:prob} with the remaining budget $B - B'$ after administering the transfers $t(x; B')$. In other words, 
\[ t(x; B) = t(x; B') + \Delta t(x; B- B').\]
The first result implies that
\[ F_{Y + t(X; B') \mid X=x} \preceq_{\text{SD}} F_{Y + t(X; B') \mid X=x'} \implies \Delta t(x; B - B') \geq \Delta t(x'; B - B').\]
The second result implies that
\[ \Delta t(x; B- B') =  t(x; B) - t(x; B')\]
for all $x \in \mathcal{X}$. Taken together, these two results imply that \eqref{eq:inc_transfer} holds.

\paragraph{Transfer is monotone decreasing in consumption.}
We show that $t_{\lambda}(x) \geq t_{\lambda}(x')$ for any $\lambda \geq 0.$ Since the optimal policy has the form in \eqref{eq:t_lambda} for some choice of $\lambda$, then the optimal policy must be monotone decreasing in $Y \mid X=x.$

We note that $L'$ exists almost everywhere and is monotone increasing because $L$ is decreasing and convex. As a result, if $Y \mid X = x \preceq_{\mathrm{SD}} Y \mid X=x'$, then $\EE[F]{L'(Y + t) \mid X=x} \leq \EE[F]{L'(Y + t) \mid X=x'}$. This implies that $G'_{x}(t) \geq G'_{x'}(t)$ for any $t \in \mathbb{R}_{+}.$ Since $G'_{x}$ is decreasing in $t$, then $(G'_{x})^{-1}$ is also decreasing. Since $G'_{x}(t) \geq G'_{x'}(t)$ for all $t$ and both are decreasing in $t$, then we also have that $(G'_{x})^{-1}(\lambda) \geq (G'_{x'})^{-1}(\lambda)$. Thus, we have that $t_{\lambda}(x) \geq t_{\lambda}(x')$. Thus, the optimal solution to \eqref{eq:prob} is monotone decreasing in consumption.

\paragraph{Cumulative policy is equal to sum of incremental policies.}
It is straightforward to see that the optimal policy under the former is given by $t_{\lambda(B)}$, where $t_{\lambda}$ is given by \eqref{eq:t_lambda} with $\lambda(B)$ being the unique choice of $\lambda$ that satisfies the budget constraint at $B$. The optimal policy under budget $B'$ is similarly given by $t_{\lambda(B')}$, where $\lambda(B')$ being the unique choice of $\lambda$ that satisfies the budget constraint at $B'$. Since $B > B'$, we have that $\lambda(B) \leq \lambda(B').$

We now consider solving
\begin{equation}
\label{eq:new_prob}
\min_{t: \mathcal{X} \rightarrow \mathbb{R}_{+}} \{ \EE[F]{L(t(X) + t_{\lambda(B')}(X) + Y)} : \EE[F]{t(X)} \leq B - B'\}. 
\end{equation}
Let 
\begin{align*}
\tilde{G}_{x}(t) &= G_{x}(t + t_{\lambda(B')}(x) ), \\
\tilde{G}_{x}'(t) &= G_{x}'( t + t_{\lambda(B')}(x)).
\end{align*}
The optimal policy that solves \eqref{eq:new_prob} is given by
\[ \tilde{t}_{\lambda}(x) = \begin{cases} (\tilde{G}_{x}')^{-1}(\lambda) & 0 < \lambda < \tilde{G}_{x}'(0) \\ 0 & \lambda \geq \tilde{G}_{x}'(0) \end{cases}.\]
for some unique choice of $\lambda$ that satisfies $\EE[F]{\tilde{t}_{\lambda}(X)} = B - B'.$ We observe that
\begin{equation}
\label{eq:new_sol}
\tilde{t}_{\lambda}(x) = \begin{cases} (G_{x}')^{-1}(\lambda) - t_{\lambda(B')}(x) & 0 < \lambda < G_{x}'(t_{\lambda(B')}(x)) \\ 0 & \lambda \geq G_{x}'(t_{\lambda(B')}(x)) \end{cases}.
\end{equation}
We show that $\lambda$ that satisfies $\EE[F]{\tilde{t}_{\lambda}(X)} = B - B'$ must be equal to $\lambda(B).$ Consider two cases: (i) $t_{\lambda(B')}(x) > 0$ and (ii) $t_{\lambda(B')}(x) = 0$. 
\begin{enumerate}
    \item In the first case, $t_{\lambda(B')}(x) > 0$. This implies that $G_{x}'(t_{\lambda(B')}(x)) = \lambda(B')$. Since $\lambda(B) \leq \lambda(B')$, then we have that $\tilde{t}_{\lambda(B)}(x) = (G_{x}')^{-1}(\lambda(B)) - t_{\lambda(B')}(x).$
    \item In the second case, $t_{\lambda(B')}(x) = 0.$ This implies $G_{x}'(t_{\lambda(B')}(x)) = G_{x}'(0).$ As a result, 
    \[\tilde{t}_{\lambda(B)}(x) = \begin{cases} (G_{x}')^{-1}(\lambda)  & 0 < \lambda < G_{x}'(0), \\ 0 & \lambda \geq G_{x}'(0). \end{cases}\]
\end{enumerate}
Combining these two cases, we have that
\[ t_{\lambda(B')}(x) + \tilde{t}_{\lambda(B)}(x) = t_{\lambda(B)}(x),\]
as desired. Taking expectations, we can confirm that $\EE[F]{\tilde{t}_{\lambda(B)}(X)} = B - B'.$ We observe that the optimal policy that solves \eqref{eq:new_prob} is given by \eqref{eq:new_sol} with $\lambda = \lambda(B)$, and $t_{\lambda(B')}(x) + \tilde{t}_{\lambda(B)}(x) = t_{\lambda(B)}(x)$. So, we have shown that our desired claim holds.

\subsection{Proof of Lemma \ref{lemm:gap}}
 Minimizing the worst-case conditional poverty rate is equivalent to minimizing an upper bound on the worst-case poverty rate, i.e.
\[ \max_{x \in \mathcal{X}} \PP[F]{Y + t(X) < c \mid X=x} = \min \{ \lambda \in [0, 1] \mid \lambda \geq \PP[F]{Y + t(X) < c \mid X=x} \quad \forall x \in \mathcal{X} \}.\]
As a result, we can rewrite \eqref{eq:wc_rate} as the following optimization problem with an auxiliary parameter $\lambda$ introduced as an upper bound on $\max_{x \in \mathcal{X}} \PP[F]{Y + t(X) < c \mid X=x}$, i.e.
\begin{equation} \label{eq:aux_param} \min_{t: \mathcal{X} \rightarrow \mathbb{R}_{+}}  \min_{\lambda \in [0, 1]} \Big\{ \lambda : \EE[F]{t(X)} \leq B,\, \quad  \PP[F]{Y + t(X) < c \mid X=x} \leq \lambda \quad \forall x \in \mathcal{X}\Big\}.\end{equation}
Note that the second constraint of \eqref{eq:aux_param} is separable. Combined with the nonnegativity constraint yields that for all $x \in \mathcal{X}$, a feasible policy $t$ must satisfy
\[ t(x) \geq (c - F_{Y \mid X=x}^{-1}(\lambda))_{+}.\]
To obtain an optimal solution to \eqref{eq:aux_param}, it is sufficient to consider $(\lambda, t_{\lambda})$ pairs, where $t_{\lambda}(x) = (c - F_{Y \mid X=x}^{-1}(\lambda))_{+}$ because a feasible pair $(\lambda, t)$, where $t(x) > (c - F_{Y \mid X=x}^{-1}(\lambda))_{+}$ yields the same objective value as $(\lambda, t_{\lambda})$, which is also feasible. Thus, we can reduce \eqref{eq:aux_param} to the following optimization problem
\[ \min_{\lambda \geq 0} \Big\{ \lambda : \EE[F]{(c - F_{Y \mid X=x}^{-1}(\lambda))_{+}} \leq B \Big\}.\]
Under Assumption \ref{assumption:cdf}, $\EE[F]{(c - F_{Y \mid X=x}^{-1}(\lambda))_{+}}$ is strictly decreasing in $\lambda$ when $0 < \EE[F]{(c - F_{Y \mid X=x}^{-1}(\lambda))_{+}} < c$, so the minimum value of $\lambda$ is obtained when the budget constraint is satisfied with equality. Recall that any optimal gap-targeting policy when $0 < B < c$ must have the form given in Lemma \ref{lemm:opt_solution_gap} and satisfy the budget constraint with equality. These forms are identical, so an optimal gap minimizing policy also minimizes the worst-case conditional poverty rate.

\subsection{Proof of Lemma \ref{lemm:opt_solution_gap}}
We compute the optimal policy when $L(z)=(c -z)_{+}$. Since $L$ is convex, the optimal policy solves
\[ \min_{t: \mathcal{X} \rightarrow \mathbb{R}_{+}} \{ \EE[F]{L(t(X)+Y)} + \lambda (\EE[F]{t(X)} - B)\}\] 
for some $\lambda \geq 0.$ The first-order stationarity condition for a given $x$ is
\begin{equation} \label{eq:first_order} \frac{d}{dt} \EE[F]{L(t+Y) \mid X=x} + \lambda = 0. \end{equation}
We note that
\begin{align*}
\frac{d}{dt} \EE[F]{L(t+Y) \mid X=x} &= \frac{d}{dt} \EE[F]{(c -t -Y) \mathbb{I}(Y < c - t)} \\
&= \frac{d}{dt} \int_{0}^{c -t} (c -t) f_{Y\mid X=x}(y)dy - \frac{d}{dt} \int_{0}^{c -t} y f_{Y \mid X=x}(y)dy \\
&= \frac{d}{dt}((c -t) \cdot F_{Y \mid X=x}(c - t)) - (c - t) \cdot f_{Y \mid X=x}(c -t) \\
&=-(c - t) \cdot f_{Y \mid X=x}(c -t) - F_{Y \mid X=x}(c -t) + (c - t) \cdot f_{Y \mid X=x}(c -t) \\
&= -F_{Y \mid X=x}(c -t).
\end{align*}

The optimal policy will either satisfy the KKT condition, simplified below \eqref{eq:first_order},
\[ F_{Y \mid X=x}(c - t) = \lambda\] 
or be equal to a boundary point $t=0$. The optimal policy is given by
\begin{equation}
t_{\lambda}(x) = \begin{cases} c- F_{Y|X=x}^{-1}(\lambda) & 0 < \lambda < F_{Y|X=x}(c) \\ 0 & \lambda \geq F_{Y|X=x}(c). \end{cases}
\end{equation}

We note that the policy must also be feasible, so $ t \geq 0.$ So, the optimal policy has the form in \eqref{eq:form_gap_policy}.

\subsection{Proof of Theorem \ref{theo:alpha_valid}}
Under Assumption \ref{assumption:X_finite} and that the maximum transfer size is $c$, the feasible set can be restricted to a compact space. Under Assumptions \ref{assumption:cdf}, \eqref{eq:prob} with $L(z) = \mathbb{I}(z < c)$ has a continuous objective. Thus, there exists a minimizer of \eqref{eq:prob} with $L(z) = \mathbb{I}(z < c)$ when the transfers are restricted to lie between $0$ and $c$. 

We note that any minimizer of \eqref{eq:prob} must satisfy the KKT conditions because \eqref{eq:prob} satisfies linear constraint qualification. We derive the KKT conditions below. We define the Lagrangian and its derivative below. 
\begin{align*}
L(t, \lambda, \nu, \mu) &= \PP[F]{t(X) + Y < c} + \lambda (\EE[F]{t(X)} - B) - \sum_{x \in \mathcal{X}} \nu_{x} \cdot t(x) + \sum_{x \in \mathcal{X}} \mu_{x}  \cdot (t(x) - c) \\
\nabla_{t(x)} L(t, \lambda, \nu, \mu) &= f_{X}(x) \cdot (-f_{Y \mid X=x}(c - t(x)) + \lambda) - \nu_{x} + \mu_{x} = 0. 
\end{align*}

Let $\lambda \geq 0, \nu_{x} \geq 0, \mu_{x} \geq 0$ for all $x \in \mathcal{X}.$ The KKT conditions of this optimization problem are as follows
\begin{align*}
f_{Y \mid X=x}(c - t(x)) - \lambda &= \frac{\mu_{x} - \nu_{x}}{f_{X}(x)} \quad \text{(Stationarity)}, \\
\mu_{x} \cdot (t(x) - c) &= 0 \quad \text{(Complementary Slackness)}, \\
\nu_{x} \cdot t(x) &= 0  \quad \text{(Complementary Slackness)}, \\
\lambda \cdot (\EE[F]{t(X)} - B) &=  0 \quad \text{(Complementary Slackness)},\\
t(x) &\geq 0 \quad \text{(Primal Feasibility)}, \\
t(x) &\leq c \quad \text{(Primal Feasibility)}, \\
\EE[F]{t(X)} &\leq B \quad \text{(Primal Feasibility)}.
\end{align*}
The optimal solution must satisfy the KKT conditions for some $\lambda \geq 0$. 

Now we show that any policy $t$ that satisfies the KKT conditions with $\lambda \geq 0$ is $\alpha$-valid for some $\alpha \geq 0$. Note that any $t$ that satisfies the KKT conditions will satisfy primal feasibility, i.e. $t(x) \in [0, c].$ 
\begin{enumerate}
    \item Suppose $t$ satisfies the KKT conditions and has $t(x) \in  \{0, c\}$ for some $x \in \mathcal{X}.$ Then we note that $t(x) \in \mathcal{T}_{\alpha}(x)$ because $\mathcal{T}_{\alpha}(x)$ contains $\{0, c\}$ for all $\alpha \geq 0$.
	\item Suppose $t$ satisfies the KKT conditions and has $0 < t(x) < c$ for some $x \in \mathcal{X}$. Then $\nu_{x} = \mu_{x} = 0$ to ensure complementary slackness is satisfied. Then we must have that $f_{Y \mid X=x}(c - t(x)) = \lambda$. So, $t(x) \in \mathcal{T}_{\alpha}(x)$ for $\lambda = \alpha$.    
\end{enumerate}
As a result, a policy $t$ that satisfies the KKT conditions for a given value $\lambda$ will be $\alpha$-valid for some choice of $\alpha \geq 0.$

\subsection{Proof of Corollary \ref{coro:two_stage}}
By Theorem \ref{theo:alpha_valid}, an optimal deterministic policy must be $\alpha$-valid for some $\alpha \geq 0.$ So, for each value $\alpha$, we can compute the policy $t_{\alpha}^{*}$. Then, the optimal policy can be obtained by computing $\argmin_{\alpha: \alpha \geq 0} \PP[F]{ t_{\alpha}^{*}(X) + Y < c}.$

\subsection{Proof of Corollary \ref{coro:knapsack}}
Recall that if a policy $t$ is deterministic and $\alpha$-valid then $t(x) \in \mathcal{T}_{\alpha}(x)$ for every $x \in \mathcal{X}$, and we can write that $t(x) = \langle \pi(x), z(x; \alpha) \rangle$ where $\pi(x): \mathcal{X} \rightarrow \{0, 1\}^{K}$ and $\sum_{k \in [K]} \pi_{k}(x) = 1.$ Since $\alpha$-valid policies have a highly structured form, the objective of \eqref{eq:two_stage} can be written as $\EE[F]{\langle \pi(X), p(X; \alpha)\rangle}$ and the constraint of \eqref{eq:two_stage} can be written as $\EE[F]{\langle \pi(X), z(X; \alpha) \rangle} \leq B$. This yields \eqref{eq:knapsack}.

\subsection{Proof of Lemma \ref{lemm:weak_equity_ranks}}
\label{sec:proof_rank_preservation}

Let $t(y; B)$ is continuous in $B$ and satisfies weak equity (Definition \ref{defi:weakly_equitable}). Suppose for the sake of contradiction that there exists two units $y_{1}, y_{2} \in \mathcal{Y}$ and budgets $0 \leq B' \leq B$ where rank preservation does not hold, i.e.
\begin{align*}
y_{1} + t(y_{1}; B') &\leq  y_{2} + t(y_{2}; B'), \\
y_{1} + t(y_{1}; B) &> y_{2} + t(y_{2}; B).
\end{align*}

Since $t(y; B)$ is continuous in $B$, $t(y_{2}; B) - t(y_{1}; B)$ is continuous in $B$. The above conditions imply that $t(y_{2}; B') - t(y_{1}; B')\geq y_{1} - y_{2}$ and $t(y_{2}; B) - t(y_{1}; B) < y_{1} - y_{2}.$ By the Intermediate Value Theorem, there exists $B_{0} \in [B', B]$ so that 
\begin{equation} \label{eq:1} t(y_{2}; B_{0}) - t(y_{1}; B_{0}) = y_{1} - y_{2}. \end{equation}
At $B_{0}$, we have that $y_{1} + t(y_{1}; B_{0}) = y_{2} + t(y_{2}; B_{0})$, so by weak equity we have that both
\begin{align*}
t(y_{1}; B)  - t(y_{1}; B_{0}) &\geq t(y_{2}; B) - t(y_{2}; B_{0}) \\
t(y_{2}; B)  - t(y_{2}; B_{0}) &\geq t(y_{1}; B) - t(y_{1}; B_{0}).
\end{align*}
for all $B \geq B_{0}.$

This implies that 
\begin{equation} \label{eq:2} t(y_{1}; B)  - t(y_{1}; B_{0}) =  t(y_{2}; B) - t(y_{2}; B_{0})\end{equation} for all $B \geq B_{0}.$ Applying both \eqref{eq:1}, \eqref{eq:2} implies 
\begin{align*}
y_{1} + t(y_{1};  B)  - y_{2} - t(y_{2}; B) &= y_{1} - y_{2} + (t(y_{1}; B) - t(y_{2}; B)) \\
&= y_{1} - y_{2} + (t(y_{1}; B_{0}) - t(y_{2}; B_{0})) \\
&= y_{1} + t(y_{1}; B_{0}) - y_{2} - t(y_{2}; B_{0}) \\
& = 0.
\end{align*}

This is a contradiction because we assumed that $y_{1} + t(y_{1}; B) > y_{2} + t(y_{2}; B).$ Thus, if $t$ is continuous in $B$ and satisfies weak equity, it must also weakly preserve ranks.

\subsection{Proof of Lemma \ref{lemm:oracle_rate_we}}
To solve this problem, we consider optimizing over a simpler class of policies $\mathcal{M}$. Denote post-transfer consumption as $s(y; B) = y + t(y; B)$. Let $\mathcal{M}$ be the class of policies where $t(y; B)$ is monotone decreasing and $s(y; B)$ is monotone increasing for every $B > 0$. We characterize the optimal policy over $\mathcal{M}$. Then, we verify that $\mathcal{W} \subset \mathcal{M}$ and that the optimal policy family over $\mathcal{M}$ also satisfies Definition \ref{defi:weakly_equitable}, so it lies in $\mathcal{W}$. Thus, the optimal policy family over $\mathcal{M}$ is also the optimal policy family over $\mathcal{W}$. 

We consider
\begin{equation}
\min_{t(\cdot; \cdot): \mathcal{Y} \times \mathbb{R}_{+} \rightarrow \mathbb{R}_{+}} \{\PP[F]{Y + t(Y; B) < c} : \EE[F]{t(Y; B)} \leq B,\, t \in \mathcal{M} \}.
\end{equation}

This is equivalent to solving the following optimization problem for each $B >0$
\begin{align*}
\label{eq:rate_min}
\min_{t: \mathcal{Y} \rightarrow \mathbb{R}_{+}, s: \mathcal{Y} \rightarrow \mathbb{R}_{+}} \{\PP[F]{Y + t(Y) < c} : \EE[F]{t(Y)} \leq B,\, s(y) = y + t(y),\, s \text{ mon. inc.}\,, t \text{ is mon. dec.} \}.
\end{align*}
Fix $B > 0$. We can further restrict to policies that only allocate transfers to units with $y < c$ because units with $y \geq c$ lie above the poverty line already. Let $\mathcal{A}$ be the set of pre-transfer consumptions of units who are lifted to the poverty line under a policy $t$ such that $s(y) = y + t(y)$ is monotone increasing, $t$ is monotone decreasing, and $t(y) = 0$ for $y \geq c$, i.e.
\[ \mathcal{A} = \{ y < c \mid y + t(y) \geq c \}.\]
Since $s$ is monotone increasing and $t$ does not allocate transfers to units with $y \geq c$, $\mathcal{A}$ must have the form $(z, c).$ An optimal policy $t^{*}$ must allocate transfers $t(y) = c - y$ for $y \in \mathcal{A}$ because this is the minimum transfer amount that raises these units to the poverty line.

Let $\lambda := \inf \mathcal{A}$. Since the optimal policy $t^{*}$ must be monotone decreasing, $t^{*}(y) \geq c - \lambda$ for $y \leq \lambda$. Since allocating transfers to these units will not decrease the poverty rate, the optimal policy $t^{*}$ will allocate the lowest possible transfers to these units provided that the feasibility constraints are satisfied. We note that the policy 
\begin{equation}
\label{eq:policy}
t(y; \lambda) = \begin{cases} c - \lambda & y \leq \lambda \\ c - y & \lambda < y \leq c \\ 0 & y > c \end{cases}.
\end{equation}
guarantees that $s(y) = y + t(y)$ is monotone increasing because
\[ s(y) =  \begin{cases} c - \lambda  + y & y \leq \lambda \\ c & y > \lambda \end{cases}.\]
 As a result, the optimal policy family over $\mathcal{M}$ must have the form \eqref{eq:optimal_oracle}.

We show that $\mathcal{W} \subseteq \mathcal{M}$. First, if $t \in \mathcal{W}$, then $t$ must be monotone decreasing for every $B>0$. This can be seen by setting $B'=0$ into Definition \ref{defi:weakly_equitable}. Second, if $t \in \mathcal{W}$, then it weakly preserves ranks (Lemma \ref{lemm:weak_equity_ranks}), so $y + t(y; B)$ must be monotone increasing for every $B > 0$.

Any policy that weakly preserves ranks has the property that $y + t(y; B)$ is monotone increasing for every $B > 0.$ Thus, if $t \in \mathcal{W}$, then $t \in \mathcal{M}$.

We can check that \eqref{eq:optimal_oracle} satisfies weak equity. Note that 
\begin{equation}
y + t(y; B) = \begin{cases} y + c - \lambda(B) & 0 \leq y < \lambda(B) \\ 
c & \lambda(B) \leq y < c \\
y & y \geq c. \end{cases}.
\end{equation}

First, we note that $\lambda(B)$ that defines the oracle policy in \eqref{eq:optimal_oracle} is decreasing in budget. To see this, we observe that the cost of a policy with form \eqref{eq:policy} is decreasing in $\lambda$. Policy cost as a function of $\lambda$ is
\begin{align*}
    \EE[F]{t(Y; \lambda)} &= \int_{0}^{\lambda} (c - \lambda)  dF_{Y}(y) dy + \int_{\lambda}^{c} (c - y)  dF_{Y}(y) dy.
\end{align*}
Using Leibniz rule, we can check that
\begin{align*} 
\frac{\partial}{\partial \lambda} \EE[F]{t(Y; \lambda)} &= (c- \lambda) f(\lambda) - F(\lambda) - (c- \lambda) f(\lambda) \\
&= - F(\lambda).
\end{align*}
Since $F$ is a cumulative density function and is positive when $\lambda > 0$, then policy cost is strictly decreasing in $\lambda$. This implies that $\lambda(B)$ is decreasing in budget $B.$

Let $B > B'$. Suppose that 
\begin{equation}
\label{eq:ineq}
    y_{1} + t(y_{1}; B') \leq y_{2} + t(y_{2}; B').
\end{equation}
We compute that
\begin{align*}
t(y; B) - t(y; B') = \begin{cases} 0 & y \geq \lambda(B') \\
\lambda(B') - y & \lambda(B) \leq y < \lambda(B') \\
\lambda(B') - \lambda(B) & 0 \leq y < \lambda(B).
 \end{cases}
\end{align*}
Note that $t(y; B) - t(y; B')$ is decreasing in $y$.

\begin{enumerate}
\item Case (i): $y_{1}, y_{2} \in [0, \lambda(B'))$. Then \eqref{eq:ineq} implies that $y_{1} \leq y_{2}.$ Since $t(y; B) - t(y; B')$ is decreasing in $y$, so $t(y_{1}; B) - t(y_{1}; B') \geq t(y_{2}; B) - t(y_{2}; B').$
\item Case (ii): $y_{1}, y_{2} \geq \lambda(B').$ Then, $t(y_{i}; B') = (c - y_{i})_{+}$. Since $\lambda(B) \leq \lambda(B'),$ then we also have that $t(y_{i}; B) = (c - y_{i})_{+}$, so $t(y_{1}; B) - t(y_{1}; B') = t(y_{2}; B) - t(y_{2}; B').$
\item Case (iii): $y_{1} \in [0, \lambda(B'))$ and $y_{2} \geq \lambda(B')$. This implies that $y_{1} \leq y_{2}$ Since $t(y; B) - t(y; B')$ is decreasing in $y$, we have $t(y_{1}; B) - t(y_{1}; B') \geq t(y_{2}; B) - t(y_{2}; B').$
\item Case (iv): $y_{2} \in [0, \lambda(B'))$ and $y_{1} \geq \lambda(B').$ In this case, we have that $c \leq y_{2} + c - \lambda(B'),$ so $y_{2} \geq \lambda(B').$ This is a contradiction because we assumed that $y_{2} \in [0, \lambda(B'))$, so this case does not occur.
\end{enumerate}

Thus, the policy family in \eqref{eq:optimal_oracle} satisfies weak equity, and it must also be the optimal solution to \eqref{eq:oracle_prob}.

\section{Technical Proofs}

\subsection{Proof of Lemma \ref{lemm:regularity}}
\label{sec:regularity}
\subsubsection{Once Differentiable}
We apply Leibniz rule to show that $G_{x}(t)$ is differentiable on $[0, \infty)$ and find an explicit form for the derivative.

\textbf{Leibniz Rule} \citep{klenke2013probability}: Let $\mathcal{T}$ be an open subset of $\mathbb{R}$ and let $\mathcal{Y}$ be a measure space. Suppose that $g:\mathcal{T} \times \mathcal{Y} \rightarrow \mathbb{R}$ satisfies the following conditions. 

\begin{enumerate} 
	\item $g(t, y)$ be a Lebesgue integrable function of $y$ for each $t \in \mathcal{T}$.
	\item For almost all $y \in \mathcal{Y}$, the partial derivative $\frac{\partial}{\partial t}g(t, y)$ exists for all $t \in \mathcal{T}$.
	\item There is a Lebesgue integrable function $\theta: \mathcal{Y} \rightarrow \mathbb{R}_{+}$ for $|\frac{\partial g}{\partial t}(t, \cdot)| \leq  \theta(y)$ almost everywhere for all $t \in \mathcal{T}.$
\end{enumerate}
Then, for all $t \in \mathcal{T},$
\[\frac{d}{dt} \int_{\mathcal{Y}} g(t, y) dy = \int_{\mathcal{Y}} \frac{\partial}{\partial t}g(t, y)dy.\]

Let $\mathcal{T} = (0, \infty)$. Recall that $F_{Y \mid X=x}$ has positive density on $\mathcal{Y}.$ We can set $g(t,y):= L(y + t) \cdot f_{Y|X=x}(y).$ Since $L$ is a real-valued decreasing function defined on $\mathbb{R}_{+}$, we have that $\sup_{z \geq 0} L(z) = L(0)$. In addition, we also have that $L$ is bounded below by $C > - \infty$. Thus, we have that $|L(y + t)| \leq \max(|L(0)|, |C|).$ We use this fact to show that $g$ is Lebesgue integrable as follows
\begin{align*}
\int_{\mathcal{Y}} |g(t, y)|dy &= \int_{\mathcal{Y}} |L(y + t) \cdot f_{Y \mid X=x}(y)| dy \\
&= \int_{\mathcal{Y}} |L(y + t)| \cdot f_{Y \mid X=x}(y) dy \\
&\leq \max(|L(0)|, |C|) \int_{\mathcal{Y}} f_{Y \mid X=x}(y) dy \\
&< \infty.
\end{align*}

Since $L$ is monotone and differentiable, for all $y \in \mathcal{Y}, \frac{\partial g}{\partial t}(t, y)= L'(y + t) \cdot f_{Y|X=x}(y)$ exists for all $t \in \mathcal{T}$.

Since $L$ is convex, differentiable, and decreasing, $L'(z) \leq 0$ and $L'$ is an increasing function. Thus, $|L'(y + t)| \leq |L'(0)| < \infty$ for all $t \in \mathcal{T}$. Let $\theta(y) := |L'(0)| \cdot f_{Y \mid X=x}(y)$. We have that $\theta(y)$ is Lebesgue integrable because 
\[ \int_{\mathcal{Y}} \theta(y) dy = |L'(0)| \int_{\mathcal{Y}} f_{Y \mid X=x}(y) dy = |L'(0)| < \infty.\] Thus, $G_{x}(t)$ satisfies the conditions of Leibniz rule, and
\begin{equation}
\label{eq:derivative}
G'_{x}(t) = - \int_{\mathcal{Y}} L'(t + y) f_{Y \mid X=x}(y)dy
\end{equation}
for $t \in \mathcal{T}.$

It remains to check that $G_{x}$ is right-differentiable at $t=0.$ Since $L$ is right-differentiable at $z=0$, we can denote $L'(0)$ as the right derivative of $L$ at $z=0$. We check that $G_{x}$ is right-differentiable at $0$.
\begin{align*}
\lim_{h \rightarrow 0^{+}} \frac{G_{x}(h) - G_{x}(0)}{h} &= -\lim_{h \rightarrow 0^{+}} \int_{\mathcal{Y}} \frac{L(y + h) - L(y)}{h} \cdot dF_{Y \mid X=x}(y) dy.
\end{align*}
By the Mean Value Theorem, we have that
\[ L(y + h) - L(y) = h \cdot L'(z(y)) \]
for $z(y) \in [y, y + h].$ As a result, we have that
\begin{align*}
\Big| \frac{L(y + h) - L(y)}{h} \cdot dF_{Y \mid X=x}(y) \Big| &= | L'(z(y)) | \cdot dF_{Y \mid X=x}(y)  \\
&\leq |L'(0)|.
\end{align*}
The last line follows from the fact that $L$ is convex and decreasing, so $L'(z) \leq 0$ and $L'$ is increasing. Thus the dominated convergence theorem can be applied to obtain that
\begin{align*}
\lim_{h \rightarrow 0^{+}} \frac{G_{x}(h) - G_{x}(0)}{h} &=  -\int \lim_{h \rightarrow 0^{+}} \frac{L(y + h) - L(y)}{h} \cdot dF_{Y \mid X=x}(y) dy \\
&= -\int L'(y) dF_{Y \mid X=x}(y) dy.
\end{align*}
Thus, $G_{x}$ is right-differentiable at $t=0.$

\subsubsection{Strict Concavity}
Since $L$ is strictly convex, it follows that $G_{x}(t)$ is strictly concave.

\subsection{Proof of Lemma \ref{lemm:fgt_regularity}}
\label{sec:fgt_regularity}

\begin{subsubsection}{Once Differentiable}
We first consider $\alpha > 1.$ Note that in this case $L$ is differentiable on $\mathbb{R}_{+}$, so it is straightforward to see that $G_{x}(t)$ is also differentiable on $\mathbb{R}_{+}$.

When $\alpha = 1$, $L$ is not differentiable on $\mathbb{R}_{+}$. Nevertheless, we have that
\[G_{x}(t) = -\int_{0}^{c - t} (c - t - y) \cdot  dF_{Y \mid X=x}(y).\]
We can apply integration by parts to see that
\begin{align*}
G_{x}(t) &= - (c - t - y) \cdot F_{Y \mid X=x}(y) \Big|_{0}^{c - t} + \int_{0}^{c - t} F_{Y \mid X=x}(y)dy \\
&= - \int_{0}^{c - t} F_{Y \mid X=x}(y)dy.
\end{align*}
By Assumption \ref{assumption:cdf}, we have that $F_{Y \mid X}$ is continuous. Then, we have that $G_{x}'(t) = F_{Y \mid X=x}(c - t).$

\end{subsubsection}

\begin{subsubsection}{Strict Concavity}
We observe that the FGT indices for $\alpha \geq 1$ are convex. In addition, they are strictly convex at $c$ in the sense defined below.
\begin{defi}[Strict convexity at a point]
\label{defi:strict_convexity_at_point}
A convex function $f: \mathbb{R} \rightarrow \mathbb{R}$ is strictly convex at a point $c$ if for all $z_{1}, z_{2} \in \mathbb{R}$ such that $z_{1} < c < z_{2}$ and $\lambda \in (0, 1)$,
\[f(\lambda z_{1} + (1 - \lambda) z_{2}) < \lambda f(z_{1}) + (1 - \lambda) f(z_{2}).\]
\end{defi}
Let $t_{1}, t_{2} \in [0, c]$ such that $t_{1} < t_{2}$. We define $y' = c - \lambda t_{1} - (1 - \lambda) t_{2}$. We note that $y' + t_{1} < c < y' + t_{2}$. Applying Definition \ref{defi:strict_convexity_at_point}, we have that
\begin{equation} \label{eq:strict_convexity} L(y' + \lambda t_{1} + (1 - \lambda) t_{2}) < \lambda L(y' + t_{1}) + (1 - \lambda) L(y' + t_{2}). \end{equation}
We note that this result also holds in a small neighborhood about $y'$. Let $\epsilon = \frac{1}{2}\cdot \min(y' + t_{2} - c, c - y' - t_{1})$. In this case, $y' + \epsilon + t_{1} < c < y' + \epsilon + t_{2}$ and $y' - \epsilon + t_{1} < c < y' - \epsilon + t_{2}$, so \eqref{eq:strict_convexity} holds when $y'$ is replaced by $y' + \epsilon$ and $y' - \epsilon$. Let $I = [y'- \epsilon, y' + \epsilon].$ We note that under Assumption \ref{assumption:cdf}, $f_{Y \mid X}(y) > 0$ on $I$.
\begin{align*}
&G_{x}(\lambda t_{1} + (1 - \lambda) t_{2}) \\
&= -\int_{\mathcal{Y}} L(y + \lambda t_{1} + (1 - \lambda) t_{2}) f_{Y \mid X=x}(y) dy \\
&= -\int_{I} L(y + \lambda t_{1} + (1 - \lambda) t_{2}) f_{Y \mid X=x}(y) dy - \int_{ \mathcal{Y} \setminus I} L(y + \lambda t_{1} + (1 - \lambda) t_{2}) f_{Y \mid X=x}(y) dy \\
&> - \int_{I} [\lambda L(y + t_{1}) + (1 - \lambda) L(y + t_{2}) ]f_{Y \mid X=x}(y) dy - \int_{ \mathcal{Y} \setminus I} L(y + \lambda t_{1} + (1 - \lambda) t_{2}) f_{Y \mid X=x}(y)  dy \\
&\geq  - \int_{I} [\lambda L(y + t_{1}) + (1 - \lambda) L(y + t_{2})] f_{Y \mid X=x}(y) dy - \int_{ \mathcal{Y} \setminus I}  [\lambda L(y + t_{1}) + (1 - \lambda) L(y + t_{2}) ] f_{Y \mid X=x}(y)  dy \\
&= \lambda G_{x}(t_{1}) + (1 - \lambda) G_{x}(t_{2}).
\end{align*}
Thus, $G_{x}$ is strictly concave on $[0, c].$
\end{subsubsection}

\subsection{Proof of Lemma \ref{lemm:nonconvex_continuity}}
Since $L$ is nonconvex there exist points $x, y \in \mathbb{R}_{+}$ and $\lambda \in (0, 1)$ such that
 \[ L(\lambda x + (1- \lambda) y) > \lambda L(x) + (1 - \lambda) L(y).\]
Let $z = \lambda x + (1 - \lambda) y.$ We aim to show that this relationship holds at three continuity points of $L$, as well.

Let $\mathcal{E}$ be the set of discontinuity points of $L$, and let $\mathcal{C} := \mathbb{R}_{+} \setminus \mathcal{E}$ denote the set of continuity points of $L$. Since $L$ is monotone, we observe that $\mathcal{C}$ must be dense in $\mathbb{R}_{+}$. For the sake of contradiction, suppose not. Then there exists some $p \in \mathbb{R}_{+}$ and $r >0$ such that the neighborhood $B_{r}(p)$ about $p$ with radius $r$ does not intersect $\mathcal{C}$, i.e. $B_{r}(p) \cap \mathcal{C} = \emptyset.$ This implies that $B_{r}(p) \subset \mathcal{E}$. However, this is a contradiction because $B_{r}(p)$ is an open set that contains an uncountable number of points, and $L$ is monotone, so $\mathcal{E}$ must be a countable set. Thus, we must have that $\mathcal{C}$ is dense in $\mathbb{R}_{+}.$ 

Since $\mathcal{C}$ is dense in $\mathbb{R}_{+}$, we can construct sequences $\{x_{n}\}, \{y_{n}\}, \{z_{n}\}$ that lie in $\mathcal{C}$ such that $x_{n} \downarrow x$, $y_{n} \downarrow y$, and $z_{n} \uparrow z.$ Since $L$ is monotone decreasing, we note that
\begin{align*} 
\lim_{n \rightarrow \infty} L(x_{n}) &= C_{x} \leq L(x),  \\
\lim_{n \rightarrow \infty} L(y_{n}) &= C_{y} \leq L(y), \\
\lim_{n \rightarrow \infty} L(z_{n}) &= C_{z} \geq L(z).
\end{align*}
Note that for $N$ sufficiently large, we have that $x_{n}  < z_{n} < y_{n}$. For $n > N$, we define
\[\lambda_{n} = \frac{z_{n} - y_{n}}{x_{n} - y_{n}}\]
and for $n \leq N$, we set $\lambda_{n} = \frac{1}{2}.$ We note that $\lambda_{n} \in (0, 1)$ and $\lambda_{n} \rightarrow \lambda.$ Note that the affine relationship $z_{n} = \lambda_{n} x_{n} + (1- \lambda_{n}) y_{n}$ is satisfied exactly for $n > N.$

We observe that
\begin{align*}
\lim_{n \rightarrow \infty} L(z_{n}) - \lambda_{n} L(x_{n}) - (1 - \lambda_{n}) L(y_{n}) &= C_{z} - \lambda C_{x} - (1 - \lambda) C_{y} \\
&\geq L(z) - \lambda L(x) - (1- \lambda) L(y) \\
&>0.
\end{align*}
Thus, for sufficiently large $n$, we have that $ L(z_{n}) - \lambda_{n} L(x_{n}) - (1 - \lambda_{n}) L(y_{n}) > 0$. Since $x_{n}, y_{n}, z_{n}$ are continuity points of $L$, $\lambda_{n} \in (0, 1)$, and $z_{n} = \lambda_{n} x_{n} + (1 - \lambda_{n}) y_{n}$, this proves the desired claim.

\subsection{Proof of Lemma \ref{lemm:smoothed_loss_diff}}
To show that $\tilde{L}_{b}$ is continuously differentiable, we show that differentiation and integration can be interchanged using Leibniz rule and obtain an expression for $\tilde{L}_{b}'$. After that, we apply the dominated convergence theorem to conclude that $\tilde{L}_{b}'$ is continuous.

We verify that Leibniz rule applies to $\tilde{L}_{b}$. First, we apply a change of variables
\[ \tilde{L}_{b}(t) = \int_{t}^{\infty} L(u) dQ_{b}(u - t) du.\]
Let $\mathcal{T}= (0, \infty).$ We note that $Q_{b}(z)$ has a positive density on $[0, \infty)$. We can show that $L(u) \cdot dQ_{b}(u - t)$ is Lebesgue integrable because
\begin{align*}
\int_{t}^{\infty} |L(u) dQ_{b}(u - t)|dz &\leq ||L||_{\infty} \int_{t}^{\infty} dQ_{b}(u - t) dz  \\
&< \infty.
\end{align*}
In addition, the partial derivative $\frac{\partial}{\partial t} L(u) dQ_{b}(u - t) = -L(u) dQ_{b}'(u - t)$, which exists for all $t \in \mathcal{T}.$ We can show that there is a Lebesgue integrable function $\theta(u)$ so that 
\[ \Big|\frac{\partial}{\partial t} L(u) dQ_{b}(u - t) \cdot \mathbb{I}(u \geq t) \Big| \leq \theta(u)\]
almost everywhere. Define $\theta(u) = |L(u)| \cdot \frac{1}{b^{2}}.$ For $z \geq 0$, we have that
\begin{align*}
\Big|\frac{\partial}{\partial t} L(u) dQ_{b}(u - t) \cdot \mathbb{I}(u \geq t) \Big|  &= | L(u) \cdot dQ_{b}'(u - t) \mathbb{I}(u \geq t)| \\
&\leq \Big|L(u) \cdot \exp(- (u - t)/b) \cdot \Big( -\frac{(u - t)}{b^{3}} + \frac{1}{b^{2}} \Big)\Big|\\
&\leq |L(u)| \cdot \frac{\exp(-(u - t)/b)}{b^{2}} \cdot \Big|1 - \frac{u - t}{b} \Big| \\
&\leq \theta(u).
\end{align*}
The last line follows because $e^{-r} |1 - r| \leq 1$ for $r \in \mathbb{R}_{+}$. We have that $\theta$ is Lebesgue integrable because 
\begin{align*}
\int_{0}^{\infty} \theta(u)  &\leq \int_{0}^{\infty} |L(u)| \cdot \frac{1}{b^{2}}  du \\
\end{align*}
and $L$ is Lebesgue integrable. Thus, we can apply Leibniz rule to conclude that
\begin{align*}
\tilde{L}_{b}'(t) &= -\int_{t}^{\infty} L(u) \cdot dQ_{b}'(u - t).
\end{align*}
We additionally note that pointwise convergence of the integrand holds, i.e. $t_{n} \rightarrow t$ implies that $ L(u) \cdot dQ_{b}'(u - t_{n}) \cdot \mathbb{I}(u \geq t_{n}) \rightarrow L(u) \cdot dQ_{b}'(u - t) \cdot \mathbb{I}(u \geq t)$. Then pointwise convergence and the existence of the Lebesgue integrable dominating function imply that $\tilde{L}_{b}'(t_{n}) \rightarrow \tilde{L}_{b}'(t)$ by the dominated convergence theorem. Thus, $\tilde{L}_{b}'(t)$ is also continuous.

\setcounter{figure}{0}
\setcounter{table}{0}
\renewcommand{\thefigure}{\Alph{section}.\arabic{figure}}
\renewcommand{\thetable}{\Alph{section}.\arabic{table}}

\section{Data}
\label{sec:data}

\subsection{Country Selection}
\label{subsec:country-selection}

This section details our criteria for including countries in the analysis, and any exceptions or caveats thereto. Generally speaking, we aim to include countries that account for as large a share of the world's poor as possible subject to various feasibility constraints. The specific criteria we impose are that countries must:
\begin{enumerate}
    \item Have an extreme poverty rate greater than 10\%, or account for more than 1\% of the world's extreme poor. This filter aims to focus our search effort on the countries that most influence aggregate costs. Globally, 94.4\% of poor households live in countries that meet this criterion.\footnote{Aggregate poverty shares stated in this section are the authors' own calculations based on poverty estimates for 2023, from the World Bank's Poverty and Inequality Platform \citep{pip, pip_stata}, accessed April 9, 2026. Country inclusion/exclusion decisions were based on estimates for 2022, from the World Poverty Clock \citep{wpc}, accessed September 25 2025.}
    \begin{itemize}
        \item We make an exception at this stage to include Côte d'Ivoire and Senegal. They are covered by the EHCVM survey series, which also covers six other West African countries already in our sample, and the surveys are harmonized in such a way that including them is unusually easy.
    \end{itemize}
    
    \item Have a recent, high-quality, nationally-representative LSMS survey or equivalent. This is a prerequisite for our analysis. Whenever possible we use the same survey as that on which estimates in the World Bank's Poverty and Inequality Platform are based. 94.4\% of poor households live in countries that additionally meet this criterion.

    \item Are estimated to have at least 1,000 extremely poor households covered by this survey, given the survey sample size and published poverty rate. This filter aims to ensure that the survey contains enough poor households for reliable statistical learning; relaxing it would require countries to conduct larger surveys (or surveys more focused in the regions containing the most poor households). 94.1\% of poor households live in countries that additionally meet this criterion.
    \begin{itemize}
        \item We make an exception at this stage to include Bangladesh (which we anticipated would contain slightly fewer than 1,000 extremely poor households) because of its size.
    \end{itemize}

    \item Have made the relevant survey data, including a measure of aggregate household consumption, publicly available, or have provided it upon request. 81.3\% of poor households live in countries that additionally meet this criterion.
    
    \item Have usable currency conversion factors available. Specifically, we remove Afghanistan and Somalia on the basis of advice from World Bank staff that reliable PPP conversion factors for the years in which the relevant surveys were conducted are not available, and we remove South Sudan because we were unable to find a CPI adjustment factor temporally aligned with the survey's currency unit.
    
\end{enumerate}
This leaves a final sample of \sampleNumCountries\ countries which collectively account for \sampleShareWorldsPoor\% of the world's extreme poor.

\subsection{Outcome Construction}
\label{subsec:outcome-construction}

We calculate consumption per capita by dividing total household consumption by household size without an adult equivalence scale adjustment, following the World Bank Poverty and Inequality Platform's methodology \citep{worldbankPIPmethodology}. Wherever possible we use the consumption aggregates provided along with the disaggregated data for this purpose. Accounting for transfers, housing, and other durable goods are among the thornier issues in measuring consumption \citep{AmendolaVecchi2022durable}, and the sources vary somewhat in the ways they do this (see Table \ref{tab:cons-agg-details-1}). That said, large majorities include the value of both cash and in-kind transfers; include an estimate of the value of housing services consumed; and include an estimate of the value of services from non-housing durable goods. 

We convert consumption aggregates from all surveys from local currency units (LCU) to 2017 PPP USD using Consumer Price Index and PPP conversion factors. In some cases we have access to precise conversion factors used by the Poverty and Inequality Platform (PIP) team to make international poverty rate estimates. In others, including when we are using a survey not used by PIP, we use factors drawn from the World Bank's development indicators \citep{developmentindicators}. This requires identifying the base year of the LCUs in which the consumption aggregate is defined. The base year is clear from survey context and/or documentation in most cases. In cases where a survey spans two years and no further information about currency base year is available, we average conversion factors across the two years. Table \ref{tab:cons-agg-details-1} describes properties of the consumption aggregates we use and indicates deflator usage.\footnote{We thank Daniel Mahler and Elizabeth Foster in particular for their generous guidance and assistance obtaining and interpreting the deflators and conversion rates required for a number of these operations.}

Table \ref{tab:surveys} displays the resulting poverty headcount rate estimates (column titled ``Survey'' under column group ``Poverty Rate'') alongside poverty rate estimates for the same country and year from PIP. Where the PIP estimate was derived from the same underlying survey as our estimate, we include it in the column labeled ``World Bank (same source);'' where it was derived from a different survey (including by interpolating between estimates from different surveys in different years) we include it in the column labeled ``World Bank (other source).'' When we use the same source we are generally able to match the corresponding PIP estimate to within a percentage point (sometimes using non-public deflators or, in Nigeria, omitting a region where survey enumeration was disrupted). In four cases we do not: Burundi (\BDIAbsDiscrepancy\ percentage points \BDIDiscrepancyDirection), Central African Republic (\CAFAbsDiscrepancy\ percentage points \CAFDiscrepancyDirection), South Africa (\ZAFAbsDiscrepancy\ percentage points \ZAFDiscrepancyDirection), and in the Democratic Republic of the Congo (\CODAbsDiscrepancy\ percentage points \CODDiscrepancyDirection). When we use a different source the average absolute difference between the estimates is \differentSurveysMeanAbsDiscrepancy\ percentage points and the average difference is \differentSurveysMeanDiscrepancy\ percentage points.

\begin{sidewaystable}
{
\centering 
\caption{Consumption Aggregate Details}
\label{tab:cons-agg-details-1}
\small
\newcounter{fnlumpy}
\newcounter{fnhousing}
\scriptsize
\begin{tabular}{lllll}
\hline
\multirow{2}{*}{\textbf{Country}} & \multicolumn{3}{c}{\textbf{Category of Consumption}} & \multirow{2}{*}{\textbf{Deflator Usage}}                                                                                                                                         \\ 
\cline{2-4}
                                  & \textbf{Grants and Transfers} & \textbf{Durable Goods\footnote{This column uses terminology introduced by \cite{AmendolaVecchi2022durable}.}} & \textbf{Non-Rental Housing\footnote{Includes any housing not rented on the market: Owned housing, family- or employer-provided housing, etc.}} &                                          \\ 
\hline
Bangladesh                        & Unspecified                        & Unspecified                    & Imputed; method unspecified                                                                                                                    & None used                           \\
Benin                             & Included                           & User cost                      & Hedonic function\footnote{A function predicting rental value based on observable housing characteristics and amenities; fit using self-reported rents of respondents living in rental housing.} & Temporal (WB)\footnote{Here ``(WB)'' indicates deflator data obtained from the World Bank which differs from, or is not included in, the corresponding survey dataset.} \\
Burkina Faso                      & Included                           & User cost                      & Hedonic function                                                                                                                               & Temporal (WB)                \\
Burundi                           & Unspecified                        & User cost                      & Self-reported rental value                                                                                                                     & Spatial                             \\
Central African Republic          & Included                           & User cost                      & Imputed; method unspecified                                                                                                                    & None used                           \\
Colombia                          & Included                           & Acquisition approach           & Imputed; method unspecified                                                                                                                    & None used                           \\
C\^{o}te d'Ivoire                 & Included                           & User cost                      & Hedonic function                                                                                                                               & Temporal (WB)                 \\
Dem.\ Rep.\ of the Congo          & Included                           & User cost                      & Hedonic function                                                                                                                               & Spatial (urban/rural)               \\
Ethiopia\footnote{Responses from Tigray Region are not included in survey microdata.} & Included & Acquisition approach           & Unspecified                                                                                                                                    & Spatial                             \\
Ghana                             & Included                           & User cost                      & Hedonic function                                                                                                                               & None used                           \\
Guinea-Bissau                     & Included                           & User cost                      & Hedonic function                                                                                                                               & Temporal (WB)                   \\
India                             & Included                           & Acquisition approach\footnote{Long-term (``lumpy'') durables excluded.} & All housing consumption excluded\footnote{All (rental or non-rental) housing consumption is excluded, per WB methodology.}              & Spatial and temporal (WB)       \\
Indonesia                         & Unspecified                        & Unspecified                    & Unspecified                                                                                                                                    & None used                           \\
Kenya                             & Food included; non-food excluded   & Unspecified                    & All housing consumption excluded\footnote{All (rental or non-rental) housing consumption is excluded, per WB methodology.}                     & None used                           \\
Liberia                           & Food included; non-food excluded   & Unspecified                    & Self-reported rental value                                                                                                                     & None used                           \\
Madagascar                        & Food included; non-food excluded   & User cost                      & Hedonic function                                                                                                                               & None used                           \\
Malawi                            & Food included; non-food excluded   & User cost\footnote{Assumes linear depreciation and disregards interest rate.} & Self-reported rental value\footnote{Outliers imputed using hedonic function.}                                       & None used                           \\
Mali                              & Included                           & User cost                      & Hedonic function                                                                                                                               & Temporal (WB)                   \\
Mexico                            & Included                           & Acquisition approach           & Imputed; method unspecified                                                                                                                    & None used                           \\
Namibia                           & Unspecified                        & Included; handling unspecified & Imputed; method unspecified                                                                                                                    & Spatial; temporal unspecified       \\
Niger                             & Included                           & User cost                      & Hedonic function                                                                                                                               & Temporal, (WB)                   \\
Nigeria\footnote{Omitting responses from state of Borno per WB methodology.} & Food included; non-food unspecified & Acquisition approach\footnote{Long-term (``lumpy'') durables excluded.} & Hedonic function                                                              & Spatial and temporal                \\
Pakistan                          & Included                           & Included; handling unspecified & Self-reported rental value                                                                                                                     & Unspecified                         \\
Rwanda                            & Included                           & User cost                      & Self-reported rental value\footnote{Outliers imputed by means unspecified.}                                                                    & Temporal                            \\
Senegal                           & Included                           & User cost                      & Hedonic function                                                                                                                               & Temporal (WB)                   \\
Sierra Leone                      & Food included; non-food unspecified & User cost                     & Hedonic function                                                                                                                               & Spatial and temporal                \\
South Africa                      & Included                           & Acquisition approach           & Average rental yield\footnote{Disaggregated by housing type and province.}                                                                     & Temporal                            \\
Sudan                             & Food included; non-food unspecified & Acquisition approach          & Self-reported rental value                                                                                                                     & None used                           \\
Tanzania                          & Food included; non-food excluded   & User cost                      & Hedonic function                                                                                                                               & Spatial and temporal                \\
Timor-Leste                     & Food included; non-food unspecified & User cost                     & Hedonic function                                                                                                                               & None used                           \\
Togo                              & Included                           & User cost                      & Hedonic function                                                                                                                               & Temporal (WB)                   \\
Uganda                            & Included                           & Acquisition approach           & Market prices (details unspecified)                                                                                                            & None used                           \\
Yemen                             & Included                           & User cost                      & Self-reported\footnote{Counterfactual rent estimated using self-reported capital value of home}                                                                                            & Spatial                             \\
Zimbabwe                          & Food included; non-food excluded   & User cost                      & Self-reported rental value                                                                                                                     & Spatial and temporal                \\
\hline
\end{tabular}\\
}
\vspace{.5em}
{\footnotesize This table lists characteristics of each consumption aggregate, including how each one handles often-tricky consumption categories.}

\end{sidewaystable}
\clearpage

\subsection{Predictor Selection}
\label{subsec:predictor-selection}

This section describes our rubric for selecting covariates from household surveys for inclusion as predictors when learning transfer policies. Our goal is to select only characteristics that are plausibly verifiable. To that end, we build a rubric from an initial list of covariates actually used in PMTs implemented in low- and middle-income countries, as described in Section \ref{sec:data-sources-uses}. In particular, we define categories of covariates, generalizing where appropriate from individual covariates that have been used. For example, if previous PMTs used the material out of which the house's walls are made, then we take that as justification for also using the material out of which other parts of the house, such as the roof, are made. The resulting categories (labelled (a), (b), (c), etc. below) are exhaustive of the categories of variables within each top-level group (e.g., household demographics) that we include. 

We also exercise judgment in excluding some variables that have been used in real-world PMTs but that we deem too difficult to plausibly verify at large scale. Examples include: Age of dwelling; indicators of food security, such as whether household members had skipped meals in the past week; use of fertilizer; use of an internet connection; income and private transfers (like foreign remittance); and ownership of a mobile phone or usage of someone else's mobile phone. We also exclude questions with unstructured free-entry text responses.

We use the resulting rubric as a guide when selecting predictors to use from each survey. The rubric, edited for clarity, is as follows:

\begin{enumerate}
	\item Household demographics
	\begin{enumerate}
		\item Counts of household members by type, including the overall number of household members, the number of male and female members, the numbers in various age-based subcategories such as children, adults, elderly adults / senior citizens, and the number of specially-abled members. If the survey does not pre-define age-based categories then we will ourselves count the number of children 17 and under, and the number of older adults 65 and older. Examples of prior usage for the number of elderly adults: 60 and over for Below Poverty Line (BPL) classification in India \citep{BPLGuidelines} or 65 and over \citep{Alatasetal2012targeting,hanna2018universal}.
		\item Household head: Age, gender, and marital status. Examples of prior usage: \citet{Alatasetal2012targeting} include married/unmarried; \citet{FernandezHadiwidjaja2018} include single/married/divorced; \citet{KiddWylde2011} include widow status.
	\end{enumerate}
	\item Human capital
	\begin{enumerate}
		\item Educational attainment of the household head. Depending on the source this may take the form of years completed, highest academic level completed, or simply a literacy indicator. Examples of prior usage: \citet{Alatasetal2012targeting}, \citet{BPLGuidelines,BPLGuidelinesUrban}.
		\item Maximum educational attainment of all adult members. Again, this may be measured in years, in level completed, etc.\footnote{We take the maximum rather than including the attainment of all members individually in order to produce a per-household predictor.} Example of prior usage: \citet{KiddWylde2011}.
		\item Maximum educational attainment of all female adult members. Example of prior usage: \citet{KiddWylde2011}.
		\item Number of children currently enrolled in school.
	\end{enumerate}
	\item Household asset: presence and number owned
	\begin{enumerate}
		\item We do \textit{not} include ownership-structure details beyond the fact of asset ownership.
		\item Dwelling characteristics such as per capita number of rooms \citep{KiddWylde2011} or floor space \citep{Alatasetal2012targeting}, material used to construct floors, walls, or roof, etc., as well as an indicator for homeownership itself. Examples of prior usage: \citet{Alatasetal2012targeting} use binary indicators for home-ownership and roof materials. \citet{hanna2018universal} consider more granular categories.
		\item Presence and physical characteristics of amenities such as: Type of latrine, water source, lighting source, drainage system, waste collection, access to electricity and gas, type of cooking fuel, cable connection. Examples of prior usage: \citet{Alatasetal2012targeting} use a simple binary indicator for availability of clean water within the house. \citet{hanna2018universal} use more granular categories of water source.
		\item Ownership of consumer durables such as appliances (e.g. radio, television, refrigerator, generator, cooker, heater, fan, air conditioner); transportation (e.g. car, bicycle, motorbike); furniture (e.g. sofa, bed etc); and devices (e.g. computers).
		\item Productive agricultural assets including land, livestock, irrigation facilities, and farm machinery. We exclude cultivation details, such as crop types and amounts, as they may not be verifiable. We include the amount of land owned. Examples of prior usage: \citet{BPLGuidelines, BPLGuidelinesUrban} include farm machinery.
		\item Productive non-agricultural assets, e.g a sewing machine.
		\item We generally do \textit{not} include financial assets, as we expect the kinds of financial assets held by poor households to be hard to verify, but we include any that they hold as part of a government scheme for which the government might plausibly hold records. Example of prior usage: \citet{BPLGuidelines,BPLGuidelinesUrban} include usage of the Kisan credit card scheme in India.
		\item We \textit{exclude} crop stores, as they may not be easy to verify.
	\end{enumerate}
	\item Livelihood activities
	\begin{enumerate}
		\item Primary sector of employment of the household head (e.g., agriculture, manufacturing, services).
		\item Primary occupation of the household head (e.g., self-employed, salaried employee, casual laborer, etc.).
		\item We do \textit{not} include the occupations of other members of the household.
		\item Ownership of enterprises.
		\item Receipt of other public transfers, including amount, if this could plausibly be verified by merging in other administrative records. We omit transfers when it is unclear whether they are administratively documented, such as child support or transfers from non-government institutions. Example of prior usage: \citet{CamachoConover2011} describes a poverty census which includes social security information.

	\end{enumerate}

    \begin{note} Note: In cases where livelihood data is encoded using coding systems like ISIC, ISCO, etc., we truncate the codes to the first two digits, to control the eventual number of categories. In cases where this data is not recorded using a coding system admitting truncation, we limit to 50 categories including a catchall ``other".\end{note}

	\item Geographic indicators
	\begin{enumerate}
		\item Urban/rural status, including any further available classifications such as peri-urban.
		\item Administrative geographic information: Which administrative division a household is in. Note that some care is required in the treatment of these indicators, as surveys may include region identifiers at a finer granularity than that at which they are representative. To obtain results which reflect expected performance across households anywhere in each country, we include geographic identifiers only at granularities at which the survey is representative. For example, if a survey were to sample households from all districts but only a subset of subdistricts, we would include district identifiers but not subdistrict identifiers as predictors.

		\item Distance to important locations such as district centers, markets, or public facilities such as post offices. Examples of prior usage: \citet{Alatasetal2012targeting} use distance to district centers and markets. \citet{hanna2018universal} use distance to a post office.
		\item Environmental conditions such as rainfall history. \citet{del2015safety} provides a few case studies that illustrate feasibility. 
	\end{enumerate}
	\item Community characteristics
	\begin{enumerate}
		\item Presence of publicly or privately provided services such as healthcare. Examples of prior usage: \citet{Alatasetal2012targeting} uses an indicator for presence of a doctor.  \citet{KiddWylde2011} uses an indicator for presence of a midwife.
		\item Presence of publicly or privately provided infrastructure such as paved roads, banking facilities, or regional government offices. Examples of prior work: \citet{Alatasetal2012targeting,KiddWylde2011} use an indicator for presence of banking facilities. \citet{KiddWylde2011} use an indicator for presence of regional government offices.
		\item Population characteristics such as headcount or population density. Example of prior work: \citet{KiddWylde2011}.
	\end{enumerate}
\end{enumerate}

\subsection{Secondary Data}
\label{subsec:secondary_data}

This section compiles, for convenient reference, the full list of secondary data sources (i.e., those other than LSMS-style survey data) referenced elsewhere in the text.

\begin{itemize}
    \item Country-level poverty data from the World Bank Poverty and Inequality Platform \citep{pip, pip_stata}:
    \begin{itemize}
        \item Poverty headcount rate and poverty gap index for each country for which we use a survey, from the year of that survey, interpolated by the authors if necessary.
        \item Poverty headcount rate and poverty gap index for all countries in the world in 2023, when necessary using interpolated data provided in the dataset.
    \end{itemize}
    \item Country-level poverty headcount data from the \href{https://worldpoverty.io/}{World Poverty Clock} \citep{wpc}.
    \item Country-level population estimates from the World Bank Development Indicators \citep{developmentindicators}.
    \item Effective currency exchange rates between local currency units (LCU) and US dollars. These rates are drawn from the World Bank's Development Indicators \citep{developmentindicators}. We use the ``DEC alternative conversion factor'' series, designed to reflect actual effective exchange rates, on advice from World Bank staff, in particular to match local currency units used in PPP conversion factors. Table \ref{tab:official_exchange_rates} documents how dollar costs for the affected countries would change if we instead used official exchange rates from the World Bank's Development Indicators \citep{developmentindicators}. 

    \item Conversion factors between LCU and USD PPP (international dollars). These factors are drawn from multiple sources, depending on data availability and purpose:
    
    \begin{itemize}
    \item The World Bank's Statistics Online portal \citep{worldbank_sol}.
    \item The World Bank's Development Indicators \citep{developmentindicators}. 
    \item Private correspondence with the World Bank (Rwanda).
    \item We fill some gaps in the World Bank dataset using data from the  IMF\footnote{Indicator name: Rate, Domestic currency per international dollar in PPP terms, ICP benchmarks 2017-2021} \citep{imf_data}: Taiwan (2017, 2021); Venezuela (2021); Yemen (2017, 2021).
    
    \item Both World Bank and IMF datasets were missing Venezuela's 2017 conversion factor at the time of access. We estimate this value by assuming that the nominal currency exchange rate and the conversion factor from LCU to USD PPP (international dollars) have the same ratio as in the nearest year with complete data. By assuming that ratio, we are able to obtain an estimate for the conversion factor based on the nominal exchange rate for that year, which is present in IMF data.
    
    \end{itemize}

    \item Country-level Consumer Price Index data from the World Bank's Statistics Online portal \citep{worldbank_sol}, the World Bank's Development Indicators \citep{developmentindicators}, and private correspondence with the World Bank (Rwanda) to adjust for inflation.
    \item Net official development assistance and official aid received, by country \citep{developmentindicators}.

    \item GDP by country and region \citep{developmentindicators}.

    \item Revenue as share of GDP by country \citep{imf_data} and by region \citep{developmentindicators}.
\end{itemize}

\clearpage

\section{Additional Exhibits}
\label{sec:additional_exhibits}

\setcounter{figure}{0}
\setcounter{table}{0}

\begin{sidewaystable}
\caption{Survey data sources}
\label{tab:survey_data_sources}
\footnotesize
\scriptsize
\begin{tabular}{lllr}
\toprule
Country & Survey Name & Citation & Survey Year \\
\midrule
Bangladesh & HIES & \cite{bgd_main_survey} & 2022 \\
Benin & EHCVM & \cite{ben_main_survey} & 2018 \\
Burkina Faso & EHCVM & \cite{bfa_main_survey} & 2018 \\
Burundi & EICVMB & \cite{bdi_main_survey} & 2020 \\
Central African Republic & EHCVM & \cite{caf_main_survey} & 2021 \\
Colombia & ENPH & \cite{col_main_survey} & 2016 \\
Congo - Kinshasa & EGI-ODD & \cite{cod_main_survey} & 2020 \\
Côte d’Ivoire & EHCVM & \cite{civ_main_survey} & 2018 \\
Ethiopia & Socio-Economic Panel Survey & \cite{eth_main_survey} & 2021 \\
Ghana & Living Standards Survey 7 & \cite{gha_main_survey} & 2016 \\
Guinea-Bissau & EHCVM & \cite{gnb_main_survey} & 2018 \\
India & Household Consumption Expenditure Survey & \cite{ind_main_survey} & 2022 \\
Indonesia & SUSENAS & \cite{idn_main_survey} & 2018 \\
Kenya & Continuous Household Survey & \cite{ken_main_survey} & 2021 \\
Liberia & HIES & \cite{lbr_main_survey} & 2014 \\
Madagascar & Permanent Household Survey & \cite{mdg_main_survey} & 2021 \\
Malawi & Fifth Integrated Household Survey & \cite{mwi_main_survey} & 2019 \\
Mali & EHCVM & \cite{mli_main_survey} & 2018 \\
Mexico & ENIGH & \cite{mex_main_survey} & 2024 \\
Namibia & Income and Expenditure Survey 2015-16 & \cite{nam_main_survey} & 2015 \\
Niger & EHCVM & \cite{ner_main_survey} & 2018 \\
Nigeria & Living Standards Survey & \cite{nga_main_survey} & 2018 \\
Pakistan & HIES & \cite{pak_main_survey} & 2018 \\
Rwanda & EICV7 & \cite{rwa_main_survey} & 2023 \\
Senegal & EHCVM & \cite{sen_main_survey} & 2018 \\
Sierra Leone & Integrated Household Survey 2018 & \cite{sle_main_survey} & 2018 \\
South Africa & Income and Expenditure Survey & \cite{zaf_main_survey} & 2010 \\
Sudan & NBHS & \cite{sdn_main_survey} & 2009 \\
Tanzania & National Panel Survey, Wave 5 & \cite{tza_main_survey} & 2020 \\
Timor-Leste & Survey of Living Standards and Extension & \cite{tls_main_survey} & 2007 \\
Togo & HSHLS & \cite{tgo_main_survey} & 2018 \\
Uganda & National Panel Survey & \cite{uga_main_survey} & 2019 \\
Yemen & HBS & \cite{yem_main_survey} & 2014 \\
Zimbabwe & PICES & \cite{zwe_main_survey} & 2017 \\
\bottomrule
\end{tabular}

\vspace{.5em}

{\footnotesize{We note that EHCVM is an abbreviation for "Enquête Harmonisée sur les Conditions de Vie des Ménages," HSHLS is an abbreviation for "Enquête Harmonisée sur les Conditions de Vie des Ménages" (Togo's WAEMU-harmonized survey, referenced in English as the Harmonized Survey on Households Living Standards), EICV7 is an abbreviation for "Enquête Intégrale sur les Conditions de Vie des Ménages," HIES is an abbreviation for "Household Income and Expenditure Survey," EICVMB is an abbreviation for "Enquête Intégrée sur les Conditions de Vie des Ménages au Burundi," EGI-ODD is an abbreviation for "Enquête par Grappes à Indicateurs des Objectifs de Développement Durable," ENPH is an abbreviation for "Encuesta Nacional de Presupuestos de los Hogares," ENIGH is an abbreviation for "Encuesta Nacional de Ingresos y Gastos de los Hogares," NBHS is an abbreviation for "National Baseline Household Survey," HBS is an abbreviation for "Household Budget Survey," PICES is an abbreviation for "Poverty, Income, Consumption and Expenditure Survey," and SUSENAS is an abbreviation for "Survei Sosial Ekonomi Nasional."}}
\end{sidewaystable}

\begin{table}
\caption{Survey characteristics} 
\label{tab:surveys}
\begin{center}
    \begin{tabular}{lrrrrr}
\toprule
&&& \multicolumn{3}{c}{Poverty Rate} \\
\cmidrule(lr){4-6}
Country & $n$ & $d$ & World Bank & World Bank & Survey \\
&&&  (same source) & (other source) & \\
\midrule
Bangladesh & 14268 & 193 & 3.6 & - & 3.6 \\
Benin & 8012 & 321 & 20.8 & - & 20.8 \\
Burkina Faso & 7010 & 379 & 31.2 & - & 31.2 \\
Burundi & 8358 & 82 & 62.1 & - & 66.8 \\
Central African Republic & 6411 & 157 & 65.7 & - & 64.1 \\
Colombia & 87201 & 109 & - & 4.9 & 2.8 \\
Congo - Kinshasa & 8993 & 64 & 78.9 & - & 84.1 \\
Côte d’Ivoire & 12992 & 366 & 11.5 & - & 11.5 \\
Ethiopia & 4959 & 148 & - & 32.0 & 35.6 \\
Ghana & 14009 & 327 & 25.2 & - & 25.2 \\
Guinea-Bissau & 5351 & 417 & 21.7 & - & 21.1 \\
India & 261696 & 40 & 2.3 & - & 2.3 \\
Indonesia & 295155 & 99 & 4.4 & - & 4.1 \\
Kenya & 16963 & 40 & 36.1 & - & 36.1 \\
Liberia & 4085 & 306 & 25.9 & - & 26.0 \\
Madagascar & 16540 & 171 & 52.3 & - & 52.3 \\
Malawi & 11434 & 152 & 70.1 & - & 70.1 \\
Mali & 6602 & 420 & 15.2 & - & 15.2 \\
Mexico & 91414 & 208 & 1.2 & - & 1.1 \\
Namibia & 10090 & 160 & 15.6 & - & 15.6 \\
Niger & 6024 & 215 & 50.9 & - & 50.9 \\
Nigeria & 21580 & 129 & 30.9 & - & 30.9 \\
Pakistan & 24809 & 134 & 4.9 & - & 4.9 \\
Rwanda & 15054 & 151 & 24.2 & - & 24.2 \\
Senegal & 7156 & 411 & 9.2 & - & 9.2 \\
Sierra Leone & 6810 & 177 & 26.1 & - & 26.1 \\
South Africa & 25328 & 102 & 18.0 & - & 15.1 \\
Sudan & 7913 & 58 & 18.6 & - & 19.1 \\
Tanzania & 4709 & 165 & - & 44.9 & 46.2 \\
Timor-Leste & 4477 & 208 & 40.8 & - & 40.8 \\
Togo & 6171 & 160 & 28.4 & - & 28.4 \\
Uganda & 3074 & 99 & - & 42.1 & 39.5 \\
Yemen & 9376 & 252 & 19.8 & - & 19.8 \\
Zimbabwe & 30155 & 222 & 34.2 & - & 34.3 \\
\bottomrule
\end{tabular}

\end{center}

\vspace{.5em}

{\footnotesize This table describes the countries and surveys used in the empirical analysis. The column $n$ is the number of households in the survey data. The column $d$ is the number of covariates that we use from the survey for learning transfer policies. The ``Poverty Rate'' columns report the headcount poverty rates we estimate (``Survey'') and those reported by the World Bank, with the latter differentiated into two cases: those in which the World Bank estimate is based on the same survey as ours, and those in which it is based on or extrapolated from other sources (\href{https://data.worldbank.org/topic/11}{World Bank's Poverty and Inequality Platform}, accessed 14 July 2025). }
\end{table}
\clearpage

\begin{table}
\caption{Cost accounting (Togo)}
\label{tab:togo_costs}
{
\centering
\begin{tabular}{lrr}
\toprule
 & \textbf{Survey} & \textbf{Satellite} \\
\midrule
\textbf{Total program cost} & \textbf{\$1,523} & \textbf{\$1,779} \\
\quad Cost of transfers & \$1,505 & \$1,777 \\
\quad \textbf{Administrative costs} & \textbf{\$18} & \textbf{\$3} \\
\qquad Screening & \$16 & \$0 \\
\qquad Training data & \$3 & \$3 \\
\bottomrule
\end{tabular} \\
}
\vspace{1em}
{\footnotesize This table lists estimated costs, in millions of nominal 2023 USD, of two hypothetical year-long programs in Togo. One is targeted using survey data, and the other using satellite imagery.}

\end{table}


\begin{sidewaystable}
{
\centering 
\caption{Comparison of Policy Costs for In-Sample Countries to Country GDP and Government Revenue.}
\label{tab:costs}
\small
\begin{tabular}{lrrrrrrr}
\toprule
Country & Reference Year & Policy Cost & GDP & Gov't Revenue & Policy Cost / GDP & Policy Cost / Gov't Revenue \\
\midrule
Bangladesh & 2022 & 4.66 $\pm$ 0.52 & 479.07 & 42.72 & 0.01 & 0.11 \\
Benin & 2018 & 1.61 $\pm$ 0.17 & 17.31 & 2.35 & 0.09 & 0.68 \\
Burkina Faso & 2018 & 2.84 $\pm$ 0.15 & 19.28 & 3.81 & 0.15 & 0.74 \\
Burundi & 2020 & 2.78 $\pm$ 0.07 & 3.12 & 0.72 & 0.89 & 3.85 \\
Central African Republic & 2021 & 2.08 $\pm$ 0.07 & 2.83 & 0.39 & 0.73 & 5.38 \\
Colombia & 2016 & 1.53 $\pm$ 0.13 & 358.93 & 99.49 & 0.00 & 0.02 \\
Congo - Kinshasa & 2020 & 31.59 $\pm$ 0.33 & 57.36 & 5.37 & 0.55 & 5.88 \\
Côte d’Ivoire & 2018 & 1.87 $\pm$ 0.26 & 71.01 & 10.41 & 0.03 & 0.18 \\
Ethiopia & 2021 & 25.57 $\pm$ 1.78 & 125.11 & 13.80 & 0.20 & 1.85 \\
Ghana & 2016 & 4.89 $\pm$ 0.25 & 71.28 & 9.37 & 0.07 & 0.52 \\
Guinea-Bissau & 2018 & 0.21 $\pm$ 0.03 & 1.89 & 0.28 & 0.11 & 0.77 \\
India & 2022 & 15.80 $\pm$ 1.98 & 3483.84 & 701.39 & 0.00 & 0.02 \\
Indonesia & 2018 & 7.93 $\pm$ 0.24 & 1264.73 & 188.71 & 0.01 & 0.04 \\
Kenya & 2021 & 9.66 $\pm$ 0.46 & 123.36 & 20.75 & 0.08 & 0.47 \\
Liberia & 2014 & 0.98 $\pm$ 0.09 & 4.15 & 1.18 & 0.24 & 0.83 \\
Madagascar & 2021 & 4.94 $\pm$ 0.14 & 16.14 & 1.79 & 0.31 & 2.76 \\
Malawi & 2019 & 4.27 $\pm$ 0.13 & 13.17 & 1.94 & 0.32 & 2.19 \\
Mali & 2018 & 1.76 $\pm$ 0.22 & 24.72 & 3.85 & 0.07 & 0.46 \\
Mexico & 2024 & 0.22 $\pm$ 0.21 & 1799.64 & 443.51 & 0.00 & 0.00 \\
Namibia & 2015 & 0.44 $\pm$ 0.03 & 14.57 & 5.15 & 0.03 & 0.09 \\
Niger & 2018 & 4.72 $\pm$ 0.21 & 15.58 & 2.83 & 0.30 & 1.67 \\
Nigeria & 2018 & 29.50 $\pm$ 1.48 & 511.75 & 43.52 & 0.06 & 0.68 \\
Pakistan & 2018 & 5.63 $\pm$ 0.60 & 432.14 & 58.06 & 0.01 & 0.10 \\
Rwanda & 2023 & 1.68 $\pm$ 0.06 & 14.33 & 3.15 & 0.12 & 0.53 \\
Senegal & 2018 & 1.11 $\pm$ 0.19 & 28.05 & 5.30 & 0.04 & 0.21 \\
Sierra Leone & 2018 & 0.75 $\pm$ 0.05 & 7.75 & 0.78 & 0.10 & 0.95 \\
South Africa & 2010 & 9.13 $\pm$ 0.52 & 583.21 & 138.96 & 0.02 & 0.07 \\
Sudan & 2009 & 3.40 $\pm$ 0.25 & 73.32 & 10.97 & 0.05 & 0.31 \\
Tanzania & 2020 & 13.98 $\pm$ 1.30 & 77.78 & 11.58 & 0.18 & 1.21 \\
Timor-Leste & 2007 & 0.14 $\pm$ 0.01 & 0.80 & 0.37 & 0.18 & 0.39 \\
Togo & 2018 & 1.50 $\pm$ 0.10 & 8.53 & 1.55 & 0.18 & 0.97 \\
Uganda & 2019 & 6.69 $\pm$ 0.46 & 42.14 & 5.67 & 0.16 & 1.18 \\
Yemen & 2014 & 4.51 $\pm$ 0.32 & 55.64 & 13.16 & 0.08 & 0.34 \\
Zimbabwe & 2017 & 2.56 $\pm$ 0.08 & 63.49 & 11.11 & 0.04 & 0.23 \\
\bottomrule
\end{tabular}

}
\vspace{1em}
\footnotesize{This table reports the cost of the gap-minimizing policy that reduces the poverty rate to \nationalTarget\% in each country, and compares it to national GDP and government revenue. ``Reference Year" corresponds to the year that the Policy Cost, GDP, and government revenue are reported and is taken to be the survey year. ``GDP'' corresponds to the country GDP in the survey year in units of billions of 2023 nominal USD. ``Policy Cost'' is the cost of the policy, in billions of 2023 nominal USD, to attain a \nationalTarget\% poverty rate in the country. We report 95\% confidence intervals for the country-specific policy cost under a fixed transfer rule. For each country, we generate 50 bootstrap samples of the evaluation set, computing the cost to attain a 1\% poverty rate on each bootstrap sample, and compute the standard error across the bootstrapped policy costs.``Gov't Revenue" corresponds to country government revenue in the survey year in units of billions of 2023 nominal USD. We obtain the survey year GDP from the World Bank (\href{https://data.worldbank.org}{data source}, accessed 13 August 25). We obtain the survey year Gov't Revenue by multiplying survey year government revenue percentages from the IMF(\href{https://www.imf.org}{data source}, accessed 14 July 25) by the country GDP in the preceding column.``Policy Cost / GDP'' is the ratio of the policy cost and GDP. ``Policy Cost / Gov't Revenue" is the ratio of policy cost and government revenue.}

\end{sidewaystable}
\clearpage




\begin{figure}
\caption{Cost Ratio of Feasible Gap-Minimizing Policy to Aggregate Poverty Gap}
\includegraphics[width=\textwidth]{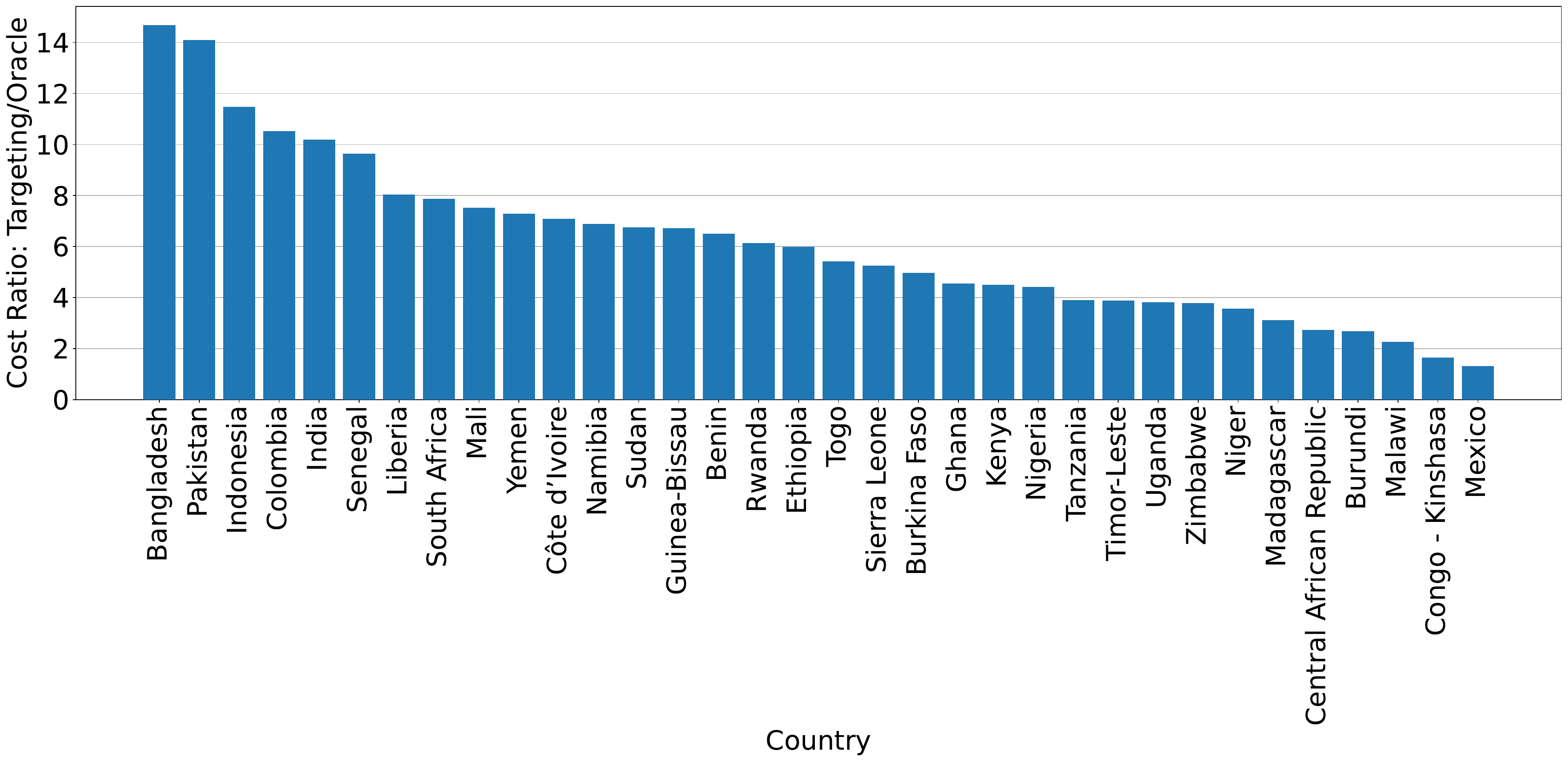}
\footnotesize{This figure reports, for each country in Table \ref{tab:surveys}, the cost ratio between the gap-minimizing policy that achieves a \nationalTarget\% post-transfer poverty rate and the aggregate poverty gap.}
\label{fig:oracle_feasible_ratio}
\end{figure}

\renewcommand{\thesubfigure}{\arabic{subfigure}}

\begin{figure}
    \centering
    \caption{Policy Cost vs. Post-Transfer Poverty Measures by Country}
    \label{fig:country1}
    \begin{subfigure}{\textwidth}
    \subcaption{Bangladesh}
    \includegraphics[width=\textwidth]{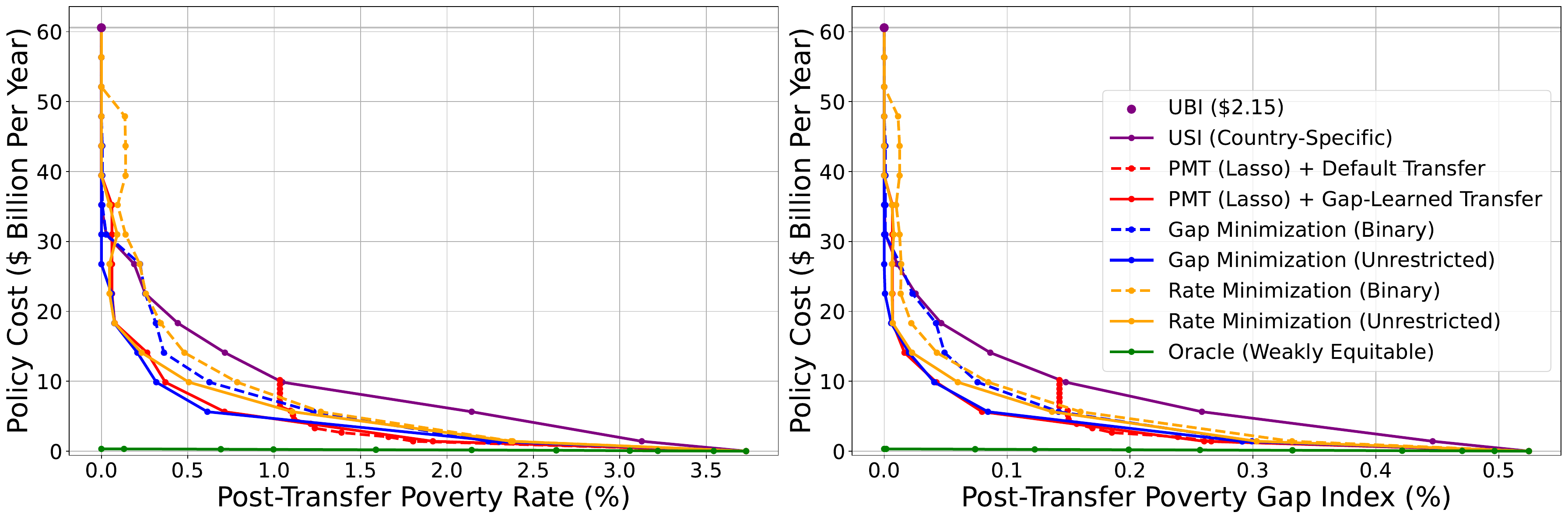}
    \end{subfigure}
    \bigskip
    \begin{subfigure}{\textwidth}
    \subcaption{Benin}
    \includegraphics[width=\textwidth]{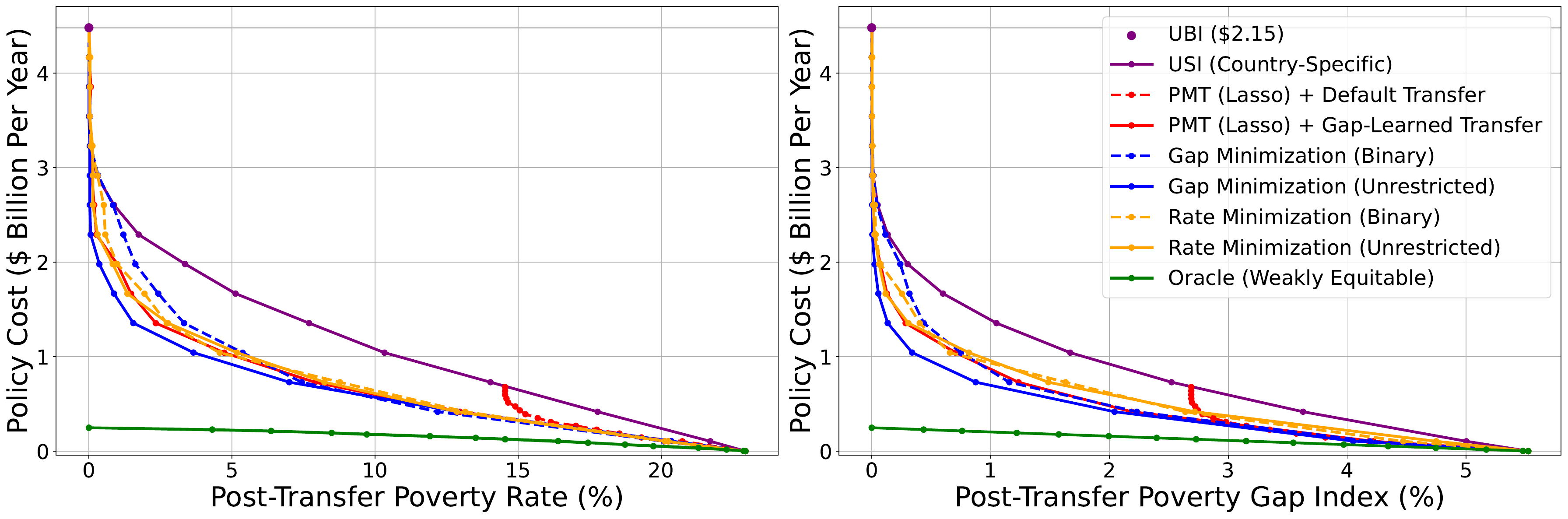}
    \end{subfigure}
    \bigskip
    \begin{subfigure}{\textwidth}
    \subcaption{Burkina Faso}
    \includegraphics[width=\textwidth]{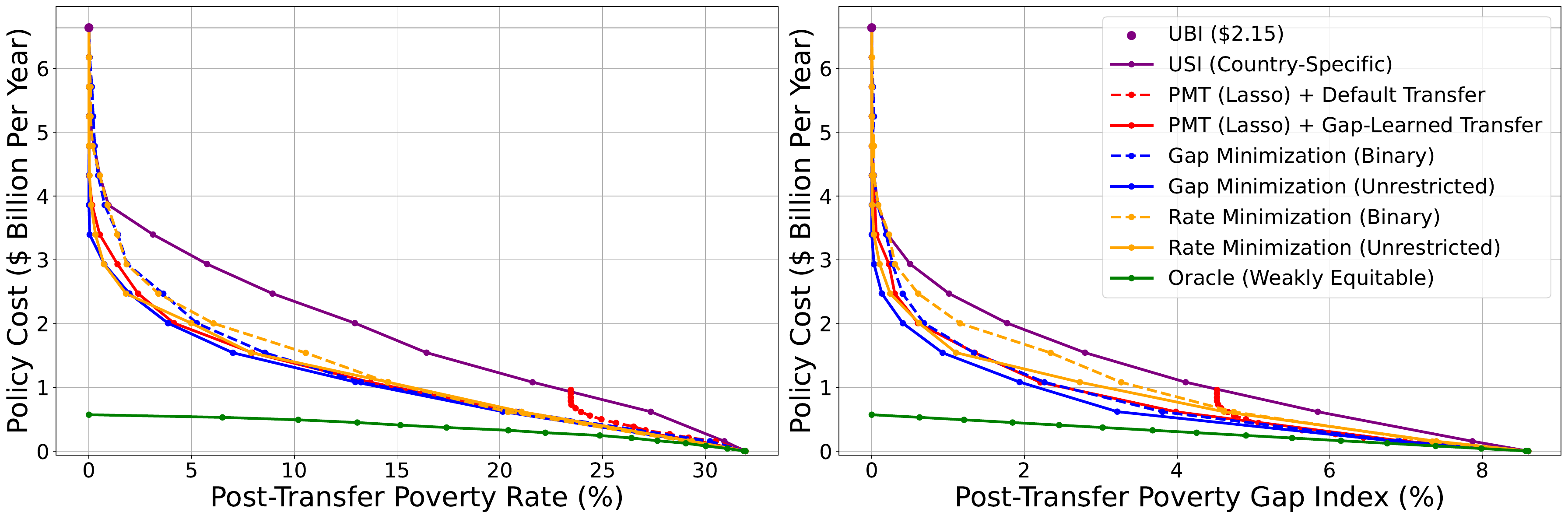}
    \end{subfigure}
\end{figure}

\begin{figure}\ContinuedFloat
    \begin{subfigure}{\textwidth}
    \subcaption{Burundi}
    \includegraphics[width=\textwidth]{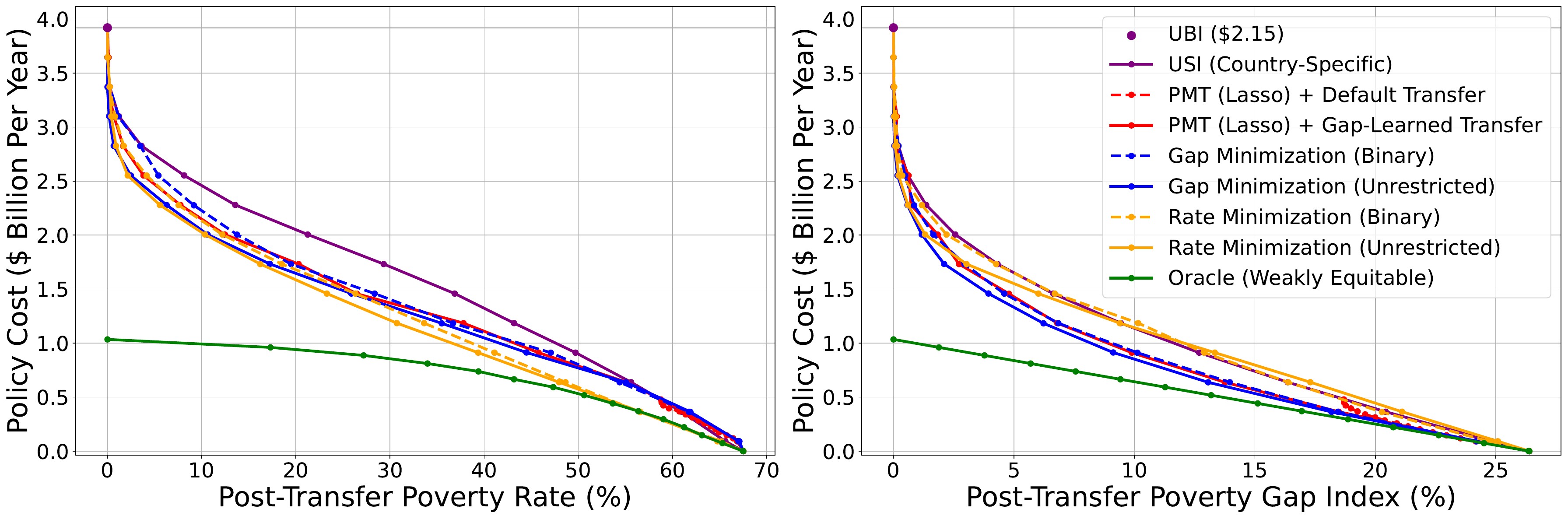}
    \end{subfigure}
    \bigskip
    \begin{subfigure}{\textwidth}
    \subcaption{Central African Republic}
    \includegraphics[width=\textwidth]{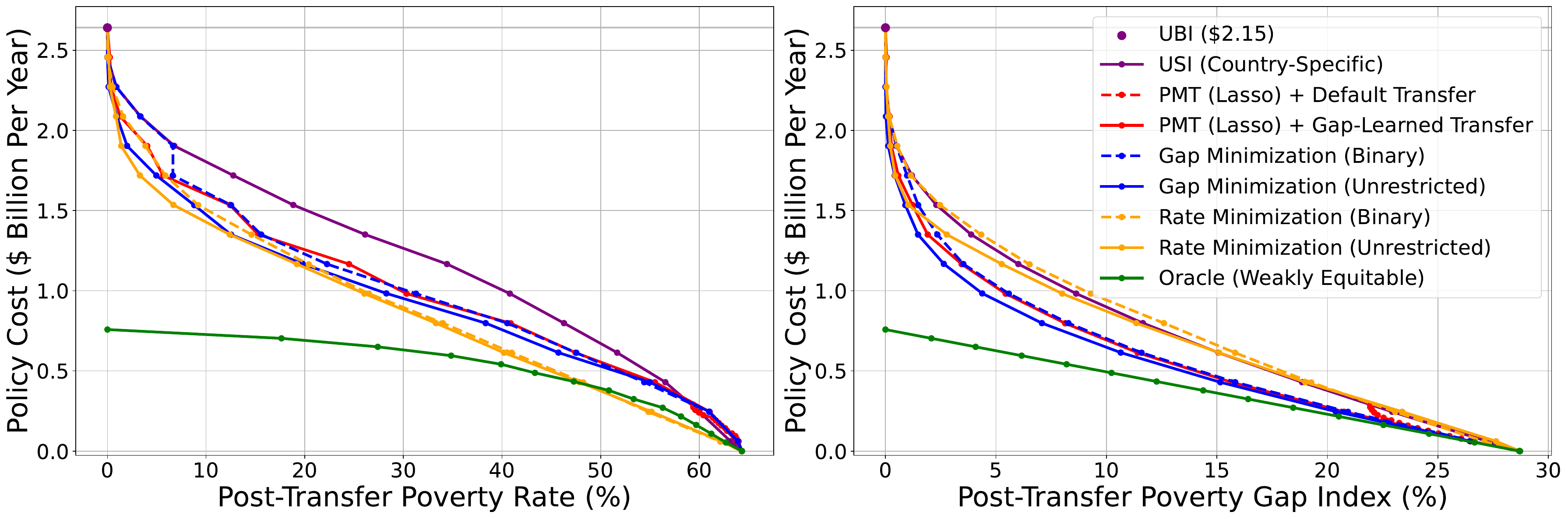}
    \end{subfigure}
    \bigskip
    \begin{subfigure}{\textwidth}
    \subcaption{Colombia}
    \includegraphics[width=\textwidth]{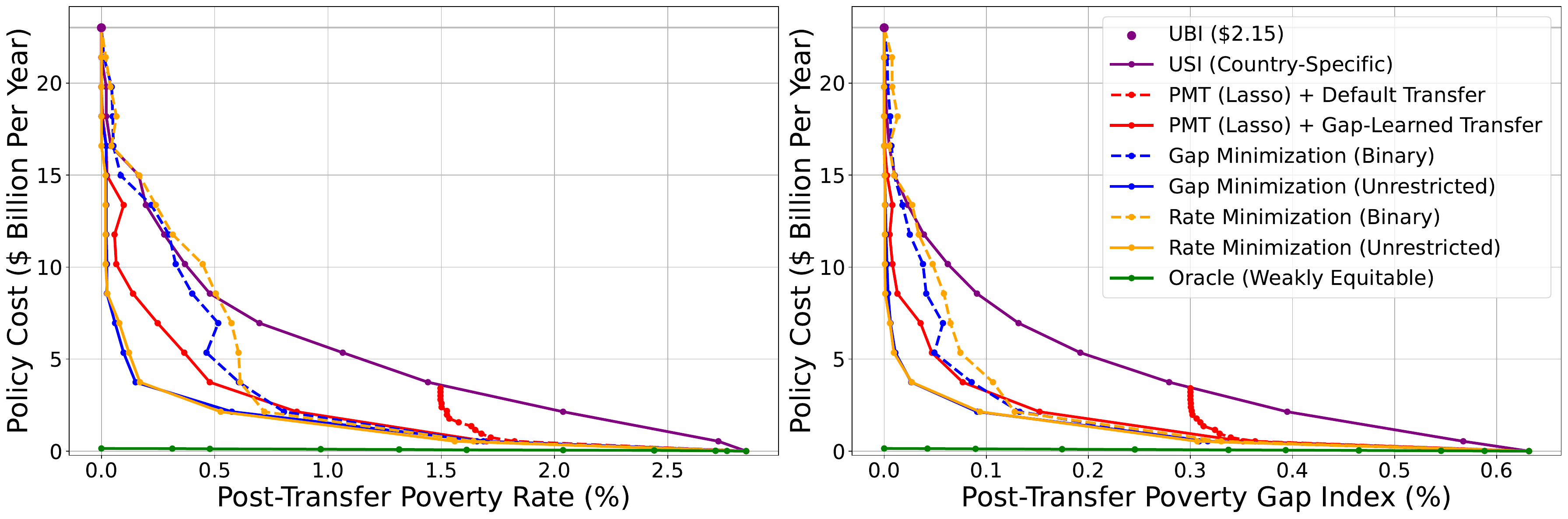}
    \end{subfigure}
\end{figure}

\begin{figure}\ContinuedFloat
    \begin{subfigure}{\textwidth}
    \subcaption{C\^{o}te d'Ivoire}
    \includegraphics[width=\textwidth]{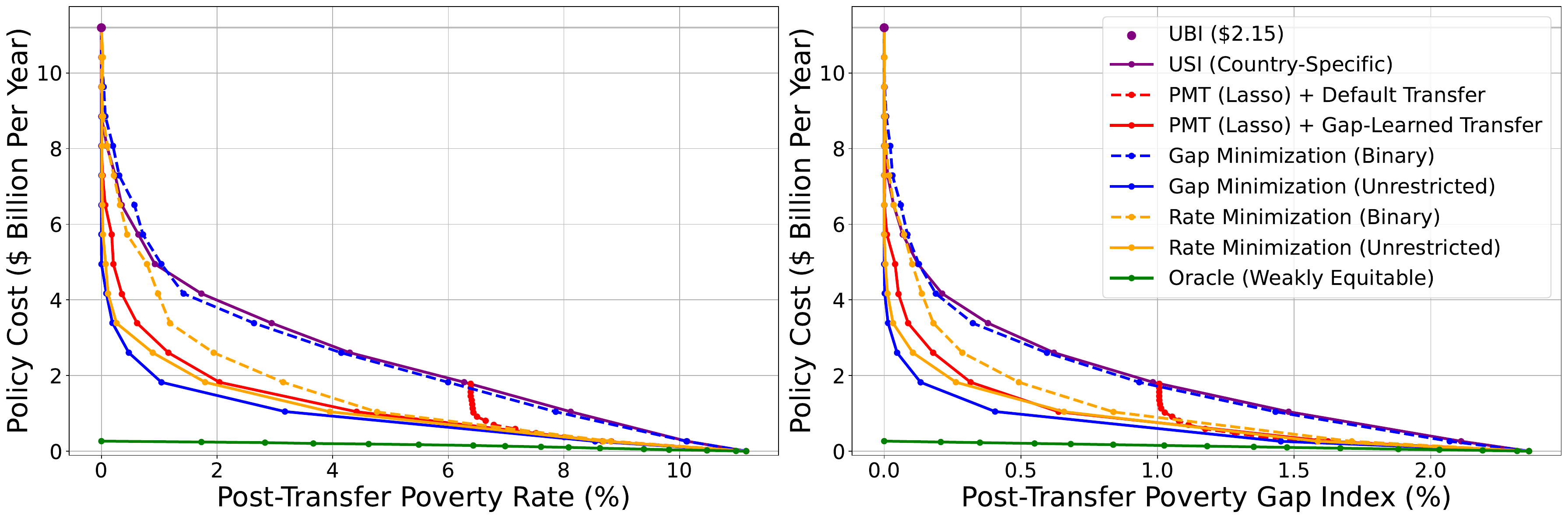}
    \end{subfigure}
    \bigskip
    \begin{subfigure}{\textwidth}
    \subcaption{Democratic Republic of Congo}
    \includegraphics[width=\textwidth]{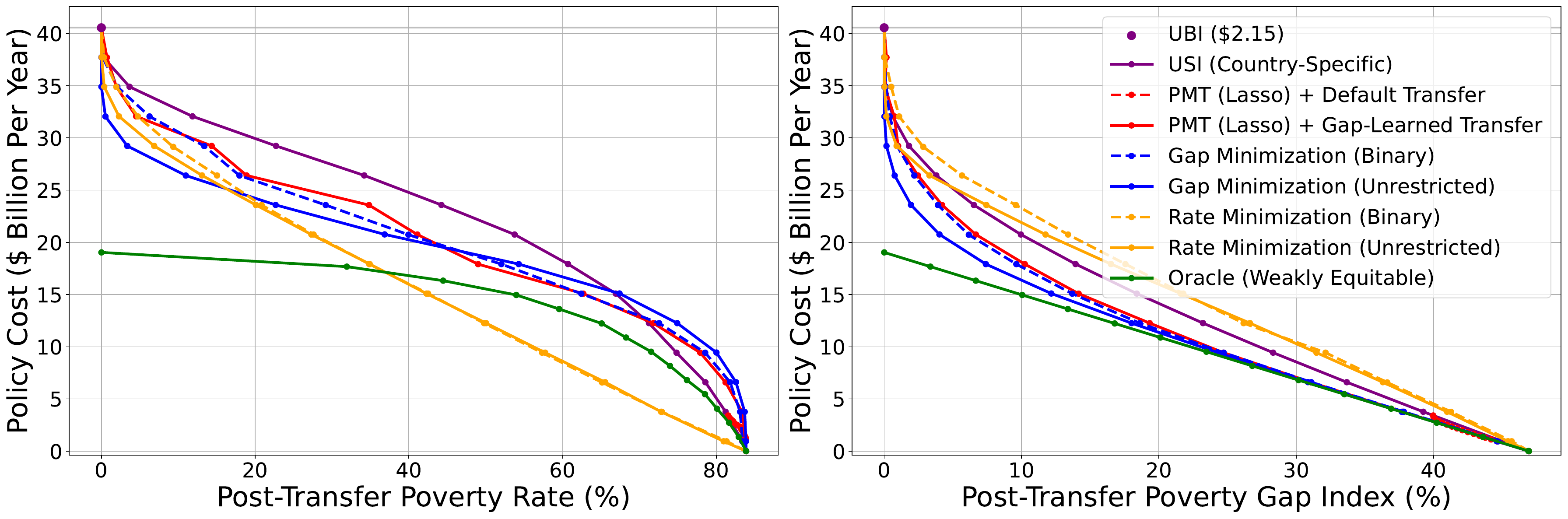}
    \end{subfigure}
    \bigskip
    \begin{subfigure}{\textwidth}
    \subcaption{Ethiopia}
    \includegraphics[width=\textwidth]{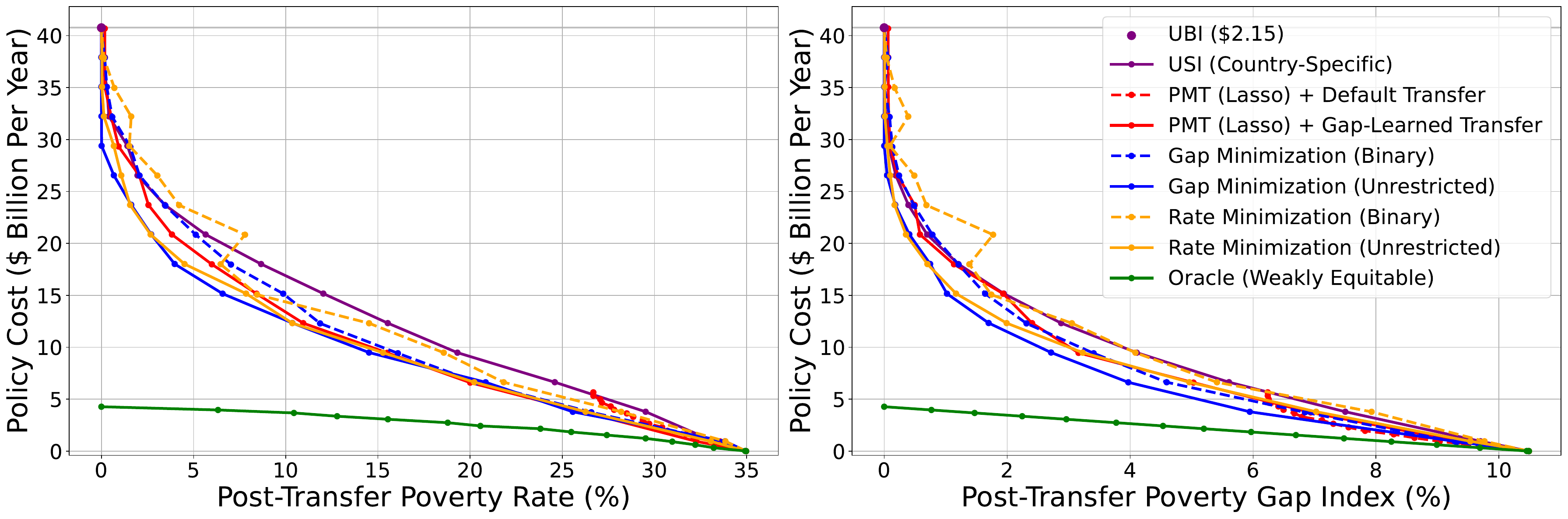}
    \end{subfigure}
\end{figure}

\begin{figure}\ContinuedFloat
    \begin{subfigure}{\textwidth}
    \subcaption{Ghana}
    \includegraphics[width=\textwidth]{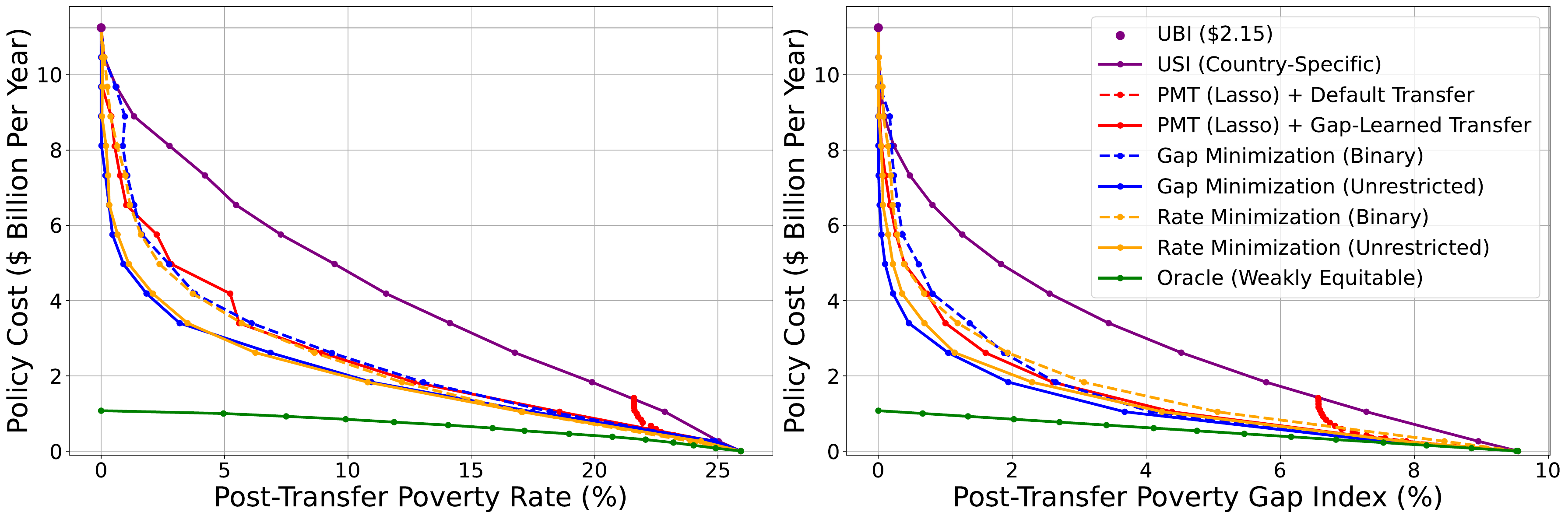}
    \end{subfigure}
    \bigskip
    \begin{subfigure}{\textwidth}
    \subcaption{Guinea-Bissau}
    \includegraphics[width=\textwidth]{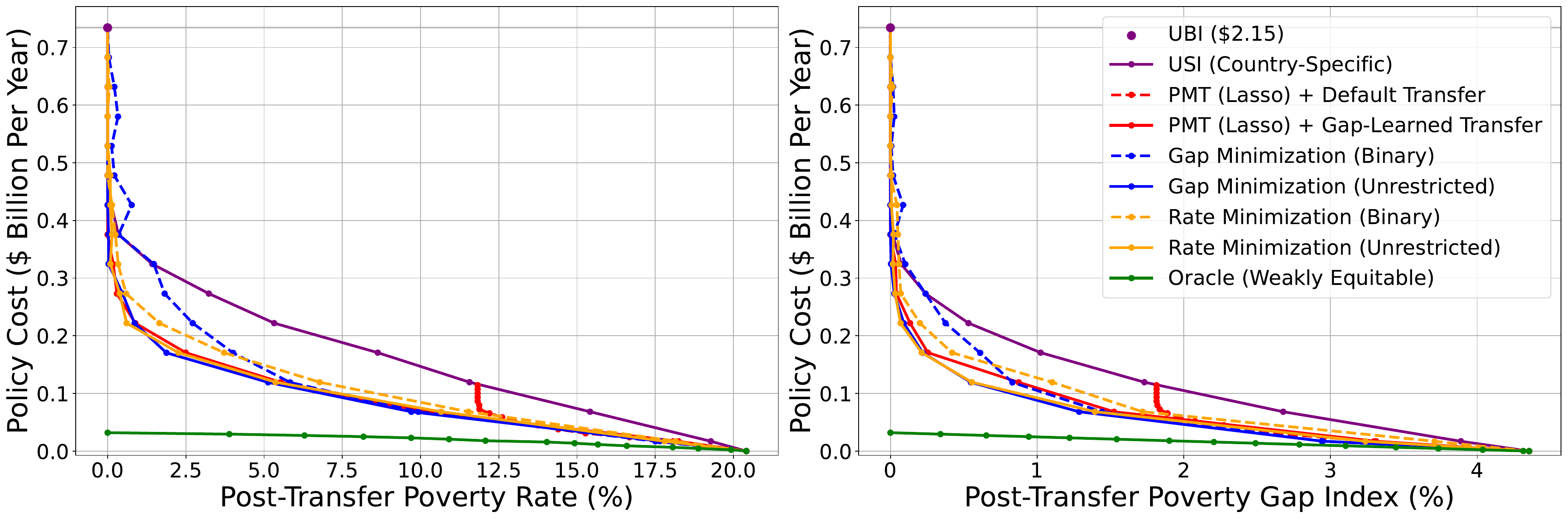}
    \end{subfigure}
    \bigskip
    \begin{subfigure}{\textwidth}
    \subcaption{India}
    \includegraphics[width=\textwidth]{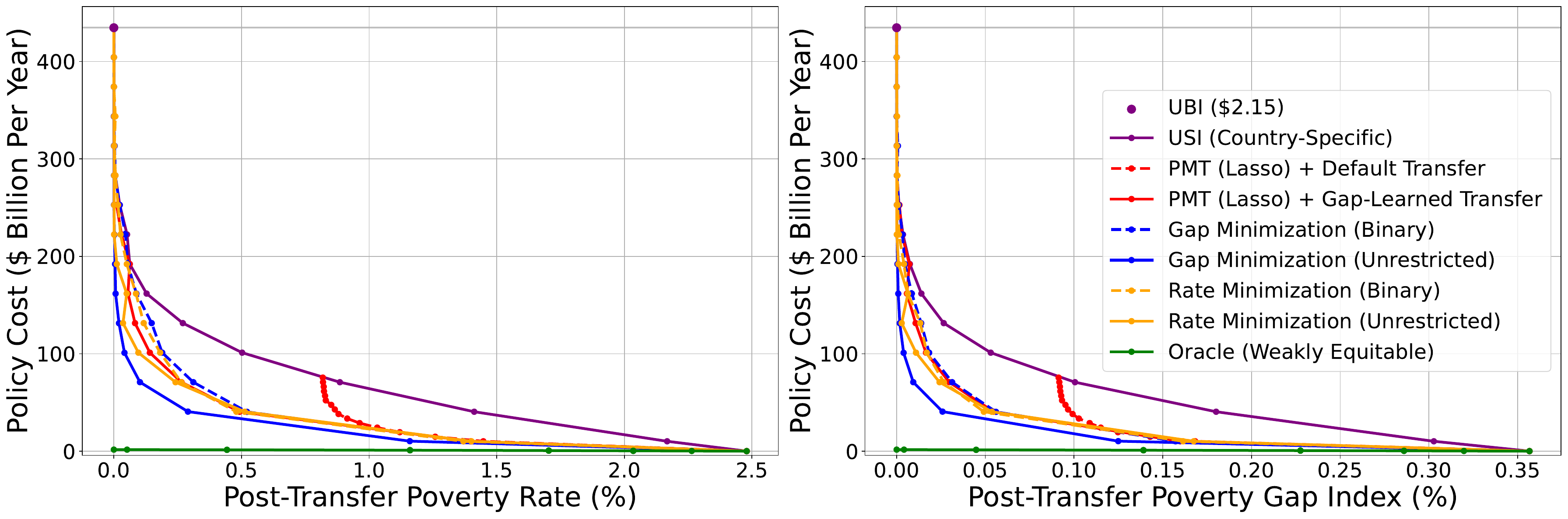}
    \end{subfigure}
\end{figure}

\begin{figure}\ContinuedFloat
    \begin{subfigure}{\textwidth}
    \subcaption{Indonesia}
    \includegraphics[width=\textwidth]{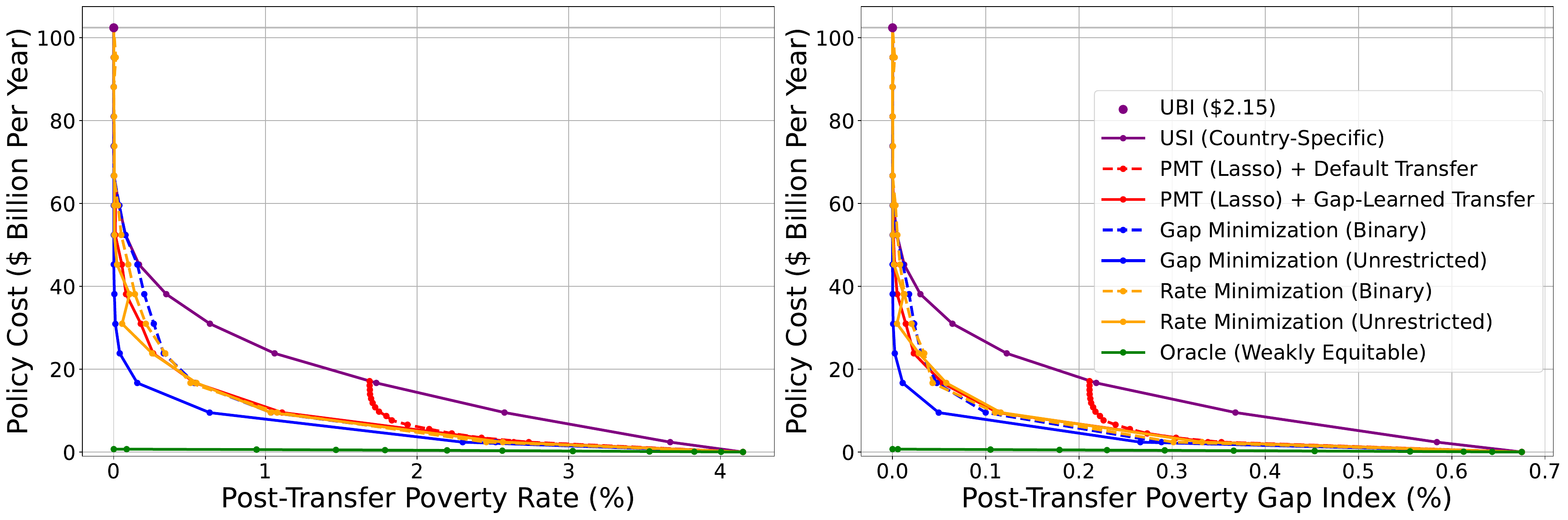}
    \end{subfigure}
    \bigskip
    \begin{subfigure}{\textwidth}
    \subcaption{Kenya}
    \includegraphics[width=\textwidth]{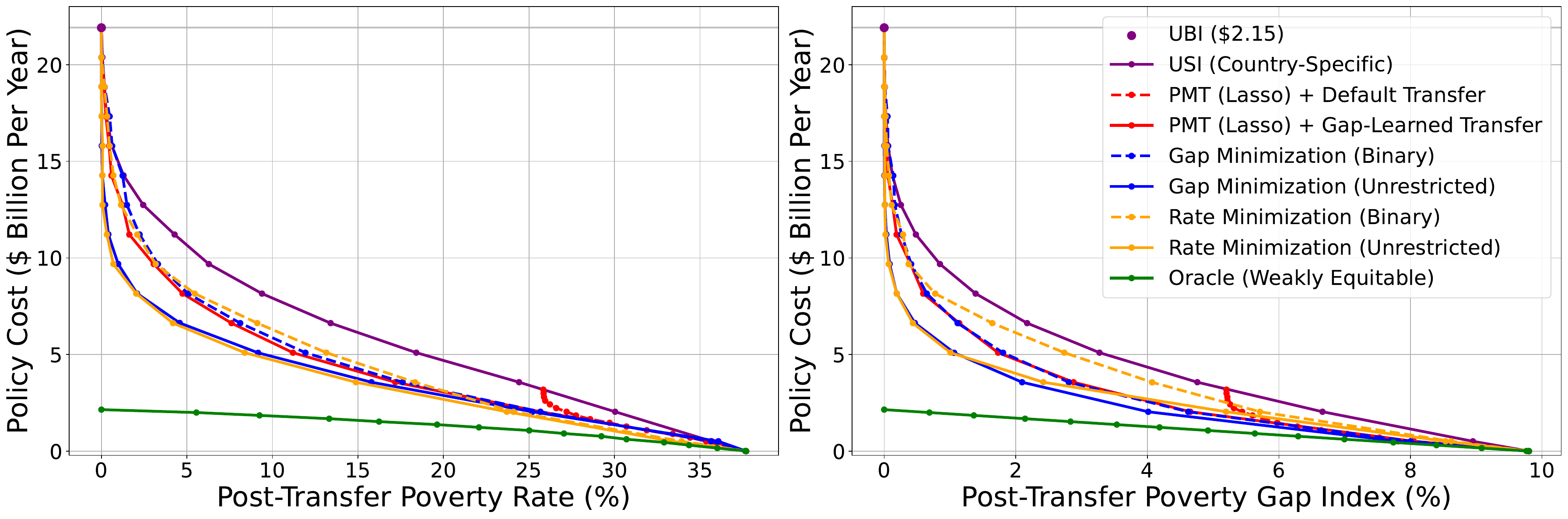}
    \end{subfigure}
    \bigskip
    \begin{subfigure}{\textwidth}
    \subcaption{Liberia}
    \includegraphics[width=\textwidth]{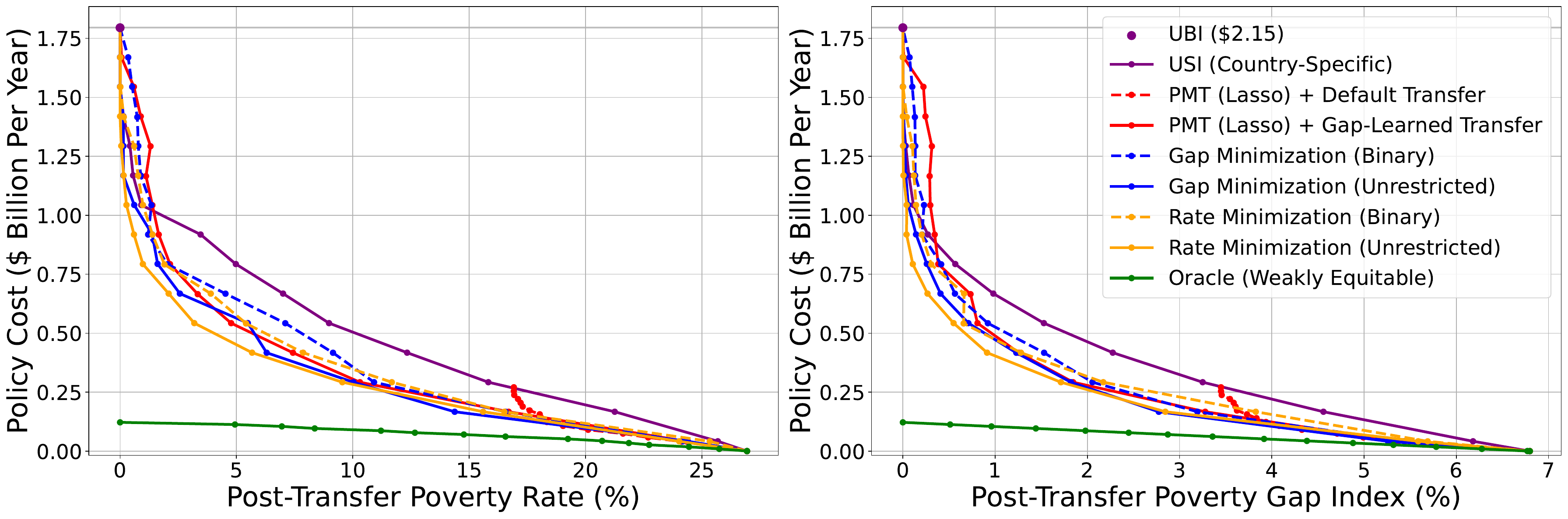}
    \end{subfigure}
\end{figure}

\begin{figure}\ContinuedFloat
    \begin{subfigure}{\textwidth}
    \subcaption{Madagascar}
    \includegraphics[width=\textwidth]{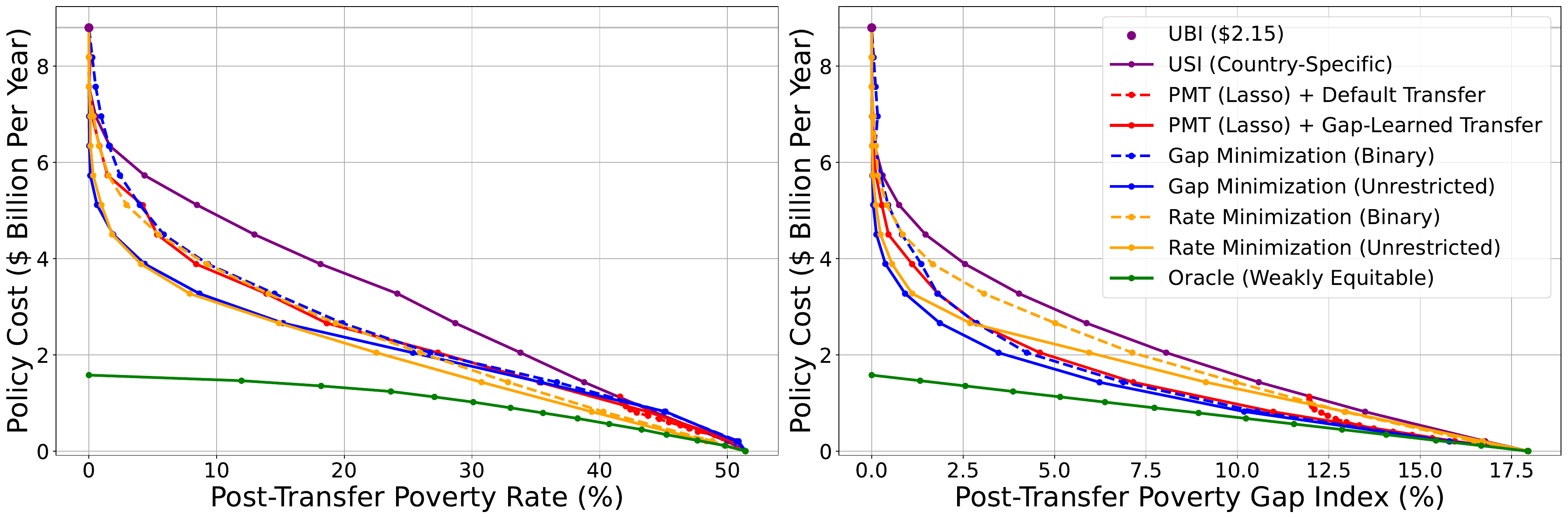}
    \end{subfigure}
    \bigskip
    \begin{subfigure}{\textwidth}
    \subcaption{Malawi}
    \includegraphics[width=\textwidth]{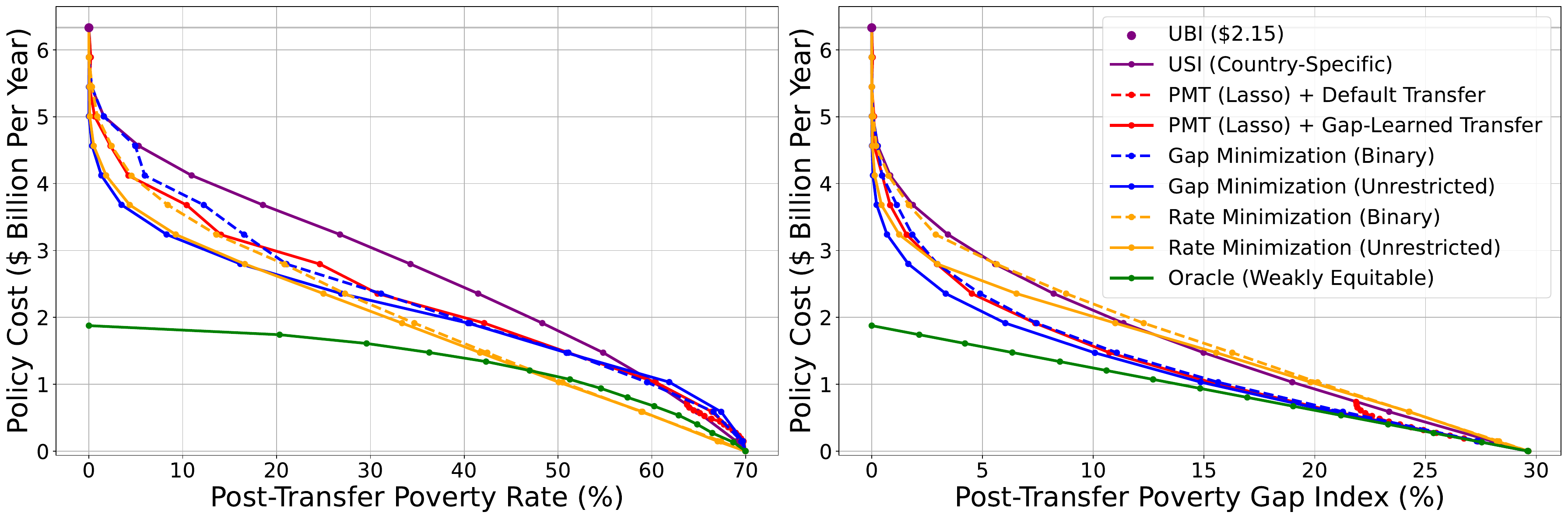}
    \end{subfigure}
    \bigskip
    \begin{subfigure}{\textwidth}
    \subcaption{Mali}
    \includegraphics[width=\textwidth]{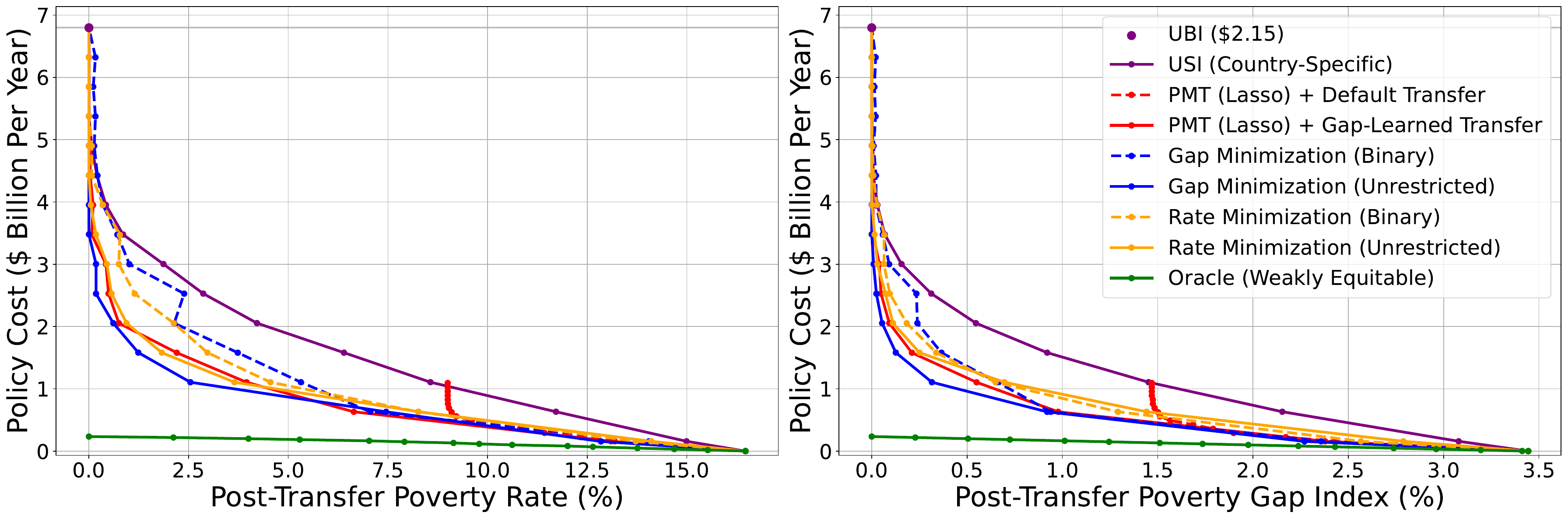}
    \end{subfigure}
\end{figure}

\begin{figure}\ContinuedFloat
    \begin{subfigure}{\textwidth}
    \subcaption{Mexico}
    \includegraphics[width=\textwidth]{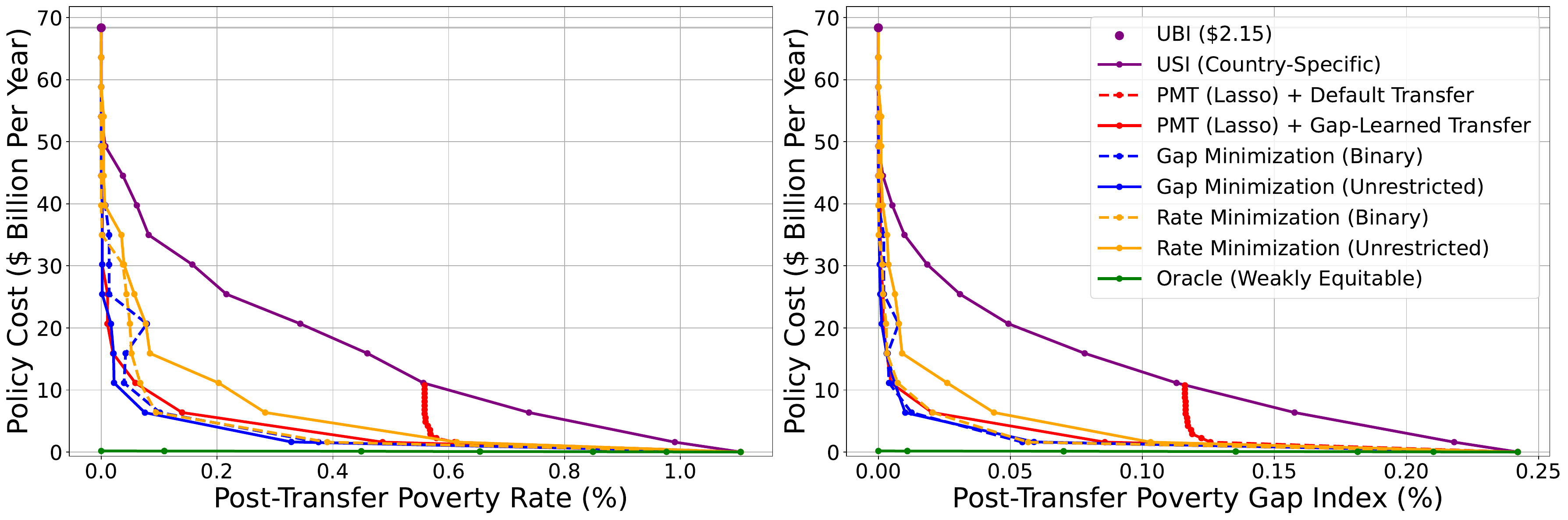}
    \end{subfigure}
    \bigskip
    \begin{subfigure}{\textwidth}
    \subcaption{Namibia}
    \includegraphics[width=\textwidth]{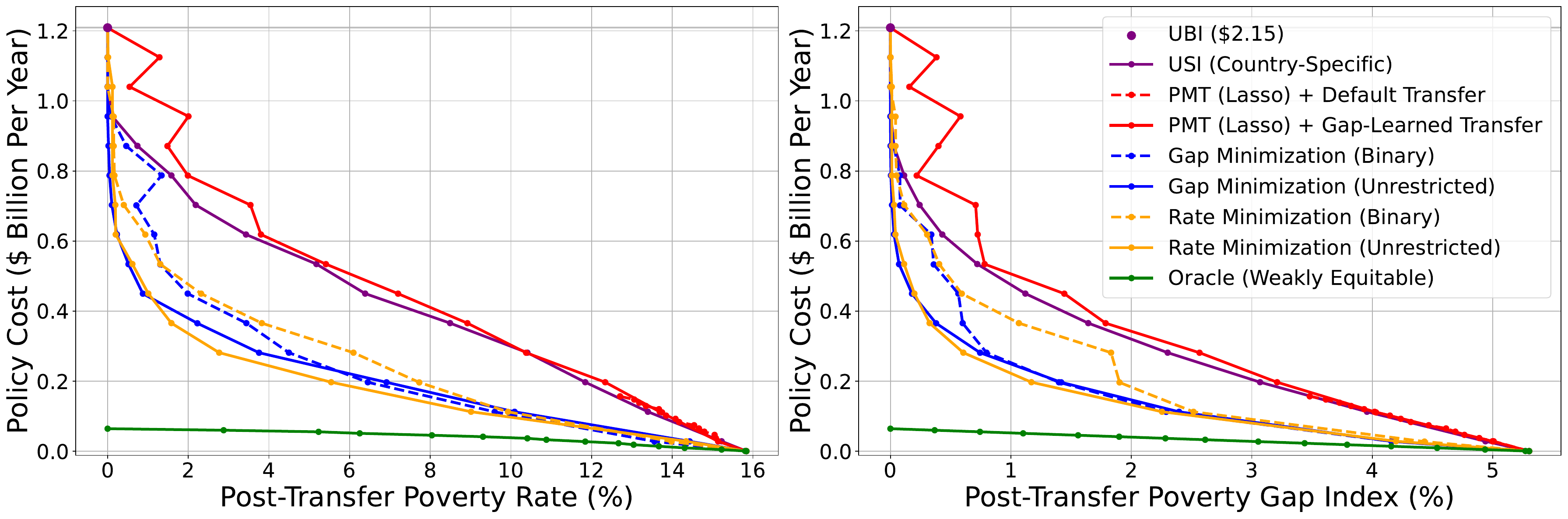}
    \end{subfigure}
    \bigskip
    \begin{subfigure}{\textwidth}
    \subcaption{Niger}
    \includegraphics[width=\textwidth]{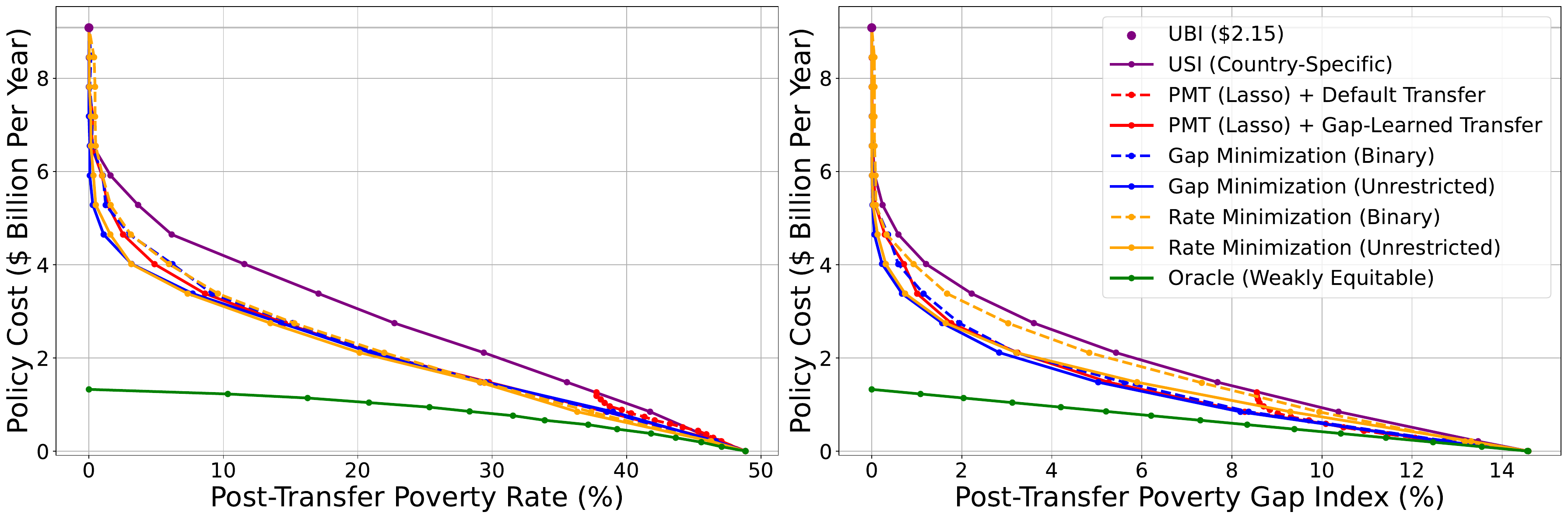}
    \end{subfigure}
\end{figure}

\begin{figure}\ContinuedFloat
    \begin{subfigure}{\textwidth}
    \subcaption{Nigeria}
    \includegraphics[width=\textwidth]{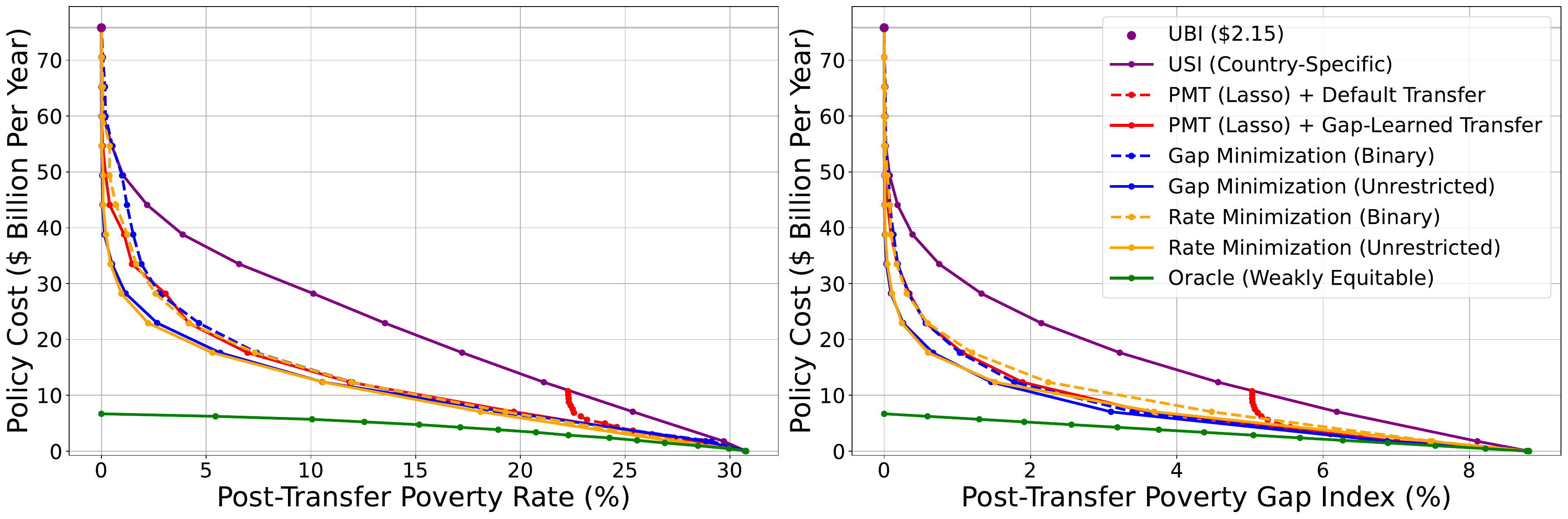}
    \end{subfigure}
    \bigskip
    \begin{subfigure}{\textwidth}
    \subcaption{Pakistan}
    \includegraphics[width=\textwidth]{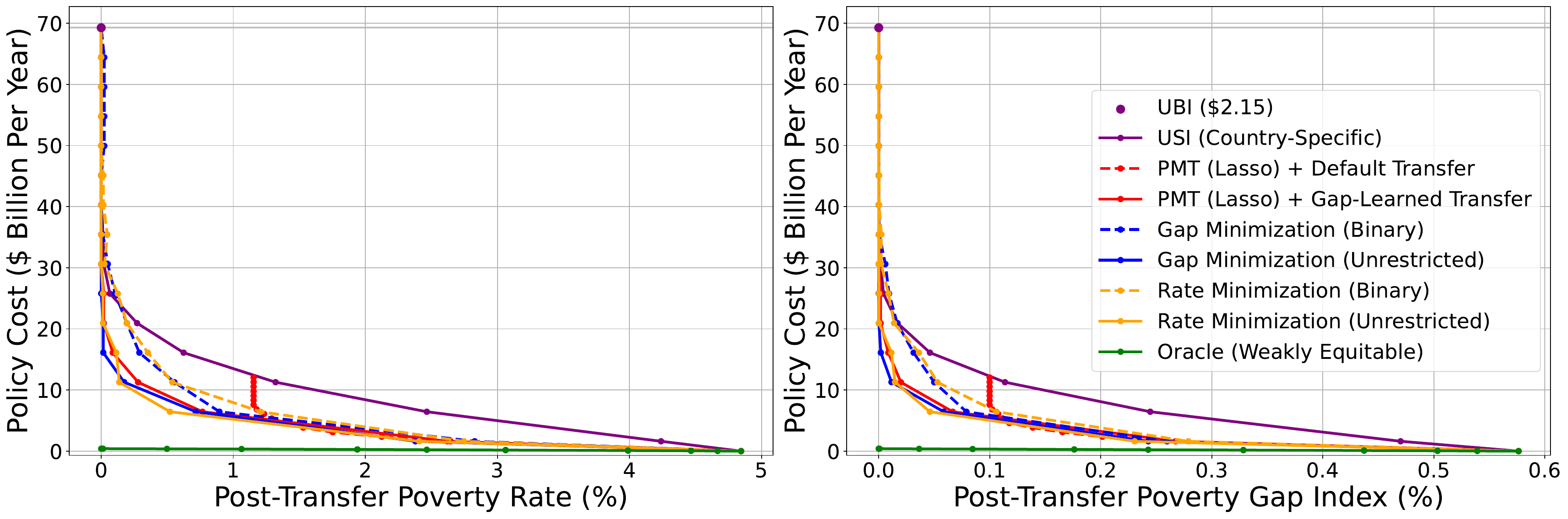}
    \end{subfigure}
    \bigskip
    \begin{subfigure}{\textwidth}
    \subcaption{Rwanda}
    \includegraphics[width=\textwidth]{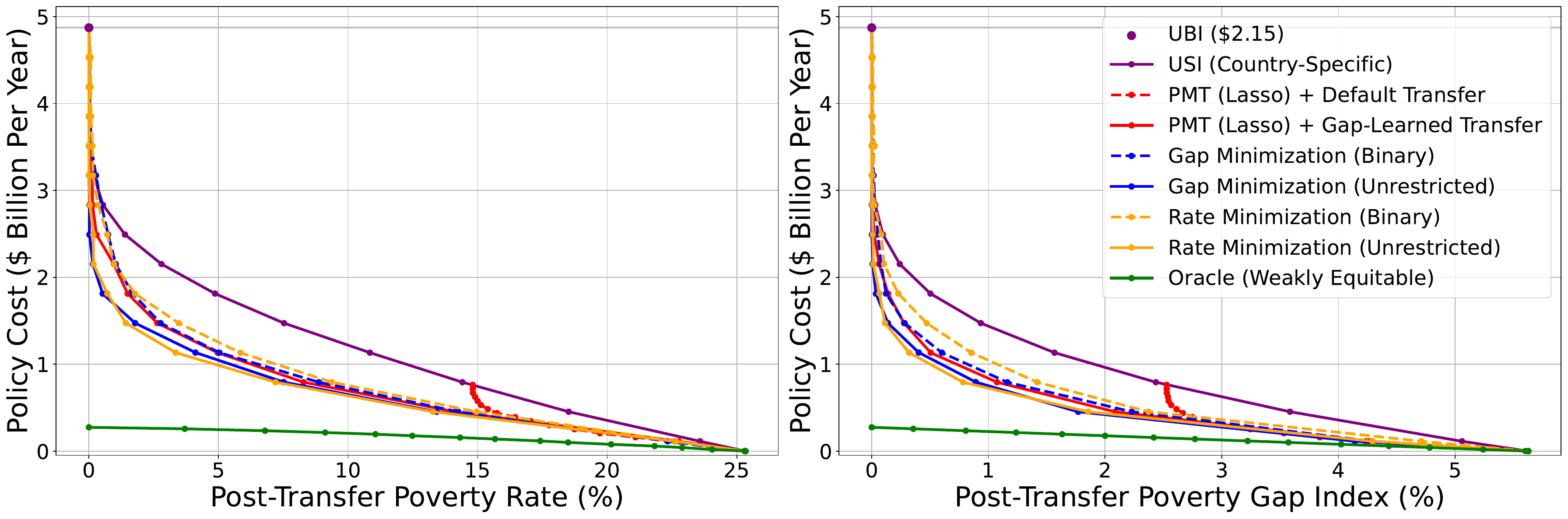}
    \end{subfigure}
\end{figure}

\begin{figure}\ContinuedFloat
    \begin{subfigure}{\textwidth}
    \subcaption{Senegal}
    \includegraphics[width=\textwidth]{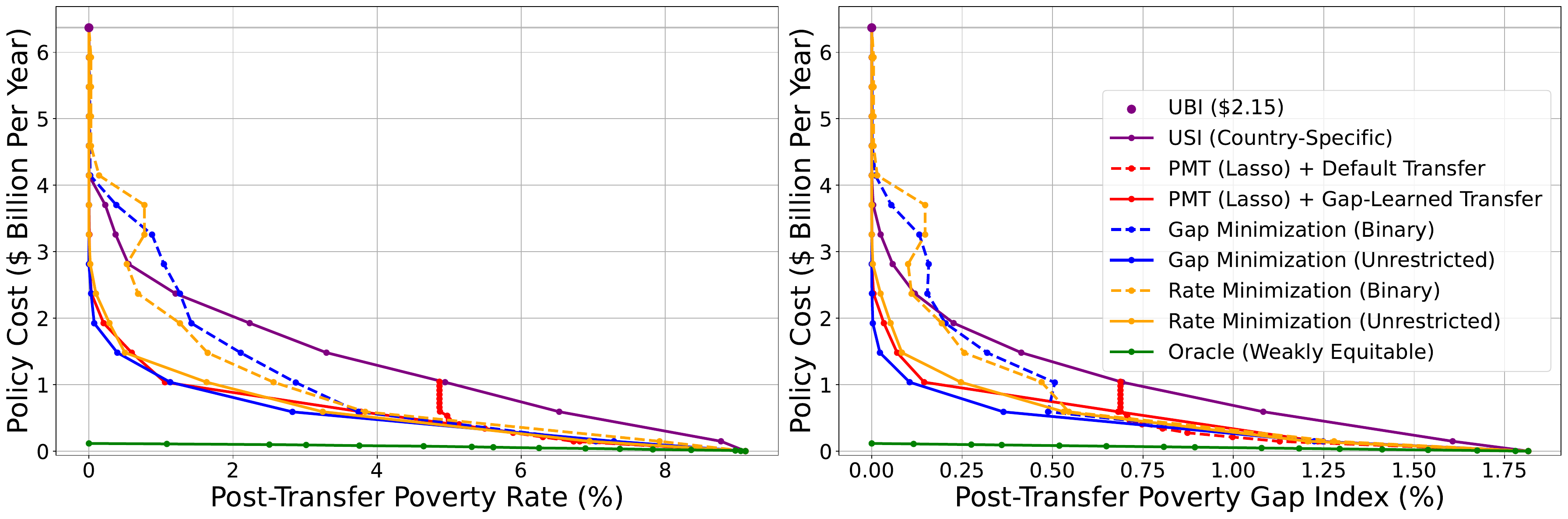}
    \end{subfigure}
    \bigskip
    \begin{subfigure}{\textwidth}
    \subcaption{Sierra Leone}
    \includegraphics[width=\textwidth]{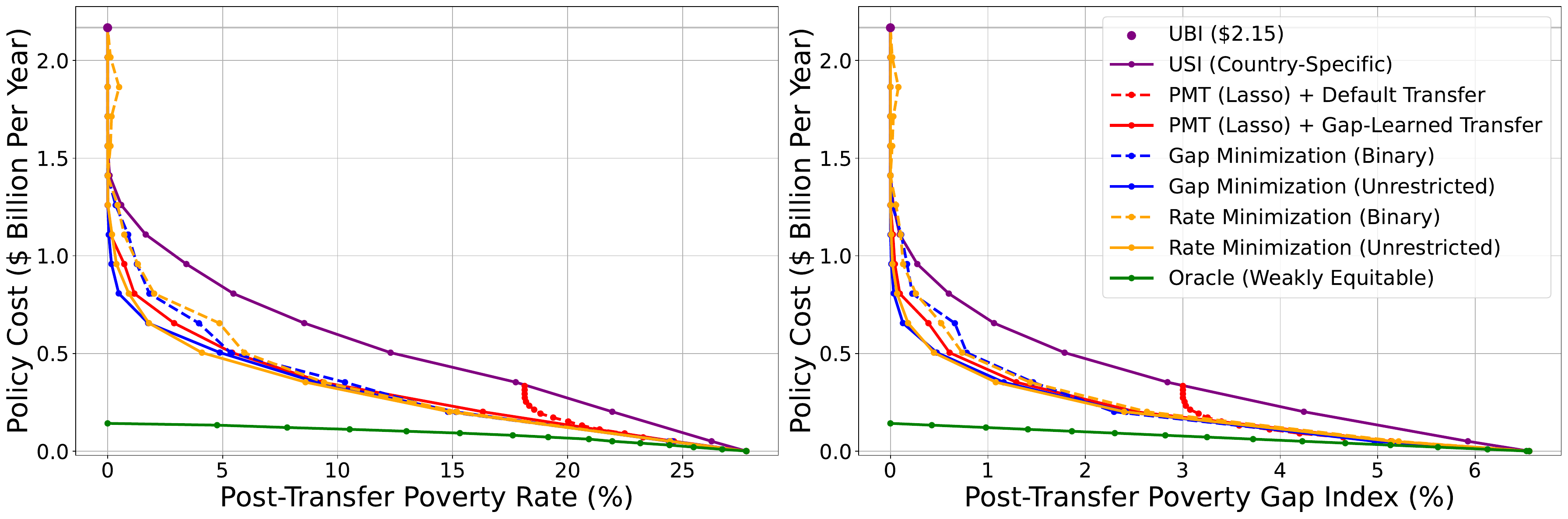}
    \end{subfigure}
    \bigskip
    \begin{subfigure}{\textwidth}
    \subcaption{South Africa}
    \includegraphics[width=\textwidth]{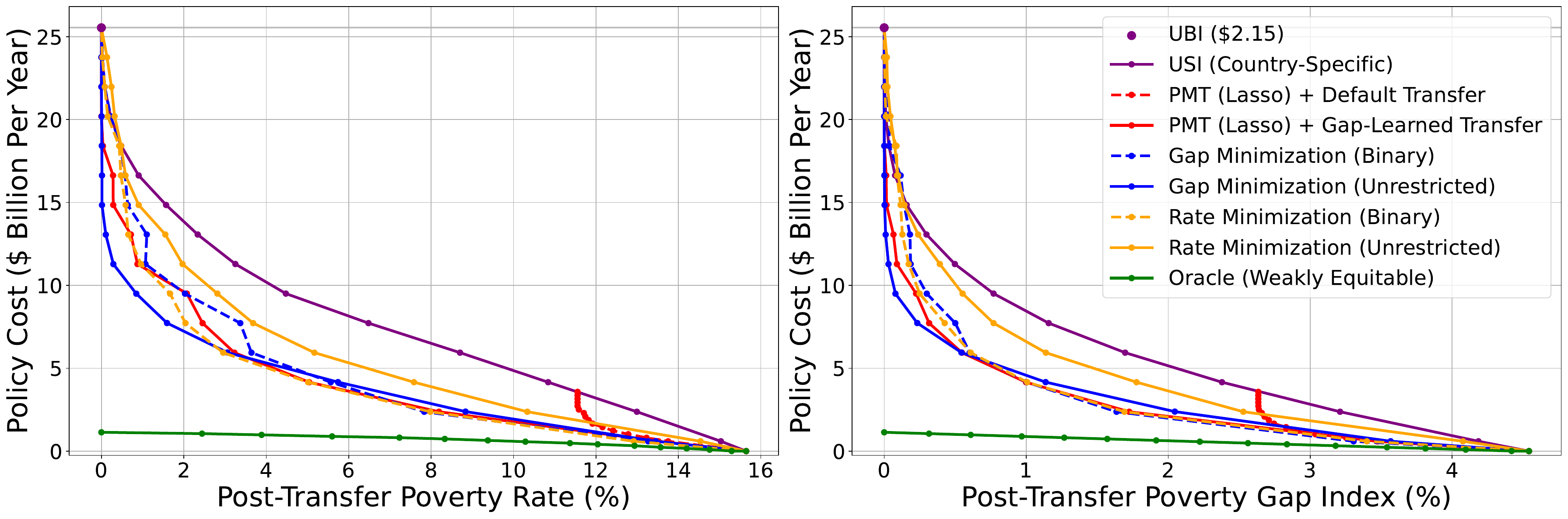}
    \end{subfigure}
\end{figure}

\begin{figure}\ContinuedFloat
    \begin{subfigure}{\textwidth}
    \subcaption{Sudan}
    \includegraphics[width=\textwidth]{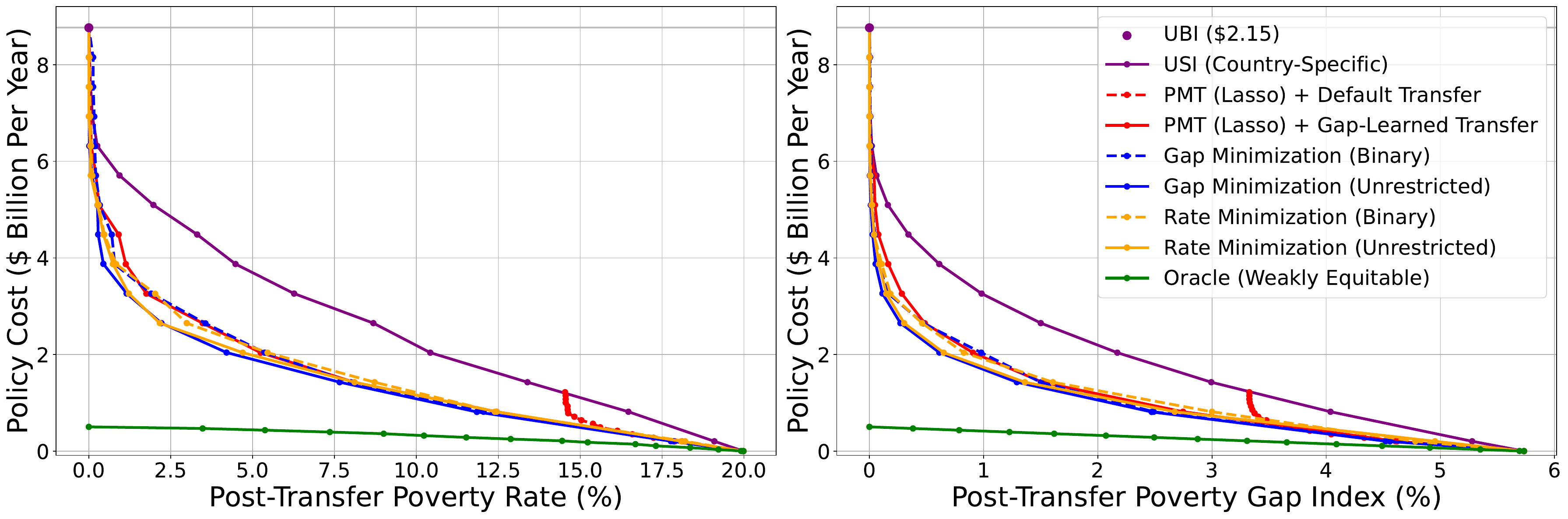}
    \end{subfigure}
    \bigskip
    \begin{subfigure}{\textwidth}
    \subcaption{Tanzania}
    \includegraphics[width=\textwidth]{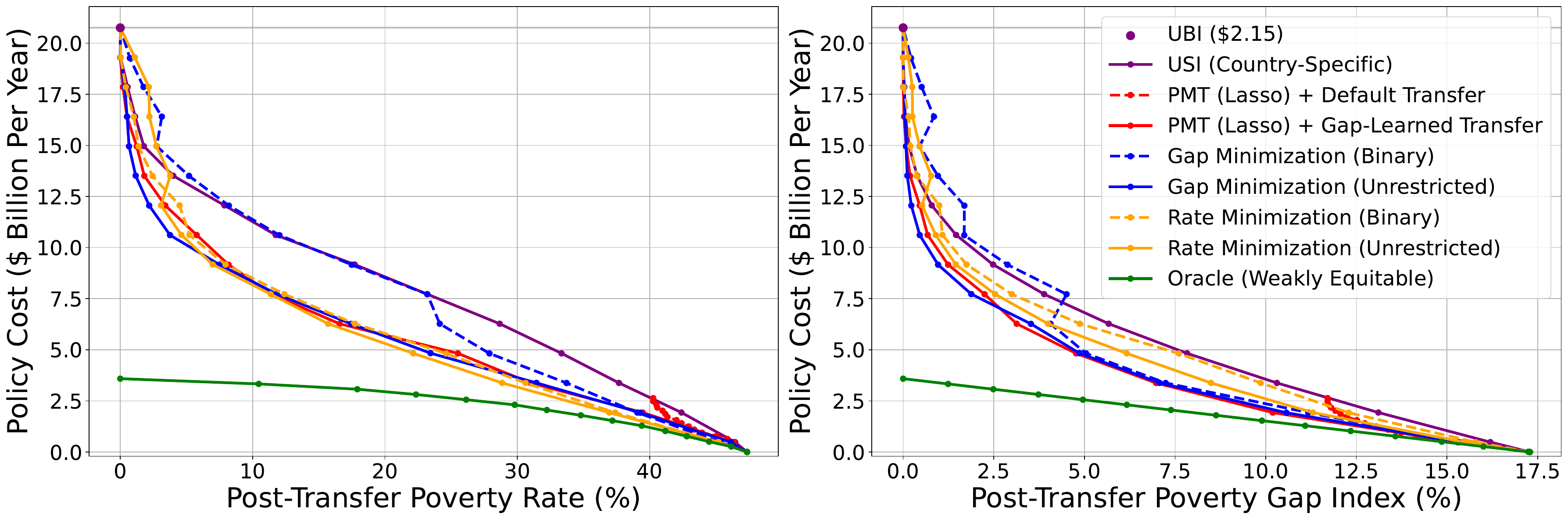}
    \end{subfigure}
    \bigskip
    \begin{subfigure}{\textwidth}
    \subcaption{Timor-Leste}
    \includegraphics[width=\textwidth]{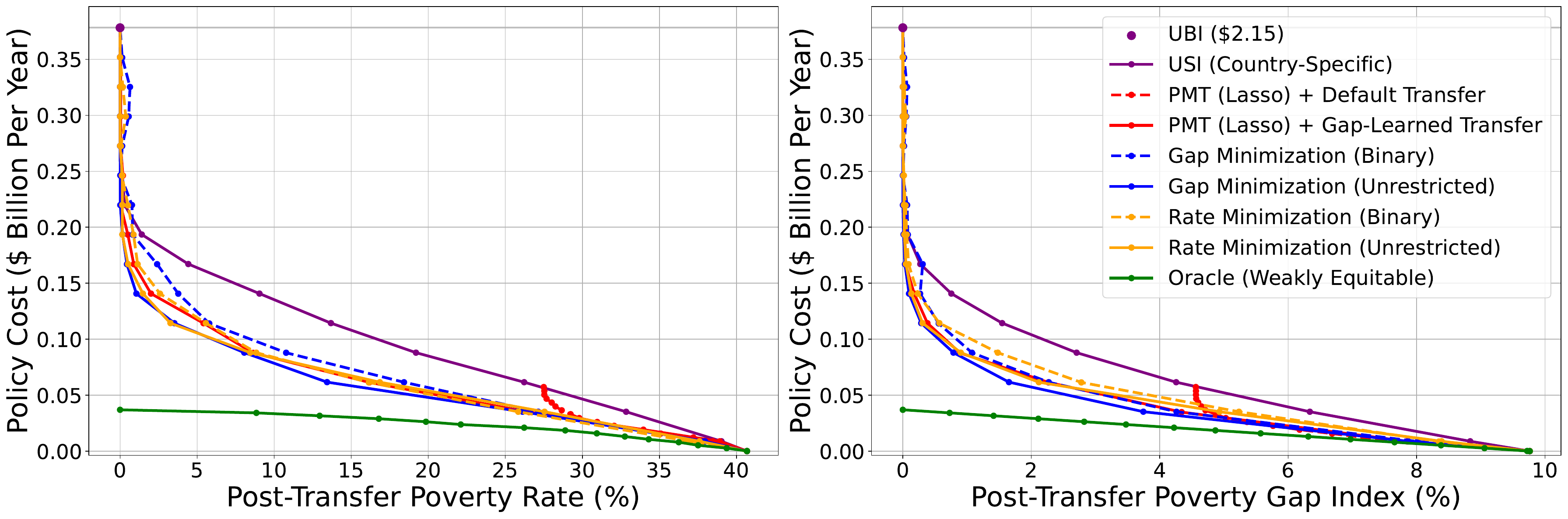}
    \end{subfigure}
\end{figure}

\begin{figure}\ContinuedFloat
    \begin{subfigure}{\textwidth}
    \subcaption{Togo}
    \includegraphics[width=\textwidth]{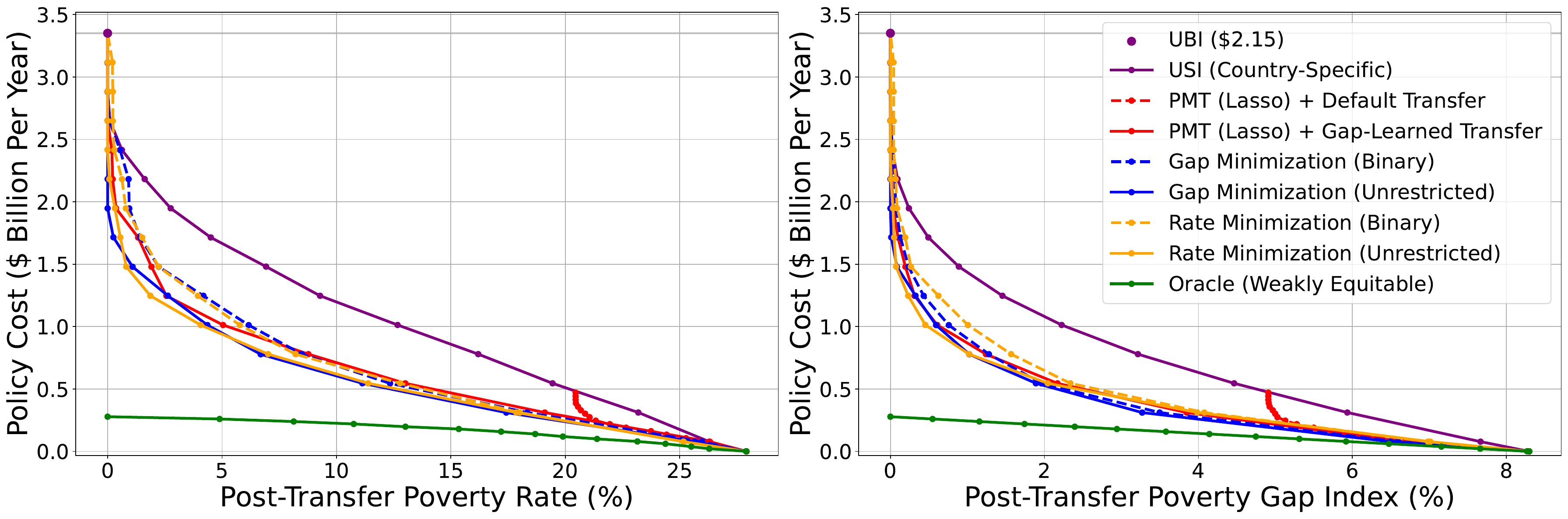}
    \end{subfigure}
    \bigskip
    \begin{subfigure}{\textwidth}
    \subcaption{Uganda}
    \includegraphics[width=\textwidth]{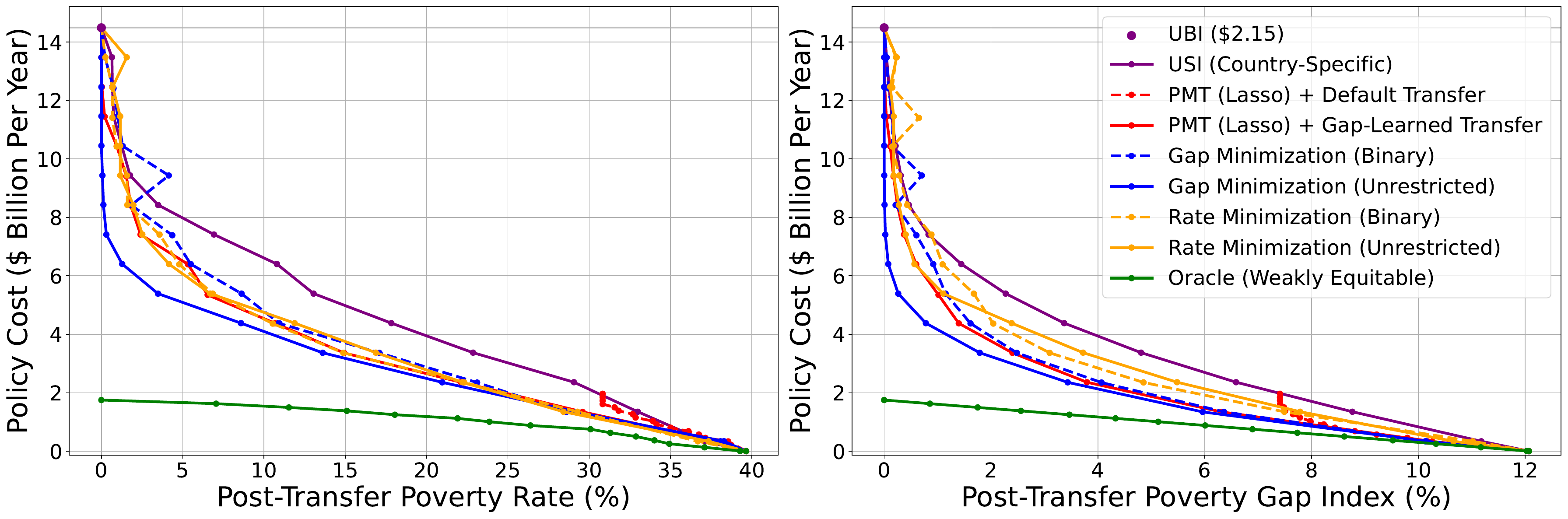}
    \end{subfigure}
    \bigskip
    \begin{subfigure}{\textwidth}
    \subcaption{Yemen}
    \includegraphics[width=\textwidth]{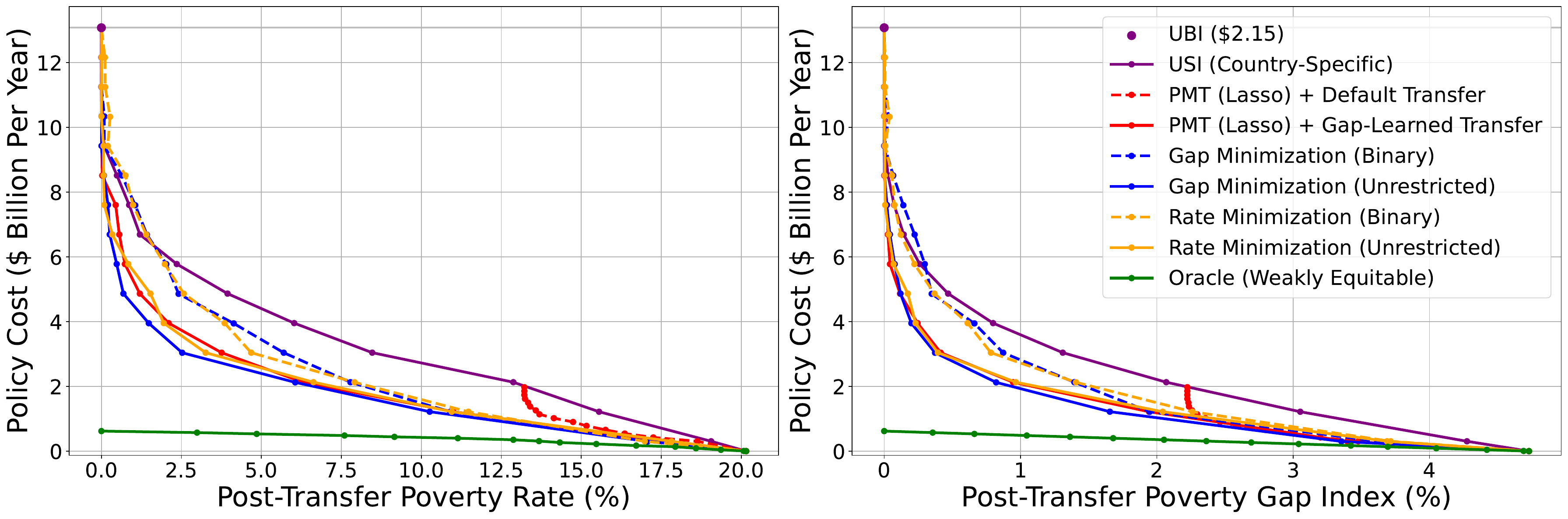}
    \end{subfigure}
\end{figure}

\begin{figure}\ContinuedFloat
    \begin{subfigure}{\textwidth}
    \subcaption{Zimbabwe}
    \includegraphics[width=\textwidth]{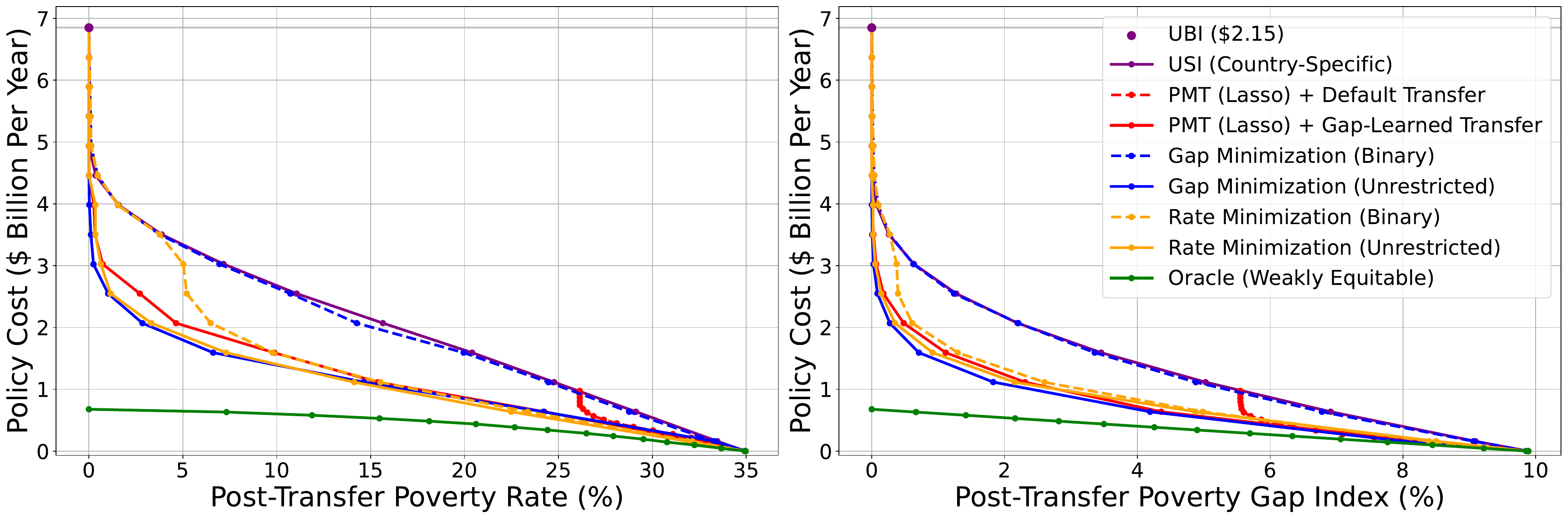}
    \end{subfigure}
    \vspace{1em}
    {\footnotesize This figure reports country-by-country results for gap minimization (binary/unrestricted), rate minimization (binary/unrestricted), PMTs learned using lasso and with default transfer sizes v.s. transfers learned to minimize the poverty objective, UBI at \$2.15 per person per day, and Universal Supplemental Income.}
\end{figure}

\begin{figure}
\caption{Comparison of Rate Minimization, Gap Minimization with Restricted Feature Set}
\label{fig:rate_gap_fewpredictors}
\includegraphics[width=\textwidth]{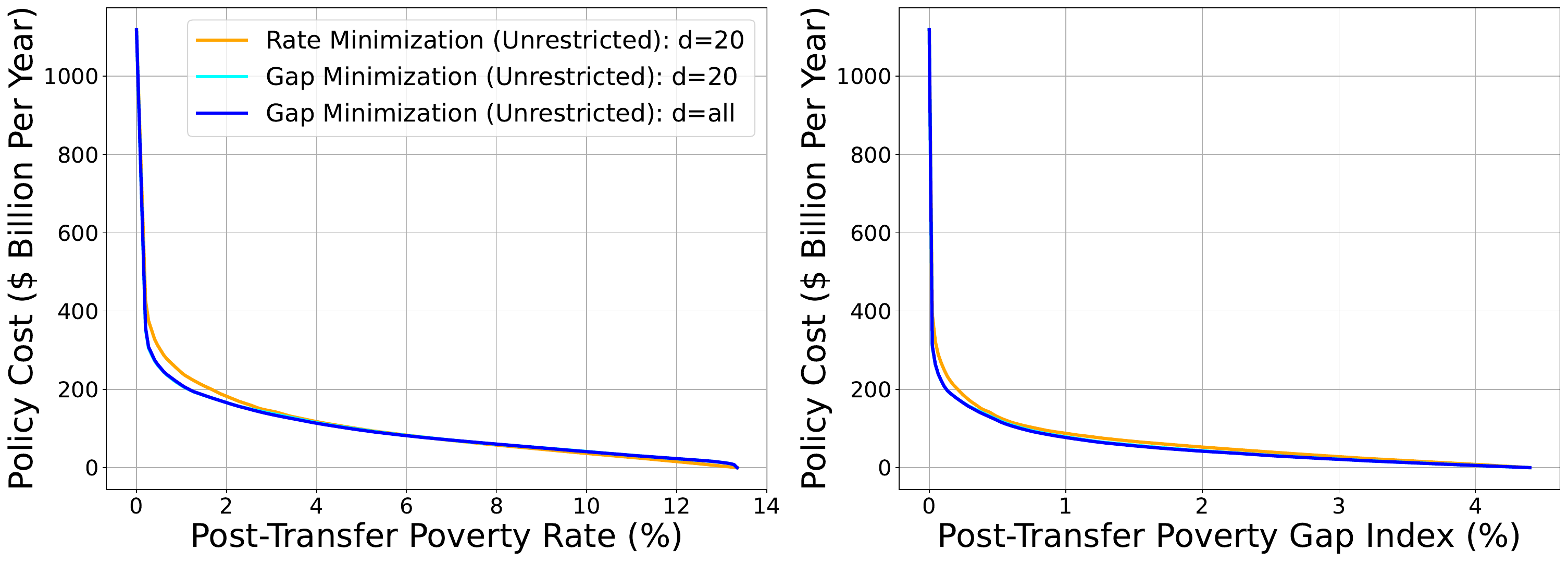}
{\footnotesize This figure reports the costs, in billions of nominal 2023 USD annually, of the indicated optimal transfer policies learned for the full sample of countries. We compare the performance of rate minimization with the restricted feature set $(d=20)$, gap minimization with the restricted feature set $(d=20)$, and gap minimization with the full feature set. We find that the policy costs of gap minimization with the full and restricted feature sets coincide.}
\end{figure}

\begin{table}
\caption{Cost ratios when using official exchange rates}
\label{tab:official_exchange_rates}
{
\centering
\begin{tabular}{lr}
\toprule
Country & Cost Multiplier Using Official FX Rate \\
\midrule
Sudan & 3.14 \\
Angola & 1.66 \\
Yemen, Rep. & 1.32 \\
Burundi & 1.16 \\
Syrian Arab Republic & 1.04 \\
Iran, Islamic Rep. & 1.03 \\
Nepal & 1.02 \\
Australia & 1.02 \\
Nauru & 1.02 \\
Haiti & 1.01 \\
India & 0.99 \\
Bangladesh & 0.98 \\
Uganda & 0.98 \\
Egypt, Arab Rep. & 0.83 \\
South Sudan & 0.75 \\
\bottomrule
\end{tabular} \\
}
\vspace{1em}
{\footnotesize This table lists how much country-level costs would change using official, rather than effective, exchange rate estimates. Only countries with usable official exchange rate data available, and where switching rates changes the cost by at least 1\%, are included.}

\end{table}

\end{document}